\documentclass[a4paper,11pt]{article}
\pdfoutput=1 % if your are submitting a pdflatex (i.e. if you have
\usepackage{jheppub} % for details on the use of the package, please
\usepackage[T1]{fontenc} % if needed

\usepackage{epstopdf}
\usepackage{graphicx}
\usepackage{epsfig}
\usepackage{dcolumn}  
\usepackage{bm}    
\usepackage{caption}
\usepackage{subcaption}
\usepackage{amssymb} 
\usepackage{tikz}
\usetikzlibrary{arrows}
\usepackage{xcolor}
\usepackage{epstopdf}
\usepackage{graphicx}
\usepackage{epsfig}
\usepackage{amsmath,bm}
\usepackage{amsfonts}  
\usepackage{braket}
\usepackage{amsmath}  
\usepackage{slashed}  
\usepackage{enumitem}
\usepackage[mathscr]{euscript}
\usepackage{tabu}
\usepackage{epsfig}
\makeatletter
\g@addto@macro\bfseries{\boldmath}
\makeatother

\DeclareMathOperator{\csch}{csch}

\newcommand{\bea}{\begin{eqnarray}}
\newcommand{\eea}{\end{eqnarray}}
\newcommand{\be}{\begin{equation}}
\newcommand{\ee}{\end{equation}}
\newcommand{\ba}{\begin{align}}
\newcommand{\ea}{\end{align}}

\def\nn{\nonumber}

\def\l1{{{1-loop}}}

\def\n1{\Bigg|_{n=1}}

\def\n{{(n)}}

\def\arccot{{\rm arccot}\;}

\usepackage[T1]{fontenc} % if needed
\usepackage{tikz}
\usepackage{amsmath,amssymb}
\usepackage{relsize}
\usepackage{latexsym}
\usepackage{leftidx}
\usepackage{xcolor}
\usepackage{diagbox}
\usepackage[T1]{fontenc}
\usepackage{array}
\usepackage{makecell}
\usepackage{csquotes}
\usepackage{tikz}
\usepackage{enumitem}
\usepackage{setspace}
\usepackage{multirow}
\usepackage{amsmath,amssymb}
\usepackage{relsize}
\usepackage{latexsym}
\usepackage{leftidx}
\usepackage{xcolor}
\usepackage{csquotes}
\usepackage{tikz}
\usepackage{enumitem}
\usetikzlibrary{decorations.markings}
\usetikzlibrary{decorations.pathmorphing}
\usetikzlibrary{decorations.markings}
\usetikzlibrary{decorations.pathmorphing}
\usepackage{pifont}
\usepackage{hyperref}

\newlength{\slength}
\newcommand{\ren}{R\'enyi\ }

\newcommand\om{\omega}

\newcommand{\bbgv}{Bogoliubov }

\newcommand{\ha}{\frac{1}{2}}

\usepackage{bookmark}

 \title{\textbf{\textsf{Precision tests of bulk entanglement:   $AdS_3$ vectors.
}}}
  \author{Rayirth Bhat$^{a}$, Justin R. David$^{a}$, Semanti Dutta$^{b}$
   }
\affiliation{\vspace{.1cm} $^{a}$Centre for High Energy Physics, \\ Indian Institute of Science,\\
C. V. Raman Avenue, Bangalore 560012, India.
\vspace{.1cm}\\
$^{b}$Department of Astrophysics and High Energy Physics, \\
S.N Bose National Centre for Basic Sciences, \\
Salt Lake, Kolkata  700106, India.
}
\emailAdd{rayirthbhat, justin@iisc.ac.in, semanti.dutta2010@gmail.com
}

\abstract{We consider  single-particle excitations of the massive Chern-Simons field of mass $M$  in $AdS_3$ and evaluate their contribution  at the first sub-leading   order in $G_N$ 
to the entanglement entropy across the Ryu-Takayanagi surface. 
Quantizing the Chern-Simons field in $AdS_3$, we evaluate the corrections to the holographic entanglement entropy using the Faulkner-Lewkowycz-Maldacena formula. The massive Chern-Simons field also obeys the equations of motion of a massive vector in $AdS_3$. The lowest-energy single-particle excitation of this field is dual to the primary operator of conformal dimension $M+1$ with spin one in the dual CFT; all other single-particle excitations are dual to its global descendants. We compare the entanglement entropy result from the FLM formula to the single-interval entanglement entropy in large-charge holographic CFT  obtained using the replica trick for the primary and its tower of holomorphic descendants. The two results agree precisely in the leading and sub-leading terms of the short interval expansion. We evaluate the contribution of the edge mode to the vacuum-subtracted entanglement and show that it vanishes, which is crucial for the FLM formula to agree with the CFT result. On taking the massless limit, the result coincides with the contribution of a $U(1)$ current to the single interval entanglement entropy. This is surprising since an earlier calculation in the literature reproduced the CFT result entirely from the edge $U(1)$ degrees of freedom on the RT surface.
}

\begin{document}
\maketitle
\flushbottom

\section{Introduction}

The Ryu-Takayanagi formula \cite{Ryu:2006bv} for holographic entanglement entropy has played an important role in our recent understanding of holography, the information paradox and emergent space-time, and introduced the ideas of quantum information in 
quantum gravity. The RT formula is classical, and it holds to the leading order in Newton's constant $G_N$. In \cite{Faulkner:2013ana}, Faulkner, Lewkowycz and Maldacena  proposed that the quantum corrections to the RT formula at  $G_N^0$ are given by 
\begin{eqnarray} \label{flm}
S_{EE}(A) = \frac{\rm {Area} ( \gamma_A) }{4G_N} + S_{\rm bulk}^{\rm EE} ( \Sigma_A) ,
\end{eqnarray}
$S_{EE}(A)$ is the entanglement entropy of the subregion $A$ in the boundary CFT. $\gamma_A$ is the minimal RT surface whose boundary is $A$. $\Sigma_A$ is the region which extends between $\gamma_A$ and $A$. Finally,  $S_{\rm bulk}^{\rm EE}$ is the entanglement entropy of all fields present in the bulk effective field theory. This formula resembles the generalised entropy of a geometry with a horizon, where now $A$ is the area of the horizon instead of the minimal surface, and $S_{\rm bulk}$ is replaced by the Von-Neumann entropy of all fields outside the black hole horizon \cite{Hawking:1971bv,Bekenstein:1973ur,Hawking:1975vcx}. The notion of the quantum extremal surface in which one performs an extremization over surfaces arose from this similarity. These concepts, as well as the generalizations of the FLM formula, have played a fundamental role in our recent understanding of the black hole information paradox  \cite{Engelhardt:2014gca,Dong:2017xht,Dong:2016eik,Faulkner:2017vdd}. 

Despite the generality of the notion of quantum extremal surface, as well as the association of entropy with any sub-region in quantum gravity, it is difficult to find situations where one can test these ideas. However, the FLM proposal, which is stated within the context of $AdS/CFT$, can be tested. This can be done 
by finding situations in which we can compare the result from the proposal to known results in the dual CFT. Such tests were initiated by \cite{Sugishita:2016iel} for $AdS_d$ with $d$ odd for scalar excitations in the bulk. In \cite{Belin:2018juv}, the vacuum-subtracted single interval entanglement entropy of the ground state of a minimally coupled scalar in $AdS_3$  evaluated using the FLM formula was shown to precisely agree with that evaluated from a primary field in large $c$ conformal field theory. Other tests involving mutual information and time-dependent quenches were done in \cite{Agon:2015ftl,Agon:2020fqs} \footnote{There is another approach to obtain quantum corrections to the RT formula, and that is to construct the background in the bulk which is dual to the replica surface on the boundary and evaluate its path integral.
Such an approach was taken for quantum corrections to the single interval entanglement entropy in $CFT_2$ on the torus, and 2 disjoint intervals in \cite{Barrella:2013wja,Chen:2013kpa,Datta:2013hba,Headrick:2015gba,Belin:2017nze}.}. In the work of \cite{Chowdhury:2024fpd}, the test done by \cite{Belin:2018juv} was generalised to single-particle excitations of the scalar as well as linear combinations of these excitations. Recently, such tests have been done in $AdS_d$ with $d>3$ for scalars, vectors and gravitons. In each of these situations, the result from the FLM formula was shown to agree with that evaluated in the CFT \cite{Colin-Ellerin:2024npf,Colin-Ellerin:2025dgq}. Such tests have been done to the leading order in the short-distance expansion of the entanglement interval, and they involve a detailed calculation of the back-reacted geometry and 
Bogoliubov coefficients, which relate single-particle excitations in the $AdS$ geometry to those of the hyperbolic black hole. They necessarily involve quantizing the fields in both these geometries. 
Therefore, these tests also yield several details about perturbative excitations in these geometries.

In this paper, we examine the case of vectors in $AdS_3$, which was not studied in \cite{Colin-Ellerin:2024npf}. However, Chern-Simons fields were studied in \cite{Belin:2019mlt}, where it was shown by methods that are different than developed in the original paper of \cite{Belin:2018juv}. This is because Chern-Simons theory in 3-dimensions is topological, and the evaluation of $S_{\rm bulk}$ is done using the edge $U(1)$ degrees of freedom on the `entanglement cut' or the RT surface. In fact, the methods of topological Chern-Simons theory were used to argue that the $S_{\rm bulk}$ can be evaluated using the $2n$ point function of a $U(1)$ current  
different from that of the boundary $U(1)$ current on a replica geometry. This mirrors the calculation on the boundary, and the results agree. 

In this paper, we would like to extend the methods developed in \cite{Belin:2018juv} and obtain the entanglement entropy for vector fields in the bulk. We consider massive vector fields in $AdS_3$.  As we will show, these fields are dual to operators with spin one and arbitrary conformal dimension and are a generalisation of operators of dimension $(1, 0)$ or $(0, 1)$ to which Chern-Simons fields are dual. This study also extends the analysis of vector fields in $AdS_d$ for $d>3$, carried out in \cite{Colin-Ellerin:2024npf}, to the case $d=3$. Our key reason to consider such fields is that giving mass to the vector breaks the 
topological nature of the  Chern-Simons theory, and therefore, we can approach the problem of studying their entanglement using the methods developed in \cite{Belin:2018juv}. The first step in the application of this method is the evaluation of the back-reacted metric and the correction to the Ryu-Takayanagi minimal area due to perturbative excitations. For the pure Chern-Simons theory, this vanishes since it does not couple to the metric. The massive Chern-Simons theory, on the other hand, couples to the metric and the methods of \cite{Belin:2018juv} can be followed. Furthermore, we take the vanishing mass limit and check whether our result agrees with that of \cite{Belin:2019mlt}. It also lays the groundwork for investigating gravitons and massless higher-spin fields in $AdS_3$, which are topological.

%Such fields break gauge symmetry and generalization of the methods developed in \cite{Belin:2018juv} 
%for scalar fields should 
%be applicable in this case.   
Massive Chern-Simons fields  obey the following  equations of motion 
\begin{eqnarray} \label{introcseq}
\epsilon^{\mu\alpha\beta} \partial_\alpha A_\beta = - M A^\mu, 
\end{eqnarray}
where $\epsilon^{\mu\nu\rho}$ is the Levi-Civita symbol in $AdS_3$ and $M>0$. It can be shown \cite{Datta:2011za} that the first-order massive Chern-Simons equations imply that the vector field obeys the second-order equations 
\begin{eqnarray}
(\nabla^2  - M^2 + 2 ) A_\mu = 0, 
\end{eqnarray}
together with the condition 
\begin{equation} \label{locond}
\nabla^\mu A_\mu =0,
\end{equation}
where $\nabla^2$ is the vector Laplacian in $AdS_3$. The massless limit is $M=0$, the shift of $2$ is due to the curvature of $AdS_3$. We have set the radius of $AdS_3$ to be unity. Therefore, studying the linearised equation (\ref{introcseq}) solves 
both the second-order massive vector equation, together with the Lorentz gauge condition. 

The massive vector field is dual to an operator of dimension $1+M$ and spin $1$ in the CFT. As we will see, the massive Chern-Simons field has only one degree of freedom. Let the single-particle states, which result from quantizing this field, be represented by
the states $a_{m, n}^\dagger |0\rangle $, where the vacuum refers to empty $AdS_3$. Then, the dictionary that relates these states to those in the CFT is  given by 
\begin{eqnarray}
a_{m, n}^\dagger|0\rangle_{\rm bulk} \leftrightarrow  (L_{-1} )^{n_1}  (  \bar{L}_{-1})^{n_2} 
\Big| 1 + \frac{M}{2}, \frac{M}{2}  \Big\rangle, 
\\ \nonumber
{\rm with} \qquad  2n +  |m| = n_1 +n_2 +1, \qquad m = n_1 - n_2 +1.
\end{eqnarray}
Here $L_{-1}, \bar L_{-1}$ are raising and lowering operators of the $SL(2, \mathbb{R})$'s of the CFT, $n_1, n_2$ take values from $0, 1, \cdots $. We therefore have the conditions,
\begin{eqnarray}
{\rm for }\; n = 0 , \;  m = 1, 2, \cdots, \qquad {\rm and \; for\; }
 n = 1, 2, \cdots, \qquad m \in \mathbb{Z}.
\end{eqnarray}
Let us first discuss the results for the single-particle ground state $a_{1, 0}^\dagger |0\rangle$. On considering this excitation, the stress tensor picks up an expectation value, due to which the
$AdS_3$ geometry back-reacts. This corrects the area term in the FLM formula  (\ref{flm}). We then proceed to evaluate $S_{\rm bulk}$. For this, we map the reduced density matrix in $AdS_3$ to 
the Rindler BTZ geometry. To complete the evaluation of the entanglement entropy, we obtain the Bogoliubov coefficients relating the state $a_{1, 0}^\dagger |0\rangle$ in $AdS_3$ to states in the Rindler BTZ. The Bogoliubov coefficients admit a short interval expansion, which enables the evaluation of the bulk entanglement entropy. Summing up the contribution due to the corrected minimal area and the contribution due to $S_{\rm bulk}$ 
we obtain 
\begin{eqnarray} \label{eeresult}
S(A) = 2 ( M+1) ( 1 - \pi x \cot \pi x ) - \frac{\Gamma( \frac{3}{2}) \Gamma( 2M +3) }{
\Gamma(  2M + \frac{7}{2} ) } ( \pi x )^{4( M + 1) } + \cdots,
\end{eqnarray}
where $2\pi x$ is the size of the interval. This result precisely agrees with the leading terms in the short-distance correction to the single interval entanglement entropy in a large $c$ CFT excited by a primary of weight $M+1$ and spin one. 

The  Chern-Simons equations (\ref{introcseq}) can be used to show that the  Proca equation for the vector field 
\begin{equation}
\nabla^\mu F_{\mu\nu} = M^2 A_\nu
\end{equation}
also holds. This then implies the all Chern-Simons fields satisfy the  Gauss law constraint  $\nabla^i F_{i 0} = M^2 A_0$, where $i $ refers to the spatial directions. Recall that the Lorentz gauge condition (\ref{locond}) also holds for these fields, which is needed for the consistency of the Proca equations. It has been shown in \cite{Blommaert:2018rsf} that a vector field which obeys the Proca equations of motion admits edge states due to the non-factorization of the Hilbert space. For the massless vector field $AdS_d$ with $d>3$, these edge states associated with the non-factorization of the Hilbert spaces in sub-regions were studied in \cite{Colin-Ellerin:2024npf} and were shown not to contribute to $S_{\rm bulk}^{EE}(\Sigma_A) $. We perform the analysis of these edge states for the massive Chern-Simons fields and show that, 
in this case too, the edge states do not contribute to entanglement. This added check makes the agreement of the FLM formula with the CFT complete \footnote{We thank the referee for requesting us to perform the analysis of the edge modes. } 

On taking the massless limit  $M\rightarrow 0$ and retaining the leading corrections  to order $ (\pi x )^4$ we obtain 
\begin{eqnarray}
S(A) = \frac{1}{6} ( 2\pi x)^2 - \frac{11}{360} ( 2 \pi x)^4 + \cdots.
\end{eqnarray}
Note that we can trust the expression only to $O(x^4)$, we can compare this result to the exact expression for the vacuum-subtracted entanglement entropy of the state excited by the holomorphic $U(1)$ current in the large $c$ CFT 
\cite{Berganza:2011mh,PhysRevLett.110.115701,Calabrese_2014,Ruggiero:2016khg,Belin:2019mlt}
\begin{eqnarray} \label{u1cftres}
S(A)^{CFT}_{U(1) } = - 2 \left[ \log ( 2 \sin \pi x) + \Psi \Big( \frac{1}{2 \sin \pi x} \Big)  + \sin \pi x \right].
\end{eqnarray}
Here $\Psi$ is the di-gamma function, expanding the above expression in small $x$, we can see that the expression obtained in (\ref{eeresult}) precisely agrees. We have performed the calculation in the massive Chern-Simons theory for which the gauge symmetry is broken, and it is non-trivial that the result 
agrees in the $M\rightarrow 0$ limit when the gauge symmetry is restored. We discuss this more in section \ref{conclus}. 

Finally, we generalize the check of the FLM formula to descendants  in the class 
\begin{eqnarray}
a_{ m, 0 }^\dagger |0\rangle_{\rm bulk} \leftrightarrow (L_{-1})^{l} \Big| 1+ \frac{M}{2}, \frac{M}{2} \Big\rangle,
\end{eqnarray}
with $l = m -1$. Using the FLM formula, we obtain 
\begin{eqnarray}
S(A) &= &2 ( M+1 + l ) ( 1 - \pi x \cot\pi x)  \\ \nonumber
&& \qquad\qquad-   \frac{\Gamma( \frac{3}{2}) \Gamma( 2M +3) }{
\Gamma(  2M + \frac{7}{2} ) }  \times \Big (  \frac{ \Gamma ( M+2 + l ) }{ \Gamma( M+2) l!} \Big)^2 
( \pi x )^{4( M + 1) }. 
\end{eqnarray}
This, again, precisely agrees with the CFT result. 

The paper is organized as follows. In section \ref{cftsection}, we review the CFT results of \cite{Chowdhury:2021qja,Chowdhury:2024fpd} briefly and present the expressions that are of importance in this work. In section \ref{massivecs}, we quantize Massive Chern-Simons fields in $AdS_3$ and derive the dictionary between single-particle excited states in the bulk and the states in the CFT. Since the massive Chern-Simons theory is a constrained system, we use the method of Dirac Brackets to quantize the theory. In section \ref{correctminarea}, we evaluate the stress tensor corresponding to the single-particle states and evaluate the back-reacted geometry and the corrected 
minimal area. In section \ref{bulkent}, we first quantize the massive Chern-Simons theory in Rindler BTZ and evaluate the necessary Bogoliubov coefficients to relate single-particle excitations in $AdS_3$ to those in Rindler BTZ. Then, using these Bogoliubov coefficients, we evaluate the bulk entanglement entropy and demonstrate that the FLM formula reproduces the CFT result.  Section \ref{conclus} contains our conclusions. The appendices contain details of the calculations needed in the main text. In particular, appendix \ref{appenb} discusses the quantization of massive Chern-Simons theory in global $AdS_3$, and appendix \ref{appenc} contains the quantization in the BTZ geometry. Finally, appendix \ref{append} contains the evaluation of the Bogoliubov coefficients relating single-particle excitations in global $AdS_3$ to Rindler BTZ.

\section{Entanglement entropy of spin-1 excited states in CFT} \label{cftsection}

The evaluation of the single interval entanglement entropy for 
excited states in  CFT  has been developed in several works \cite{PhysRevLett.110.115701,Berganza:2011mh,Calabrese_2014,Ruggiero:2016khg,Sarosi:2016oks}. The formalism introduced in \cite{Chowdhury:2021qja,Chowdhury:2024fpd} makes it easy to evaluate the entanglement entropy for descendants as well as linear combinations of operators. 

We first briefly review the approach in \cite{Chowdhury:2021qja,Chowdhury:2024fpd} and summarise the result for a primary of dimensions $(h, \bar h)$. We consider the theory on the Euclidean cylinder and place the operator of interest at the infinite past. This creates the state $|{\cal O} \rangle$. 
We then wish to examine the change in entanglement entropy compared to the vacuum of an interval on the cylinder at the time $t=0$ of length $2\pi x$. The reduced density matrix for such a state is given by 
\begin{equation}
\rho_{\cal O} = {\rm Tr}_{[0, 2\pi x]} \Big(  |{\cal O} \rangle\langle {\cal O }| \Big).
\end{equation}
Then $S(\rho) $, the vacuum-subtracted single interval entanglement entropy of this state, is given by 
\begin{eqnarray}
S(\rho) =\lim_{n\rightarrow 1} S_n(\rho_{\cal O} ) , \qquad 
S_n(\rho_{\cal O}) = \frac{1}{1-n} \log \Big( \frac{{\rm Tr} \rho_{\cal O}^n }{ {\rm Tr} \rho_{(0)}^n } \Big) , 
\end{eqnarray} 
$\rho_{(0)}$ refers to the density matrix without any operator insertions. Using the  path integral to implement the replica trick and  using conformal mappings, we can relate the ratio of density matrices to the $2n$-point function involving the operator ${\cal O}$ and its conjugate ${\cal O}^*$ on the uniformized plane
\begin{eqnarray} \label{2nptfn}
\frac{{\rm Tr} \rho_{\cal O}^n }{ {\rm Tr} \rho_{(0)}^n }  = 
\frac{1}{ \Big( \langle {\cal O}  |  {\cal O} \rangle \Big)^n }
\Big\langle \prod_{k=0}^{n-1} w\circ {\cal O} (w_k) 
\prod_{k'=0}^{n-1} \hat w  \circ {\cal O}^* (\hat w_{k '} )
\Big\rangle.
\end{eqnarray}
The expression  $w\circ {\cal O}$ refers to the action of the conformal transformation $w(z)$ on the operator ${\cal O}$, with
\begin{equation}
w(z) = \left( \frac{z-u}{ z-v} \right)^{\frac{1}{n} } , \qquad u = e^{2\pi i x}, \qquad v = 1.
\end{equation}
For a primary of weight $(h, 0)$, the action of the conformal transformation is given by 
\begin{eqnarray} \label{holmap}
w\circ {\cal O} (z) = \Big( \frac{\partial w}{\partial z} \Big)^h {\cal O} ( w(z) ). 
\end{eqnarray}
In (\ref{2nptfn}), we have labelled the positions of the operators after the conformal transformation as $w_k$. This is the position of the operator located at $t\rightarrow-\infty$ on the $k$-th 
branched cylinder mapped to the uniformized plane,  and is given by 
\begin{eqnarray}\label{wkloc}
w_k = e^{ \frac{2\pi i ( k +x) }{n} }  = \lim_{z\rightarrow 0_k} \Big( \frac{ z-u}{ z-v } \Big)^{\frac{1}{n} }.
\end{eqnarray}
Note that the map (\ref{holmap}) takes each copy of the cylinder to a wedge on the uniformized plane. Similarly the $\hat w(\hat z) $ is given by 
\begin{eqnarray} \label{aholmap}
\hat \omega (\hat z) =   \left( \frac{\frac{1}{\hat z}-u}{\frac{1}{\hat z} -v} \right)^{\frac{1}{n} }.
\end{eqnarray}
This takes the operator ${\cal O }^*$ placed at $t\rightarrow +\infty$ on the branched cylinder to a wedge on the plane. The position on the $k$-th wedge is given by 
\begin{eqnarray} \label{hatwkloc}
\hat \omega_k =  \lim_{\hat z \rightarrow \hat 0_k }  \left( \frac{\frac{1}{\hat z}-u}{\frac{1}{\hat z} -v} \right)^{\frac{1}{n} } = e^{\frac{2\pi i k }{n} }. 
\end{eqnarray}
From (\ref{wkloc}) and (\ref{hatwkloc}), we see that the operators on a given wedge are on the unit circle, separated by an 
arc distance of $2\pi x$. The norm  $\langle {\cal O}  |  {\cal O} \rangle$ used to normalize the correlator in (\ref{2nptfn}), can be obtained, by taking the $2$-point function of ${\cal O }$ and ${\cal O}^*$ on the plane using the maps in (\ref{holmap}), (\ref{aholmap}) but with $n =1$. 
 
The $2n$-point functions of arbitrary operators in a CFT are generally unknown. However, a short interval expansion can be set up. The leading contribution to the short-distance expansion is given by factorizing the $2n$-point functions as products of $n$ 2-point functions on the same wedge. The sub-leading contribution is given by all possible  factorizations of the the $2n$-point function as a $4$-point function together with products of $n-2$, 2-point functions on the same wedge. For more details of the replica method, the maps and  the resulting calculations, please see
\cite{Chowdhury:2021qja,Chowdhury:2024fpd}.
 
Adding up the leading and sub-leading contributions systematically, the result for the single interval entanglement entropy for a primary of weight $(h, \bar h)$ is given by 
\begin{eqnarray} \label{gscft}
S ( \rho_{| h, \bar h\rangle } )   = 2( h + \bar h ) ( 1- \pi x \cot \pi x ) - 
\frac{\Gamma( \frac{3}{2} ) \Gamma \big( 2 (h + \bar h )  +1 \big) }{\Gamma\big(  2 (h +  \bar h) + \frac{3}{2} \big) } (\pi x)^{ 4(h + \bar h) } + \cdots.
\end{eqnarray}
To obtain this result, the sub-leading term is evaluated using the generalized free field expression for the $4$-point function, which is valid in a large $c$ CFT. In a generalized free field theory, the leading operator, which contributes to the exchange in the $4$-point function, is the composite operator ${\cal O}_p = {\cal O }^2$ of dimension $2h + 2\bar h$ and the OPE coefficients for this exchange are given by 
\begin{eqnarray}
C_{{\cal O }{\cal O} {\cal O}_p} C^{{\cal O}_p}_{\; {\cal O}{\cal O} } = 2.
\end{eqnarray}
The vacuum subtraction removes the contribution from the identity, its descendants, and the stress tensor block, which are suppressed since $h, \bar h << c$. Please see \cite{Chowdhury:2024fpd}, in particular the analysis leading up to equation (2.51) in which the result for the single interval entanglement entropy of the primary $|h, h\rangle$ is given. This analysis is easily generalized for the case in which left and right weights do not coincide. 
 
We can also calculate the single interval entanglement entropy of the global descendants $ (L_{-1})^l | h, \bar h \rangle $ Using the results in \cite{Chowdhury:2024fpd} \footnote{Equation (2.63) of 
\cite{Chowdhury:2024fpd} contains the result of $S(\rho_{(L_{-1})^l|h,  h \rangle} )$, which can be easily generalized for the situation in which holomorphic and anti-holomorphic weights do not coincide. }, we obtain 
\begin{eqnarray} \label{maincftres}
S ( \rho_{(L_{-1})^l  |h, \bar h \rangle} ) &=& 2 ( h + \bar h + l ) ( 1 - \pi x \cot \pi x )  \\ \nonumber
&& - 
\frac{ \Gamma( \frac{3}{2} ) \Gamma \big( 2 ( h +  \bar h )+ 1\big ) }
{ \Gamma \big( 2 (h + \bar h )+ \frac{3}{2} \big) }
\times \Big( \frac{ \Gamma ( 2h + l ) }{ \Gamma( 2h ) l !} \Big)^2 ( \pi x ) ^{ 4 ( h + \bar h ) }+ \cdots. 
\end{eqnarray}

\section{Massive Chern-Simons  theory in $AdS_3$} \label{massivecs}

In this section, we will use the Dirac quantization procedure to canonically quantize the massive Chern-Simons theory in $AdS_3$.  We solve the wave equations and study the spectrum of excitation of the massive Chern-Simons field in $AdS_3$.  We then obtain the quantum numbers of these excitations under the isometries of $AdS_3$. This enables us to map the excitations in the bulk 
to that of the dual CFT. 

\subsection{Dirac quantization of massive Chern-Simons theory} \label{diracquant}

The action of the massive Chern-Simons theory in $AdS_3$ is given by 
\begin{eqnarray}
S =   \int d^3 x {\cal L }  =
- \int d^3 x \sqrt{-g} \left( \epsilon^{\mu \nu \rho} A_\mu \partial_\nu A_\rho + M A_\mu A^\mu \right). 
\end{eqnarray}
Here we have chosen to work with $M>0$, $\epsilon$ is the Levi-Civita symbol in curved space and is defined by 
\begin{eqnarray}
\epsilon^{\mu \nu \rho} = \frac{\tilde \epsilon^{\mu \nu \rho} }{\sqrt{-g} }, \qquad \tilde \epsilon^{t\rho \varphi}  =1.
\end{eqnarray}
Here we have chosen $t, r,\varphi $ to be the time, radial and the angular directions in global $AdS_3$. The Chern-Simons term is independent of the metric. The global $AdS_3$  metric is given by
\begin{eqnarray}
ds^2 = - ( 1+ r^2) dt^2 + \frac{dr^2}{ 1 + r^2 } + r^2 d\varphi^2. 
\end{eqnarray}
The equations of motion from the Chern-Simons action are given by 
\begin{eqnarray} \label{cseq}
\epsilon^{\mu\alpha\beta} \partial_{\alpha} A_\beta = - M A^\mu. 
\end{eqnarray}
It is useful to write the equations out in components
\begin{eqnarray} \label{cherneq}
\frac{g_{tt} }{ \sqrt{-g} }\Big( \partial_r A_{\varphi} - \partial_{\varphi} A_r \Big) = - M A_t, 
\\ \nonumber
\frac{g_{rr}}{\sqrt{-g} } \Big( \partial_{\varphi} A_t - \partial_t A_{\varphi} \Big) = - M A_r, 
\\ \nonumber
\frac{g_{\varphi\varphi} }{\sqrt{-g} } \Big( \partial_t A_r - \partial_r A_t \Big) = 
- M A_{\varphi }.
\end{eqnarray}
It is clear from the equation in (\ref{cseq}) that we have the transverse constraint
\begin{eqnarray} \label{transc}
\nabla_\mu A^\mu = 0. 
\end{eqnarray}
Here $\nabla_\mu$ refers to the covariant derivative. Note that the massive Chern-Simons theory is not invariant under $U(1)$ gauge transformations. Gauge symmetry is restored in the $M\rightarrow 0$ limit. 

The first step to canonically quantize this system is to determine the canonical momenta, which are given by 
\begin{eqnarray} \label{csmomenta}
\Pi^t &=&  \frac{\partial {\cal L} }{\partial ( \partial_t A_t ) } = 0, \\ \nonumber
\Pi^r &=&  \frac{\partial {\cal L} }{\partial ( \partial_t A_r) }  = -A_{\varphi}, \\ \nonumber
\Pi^\varphi &=&  \frac{\partial {\cal L} }{\partial ( \partial_t A_\varphi ) } = A_{r}.
\end{eqnarray}
The first constraint on the momentum and the relations between the coordinates and the momenta imply that the system is constrained; these are the primary constraints since they hold without the equations of motion. The equal-time Poisson bracket of this system is given by 
\begin{eqnarray} 
\{ A_t (t, r, \varphi), \Pi^t (t, r'\varphi') \} = \delta ( r-r') \delta ( \varphi-\varphi') , \\ \nonumber
\{ A_r (t, r, \varphi), \Pi^r (t, r'\varphi') \} = \delta ( r-r') \delta ( \varphi-\varphi') , \\ \nonumber
\{ A_\varphi (t, r, \varphi), \Pi^\varphi (t, r'\varphi') \} = \delta ( r-r') \delta ( \varphi-\varphi'). 
\end{eqnarray}
The  n\"{a}ive  Hamiltonian is obtained by the Legendre transform and is given by 
\begin{eqnarray}
{\cal H}_n = \int d^2 x \left( - 2 A_t ( \partial_r \Pi^r + \partial_\varphi \Pi^\varphi) + M \sqrt{-g} A_\mu A^\mu 
\right).
\end{eqnarray}
But since the system has primary constraints, the Hamiltonian is not unique since we can add the primary constraints to it. We consider the following Hamiltonian, which coincides with the  n\"{a}ive Hamiltonian when the constraints are imposed
\begin{eqnarray}
{\cal H} &=& \int d^2 x \left( - 2 A_t ( \partial_r \Pi^r + \partial_\varphi \Pi^\varphi )+ M \sqrt{-g} A_\mu A^\mu 
\right)  \\ \nonumber
 && + \int d^2x \Big[  u_t \Pi^t  + u_r ( \Pi^r + A_\phi) + u_\varphi( \Pi^\varphi - A_r)  \Big].
\end{eqnarray}
We have introduced the Lagrange multipliers $u_t, u_r, u_\varphi$, which can be determined by demanding that Hamilton's equations of motion coincide with those obtained from the Lagrangian. 

Let us now consider each of the fields and obtain their evolution with the Hamiltonian. Consider first 
\begin{eqnarray}
\{A_t, {\cal H}  \} = \partial_t A_t =  u_t.
\end{eqnarray}
Therefore, we must have 
\begin{eqnarray}
u_t = -g_{tt} \left( \frac{1}{ \sqrt{-g} } \partial_r ( \sqrt{-g} g^{rr} A_r )  +  
g^{\varphi \varphi}\partial_{\varphi } A_\varphi 
\right). 
\end{eqnarray}
This ensures the transverse field constraint in (\ref{transc}) is satisfied. Let us look at the evolution of the momentum $\Pi^t$,
\begin{align} \label{secondary}
\{ \Pi^t, {\cal H} \} & =  \partial_t \Pi^t=0, \\ \nonumber
&= 2 ( \partial_r \Pi^r + \partial_\varphi \Pi^\varphi) - 2 M  \sqrt{g} g^{tt} A_t. 
\end{align}
This is a secondary constraint since it needs to be held so that the primary constraint $\Pi^t =0$ remains true under time evolution. The secondary constraint coincides with the first Chern-Simons equations of motion in (\ref{cherneq}) when we use the equations in (\ref{csmomenta}) relating the co-ordinates and the momenta. Now, let us examine the spatial directions. 
\begin{eqnarray}
\{ A_r, {\cal H} \} &=& \partial_t A_r , \\ \nonumber
&=& 2 \partial_r A_t + u_r. 
\end{eqnarray}
Requiring this equation to agree with the last equation of motion in (\ref{cherneq}), we obtain 
\begin{eqnarray} \label{defur}
u_r = -\partial_r A_t - M g^{\varphi\varphi}\sqrt{-g} A_{\varphi}.
\end{eqnarray}
Considering the evolution of the corresponding momentum, we obtain, 
\begin{eqnarray}
\{ \Pi^r, {\cal H} \} &=& \partial_t \Pi^r , \\ \nonumber
&=& - 2 M \sqrt{-g} A_r g^{rr} + u_\varphi.
\end{eqnarray}
This results in 
\begin{eqnarray} \label{defuphi}
u_\varphi =  M \sqrt{-g} g^{rr} A_r - \partial_{\varphi} A_t . 
\end{eqnarray}
To ensure that the 2nd equation in (\ref{cherneq}) is satisfied, let us examine the fields in the angular directions.
\begin{eqnarray}
\{A_\varphi, {\cal H} \} &=&  \partial_t A_{\varphi} , \\ \nonumber
&=& 2 \partial_{\varphi} A_t + u_{\varphi }.
\end{eqnarray}
It is clear that substituting $u_{\varphi}$ from (\ref{defuphi}) ensures that the  2nd equation of motion in (\ref{cherneq}) holds.  It is clear that though $u_{\varphi}$ is overdetermined, it admits a consistent solution. Finally, we examine
\begin{eqnarray}
\{ \Pi^{\varphi}, {\cal H} \} &=& \partial_t \Pi^{\varphi}, \\ \nonumber
&= & - 2 M \sqrt{-g} g^{\varphi\varphi} A_\varphi - u_r. 
\end{eqnarray}
Again, $u_r$ from (\ref{defur}) ensures that the 3rd equation of motion is satisfied. It is easy to check that the $2$ constraints in (\ref{csmomenta}) do not generate further secondary constraints 
once the Lagrange multipliers are determined as in (\ref{defur}), (\ref{defuphi}). We therefore see that these constraints are invariant under time evolution
\begin{eqnarray}
\{ \Pi^r  + A_\varphi, {\cal H} \} = \{\Pi^\varphi - A_{r}, {\cal H} \} =0.
\end{eqnarray}
It can also be verified that the transverse condition (\ref{transc}) is preserved by the Hamiltonian evolution
\begin{eqnarray}
\{\nabla_\mu A^\mu, {\cal H} \} =0.
\end{eqnarray}
Now that we have found the primary and secondary constraints  (\ref{csmomenta}), (\ref{secondary}), we can find the Dirac bracket. Let us label the constraints as 
\begin{eqnarray}
C_1 &=& \Pi^ t , \\ \nonumber
C_2 &=&  \partial_r \Pi^r + \partial_\varphi\Pi^\varphi  -M \sqrt{-g} g^{tt} A_t,  \\ \nonumber
C_3 &=&  \Pi^r + A_{\varphi}, \\ \nonumber
C_4 &=&  \Pi^{\varphi}  -A_r.
\end{eqnarray}
The matrix that lists the Poisson bracket of the constraints is given by 
\begin{eqnarray}
{\cal C} =  \left( 
\begin{array}{cccc}
0 &  M \sqrt{-g} g^{tt}  & 0 & 0 \\ 
-M \sqrt{-g} g^{tt} & 0 & -\partial_\varphi & \partial_r  \\ 
0& -\partial_\varphi & 0 & 2 \\ 
0 & \partial_r & -2 & 0 
\end{array}
\right) \delta( r-r') \delta( \varphi - \varphi'). 
\end{eqnarray}
Here the matrix is defined by ${\cal C}_{ij} = \{ C_i, C_2 \}$, the $i, j$ subscript on ${\cal C}$ also refers to the positions. We would also need the inverse, which is given by 
\begin{eqnarray} \label{invcons}
{\cal C}^{-1} = \left( 
\begin{array}{cccc}
0 & -\frac{g_{tt} }{M \sqrt{-g} } &\frac{ g_{tt} \partial_r  }{2M\sqrt{-g} } & \frac{ g_{tt}  \partial_{\varphi}}{ 2M\sqrt{-g} }  \\ 
\frac{g_{tt}}{M \sqrt{-g} } & 0 & 0 & 0 \\
\frac{ g_{tt} \partial_r}{2M \sqrt{-g} } & 0 &  0 & - \frac{1}{2} \\
\frac{g_{tt} \partial_{\varphi}}{2M\sqrt{-g} } &0 & \frac{1}{2} & 0 
\end{array}
\right) \delta( r- r') \delta( \varphi - \varphi'). 
\end{eqnarray}
We can now use this to find the Dirac bracket between the fields. 
Since there are $4$ constraints, in a $6$ dimensional phase space, we expect that the phase space reduces to $2$ dimensions. It is convenient to choose $A_r, A_{\varphi}$ to be the independent coordinates in this reduced phase space. Therefore, we evaluate the Dirac bracket between these fields
\begin{eqnarray} \label{dbarap}
\{ A_{r}, A_{\varphi} \} _{\rm DB}=  \{ A_{r}, A_{\varphi} \} 
 - \{A_r , {\cal C}_i\} {\cal C}^{-1}_{ij} \{ {\cal C}_j , A_{\varphi} \}. 
\end{eqnarray}
Here, the repeated indices $i, j$ also imply integrations over the corresponding positions along with the sum over the indices.
Evaluating this, we obtain 
\begin{eqnarray} \label{araphdb}
&& \{ A_{r} (t, r, \varphi ) , A_{\varphi} (t, r', \varphi' \} _{\rm DB} = - \frac{1}{2} \delta( r-r') \delta ( \varphi - \varphi'), 
\\ \nonumber
&& \{ A_{r} (t, r, \varphi ) , A_{r} (t, r', \varphi' \} _{\rm DB}= 
\{ A_{\varphi} (t, r, \varphi ) , A_{\varphi} (t, r', \varphi' \} _{\rm DB}  =0.
\end{eqnarray}
We will promote this Dirac bracket to be the fundamental commutation relation between the quantum field and treat all other fields as dependent on the independent phase space variables $A_r, A_{\varphi}$. 

It is useful for later purposes to obtain the Dirac bracket between the variables $A_-, A_+$ defined as
\begin{eqnarray}
A_{\pm } = A_t \pm A_{\varphi}.
\end{eqnarray}
We can proceed either by using the definition of the Dirac bracket 
in terms of the constraint matrix as in  (\ref{dbarap}) or by writing $A_{\pm}$ in terms of $A_r, A_\varphi$ and then using the Dirac bracket in (\ref{araphdb}).  Both lead to the same results, here we proceed using the latter method. Consider the equal time commutation relation between $A_{-}$, 
\begin{eqnarray}
\{ A_-(t, r, \varphi) , A_-( t, r', \varphi' ) \}_{\rm{DB}}  &=& 
\{ A_t ( x) - A_\varphi(x) ,  A_t ( x') - A_\varphi(x')\}_{\rm{DB}},  
\end{eqnarray}
where $x, x'$ stand for 3 coordinates defined on the LHS of the equation. Substituting the expressions for $A_t, A_{\varphi}$ from the equations of motion (\ref{cherneq}), we obtain 
\begin{eqnarray} 
\{ A_-(t, r, \varphi) , A_-( t, r', \varphi' ) \}_{\rm{DB}}   
&=&  -  \left\{ \frac{1+r^2}{M r}\Big(  \partial_r A_\varphi (x) - 
\partial_\varphi A_r (x)  \Big) ,   A_\varphi( x') \right\}_{\rm DB} \\ \nonumber
& &  - \left\{ A_{\varphi} ( x) , 
\frac{1+r^{\prime 2} }{M r'}\Big(  \partial_{r'} A_\varphi (x') - 
\partial_\varphi A_r (x') \Big)   \right\}_{\rm DB}.
\end{eqnarray}
Here we have used the fact that Dirac brackets $\{ A_t(x), A_t(x')\}_{\rm DB}$ as well as $\{ A_\varphi(x), A_\varphi(x')\}_{\rm DB}$ vanish as can be shown using the definition of the Dirac bracket and the inverse of the constraint matrix in (\ref{invcons}). Now using the commutation relation in (\ref{araphdb}), we obtain
\begin{eqnarray}  \label{amamdb}
\{ A_-(t, r, \varphi) , A_-( t, r', \varphi' ) \}_{\rm{DB}}    = - \frac{1+r^2}{M r}\delta(r-r') \partial_\varphi \delta ( \varphi - \varphi').
\end{eqnarray}
Performing similar manipulations, we obtain 
\begin{eqnarray} \label{apapdb}
\{ A_+(t, r, \varphi) , A_+( t, r', \varphi' ) \}_{\rm{DB}}    =  \frac{1+r^2}{M r}\delta(r-r') \partial_\varphi \delta ( \varphi - \varphi'),
\end{eqnarray}
and 
\begin{eqnarray} \label{apamdb}
\{ A_+(t, r, \varphi) , A_-( t, r', \varphi' ) \}_{\rm{DB}}    =   0.
%\frac{1+r^2}{M r}\delta(r-r') \partial_\varphi \delta ( \varphi - \varphi')
\end{eqnarray}

\subsection{Construction of single-particle vector excitations} \label{conssingle}

The equations of motion for the massive Chern-Simons fields
(\ref{cherneq}) are coupled first-order equations. It is possible to decouple them in $AdS_3$ and obtain their solutions using the methods developed in \cite{Datta:2011za} for the BTZ background. The strategy is to consider the action of the vector Laplacian on components along the boundary $A_t, A_{\varphi}$ and obtain decoupled equations for the linear combinations $A_t \pm A_{\varphi}$. These equations can be solved in terms of hypergeometric functions. Once solved, we can use the first-order equations (\ref{cherneq}) to fix the constants of integration and obtain $A_r$. 

Let us begin with the action of the vector Laplacian on $A_t$. Using standard manipulations, we can write 
\begin{eqnarray} \label{mani1}
\nabla^2 A_t = \frac{1}{\sqrt{-g}}  \partial_\alpha ( \sqrt{-g} g^{\alpha \beta} \partial_\beta A_t) 
- \frac{1}{\sqrt{-g} } \partial_\alpha ( \sqrt{-g} g^{\alpha\beta} \Gamma_{\beta t}^\sigma) A_\sigma
- 2 \Gamma_{\alpha t }^\sigma  g^{\alpha\beta } \partial_\beta A_\sigma 
+ g^{\alpha\beta} \Gamma_{\alpha t}^\gamma \Gamma_{\beta \gamma}^\sigma A_\sigma.  \nonumber
\\
\end{eqnarray}
From the expressions for the Christoffel symbols for the global $AdS_3$ metric given in  (\ref{chrisglobal}), we see that the only components that contribute are 
\begin{eqnarray}
\nabla^2 A_t &=& \frac{1}{\sqrt{-g}}  \partial_\alpha ( \sqrt{-g} g^{\alpha \beta} \partial_\beta A_t) 
- \frac{1}{r} \partial_r ( r g^{rr} \Gamma^t_{rt} ) A_t 
-2 \Gamma^t_{rt}g^{rr} \partial_r A_t   \\ \nonumber
& & - 2\Gamma^r_{tt}g^{tt} \partial_t A_r 
+ \Gamma^t_{rt} g^{rr} \Gamma^t_{rt} A_t 
+ \Gamma^r_{tt} g^{tt} \Gamma^t_{tr} A_t.
\end{eqnarray}
Substituting the values of the Christoffel symbols from the appendix, we obtain 
\begin{eqnarray} \label{man1}
\nabla^2 A_t = \Box A_t  - 2 A_t - 2r ( \partial_r A_t - \partial_t A_r ), 
\end{eqnarray}
where $\Box$ refers to the scalar Laplacian
\begin{eqnarray}
\Box \psi = \frac{1}{\sqrt{-g} } \partial_\mu ( \sqrt{ -g} g^{\mu\nu} \partial_\nu \psi ). 
\end{eqnarray}
In (\ref{man1}), we can use the 3rd Chern-Simons equations of motion from (\ref{cherneq}) to arrive at
\begin{equation} \label{simpeq1}
\nabla^2 A_t = \Box A_t - 2 A_t  -2M A_\varphi.
\end{equation}
Using similar manipulations, we arrive at the equation 
\begin{equation}\label{simeq2}
\nabla^2 A_\varphi = \Box A_\varphi - 2 A_\varphi - 2M A_t.
\end{equation}
Equations (\ref{simpeq1}) and ( \ref{simeq2}) are the same equations obtained in \cite{Datta:2011za} for the massive Chern-Simons field in the rotating BTZ background, which is also locally $AdS_3$ \footnote{See equations (A.4), ( A.5) of \cite{Datta:2011za}. Note that subscripts $+, -$ of \cite{Datta:2011za} correspond to $t, \varphi$ in this paper when the angular momentum of the BTZ black hole is set to zero.}. The Chern-Simons equations (\ref{cseq}) imply that the vector field satisfies the wave equation. To see this, we can substitute for $A_\beta$ on the right hand side of (\ref{cseq}) to obtain 
\begin{eqnarray}
\nabla^2 A_\mu = ( M^2 - 2 ) A_\mu. 
\end{eqnarray}
To reduce the equation to the above form, we need to use the transverse condition (\ref{transc}), as well as the curvature of $AdS_3$ \cite{Datta:2011za}. Using the wave equation, together with the relations (\ref{simpeq1}) and (\ref{simeq2}), we obtain 
\begin{eqnarray}
\Box A_t - M^2 A_t -2 M A_\varphi =0, \\ \nonumber
\Box A_\varphi - M^2 A_\varphi - 2M A_t =0.
\end{eqnarray}
We can now decouple these equations by considering the linear combinations 
\begin{eqnarray}
A_+ = A_t+ A_{\varphi} , \qquad A_- = A_t - A_{\varphi},
\end{eqnarray}
which results in the equations
\begin{eqnarray} \label{2ndorder}
( \Box  -( M + 1)^2 +1 ) A_+ =0, \\ \nonumber
( \Box  -( M - 1)^2 +1 ) A_- =0.
\end{eqnarray}
Note that these are equations that involve the scalar Laplacian; we can expand the solutions in terms of modes. The details of this are provided in appendix \ref{appenb}.
\begin{eqnarray} \label{aminusr}
A_- &= &\sum_{n=1, m =-\infty }^\infty \left(  R_{-}^{(m, n )} a_{m, n} e^{ - i\Omega_{(m, n ) } t }
e^{i m \varphi} 
+  R_{-}^{(m, n ) *} a_{m, n}^\dagger e^{  i\Omega_{(m, n )}  t } e^{-i m \varphi} \right)   \\ \nonumber
&&  +\sum_{m = 1}^\infty 
\left(  R_{-}^{(m, 0 )} a_{m, n} e^{ - i\Omega_{(m, 0 ) } t }
e^{i m \varphi} 
+  R_{-}^{(m, 0 ) *} a_{m, n}^\dagger e^{  i\Omega_{(m, 0 ) } t } e^{-i m \varphi} \right), 
\end{eqnarray}
and 
{\small 
\begin{eqnarray}\label{aminusm} 
%A_ -^{(m, n)} &=&  R_-^{(m, n)} e^{- i \Omega_{m \,n} t} e^{ i m \varphi} \\ \nonumber
R_{-}^{(m, n)} &=& C_-^{(m,n)} r^{|m| } (1+ r^2)^{ \frac{\Omega_{(m, n)}}{2} } 
{}_2 F_1\Big( \frac{1}{2} ( 2 + \Omega_{(m, n)} + |m|  -M ), 
\frac{1}{2} ( \Omega_{(m,  n)} + |m|  +M ), 1+ |m|, -r^2 \Big)  \nonumber
\\ & & {\rm for }   \qquad n = 1, \cdots   \qquad m \in \mathbb{Z},  \\ 
\nonumber
& & {\rm for} \qquad n = 0,  \qquad m = 1, 2, 3, \cdots
\end{eqnarray}  }
with
\begin{eqnarray}
\Omega_{(m, n)} = 2 n + |m| +M. 
\end{eqnarray}
The constants $C_-^{(m,n)} $ has been determined in (\ref{finalcminus}). A similar expansion exists for the field $A_+$ and is given by
\begin{eqnarray} \label{aplusrm}
A_+ &=&\sum_{n=1, m =-\infty }^\infty \left(  R_{+}^{(m, n )} a_{m, n} e^{ - i\Omega_{(m, n ) } t }
e^{i m \varphi} +  R_{+}^{(m, n )*} a_{m, n}^\dagger e^{  i\Omega_{(m, n ) } t } e^{-i m \varphi} \right),
\end{eqnarray}
and
{\small 
\begin{eqnarray}\label{aplusm}
A_+^{(m ,n)} &=&  R_+^{(m, n)} e^{- i \Omega_{(m ,n)} t} e^{ i m \varphi}, \\ \nonumber
R_{+}^{(m, n)} &=& C_+^{(m, n)} r^{|m| }  (1+ r^2)^{ \frac{\Omega_{(m, n) }}{2} } 
{}_2 F_1\Big( \frac{1}{2} (   \Omega_{(m, n)} + |m|  -M ), 
\frac{1}{2} ( 2+ \Omega_{(m, n)} + |m| +M ), 1+ |m|, -r^2 \Big) , \\ \nonumber
&& n = 1, 2, \cdots, \qquad m \in \mathbb{Z}.
\end{eqnarray} }
The constants $C_+^{(m, n) }$ are  determined from $C_-^{(m, n )}$ using the relations in (\ref{goodratio}), (\ref{eqcoeff}). Given the mode expansions of $A_-, A_+$, we can use the Chern-Simons equation for $A_r$ to find its mode expansion. Note that as shown in the appendix, the mode expansion for $A_+, A_-$ and $A_r$ solve all the 3 first order Chern-Simons equations. We can now quantize the fields by promoting them to operators. Then, requiring that these fields satisfy the commutation relations that follow from the 
Dirac bracket in (\ref{araphdb}) results in the oscillator algebra 
\begin{eqnarray} \label{oscalgeb}
[a_{m, n}, a_{m', n'}^\dagger] = \delta_{m, m'} \delta_{n, n'}
\end{eqnarray}
This is demonstrated in appendix \ref{appenb}. The normalization constants $C_-^{(m, n ) }$ are in fact determined by demanding the above canonical normalized oscillator algebra. In this paper, we will be dealing with the class of states having quantum numbers $n=0, m =1, \cdots$. Let us write down the corresponding normalization for reference
\begin{eqnarray} \label{neqzerom}
C_-^{(m, 0)} = \sqrt{  \frac{1}{\pi} \frac{ \Gamma( M + m +1) }{ (m-1) ! \Gamma( M+1)  }  }, 
\quad\qquad n=0, m \geq 1 .
\end{eqnarray}
In table \ref{low lying states}, we list out the few low-lying wave functions of all the 3 vector components explicitly

\begin{table}[!h]
\resizebox{\textwidth}{!}{%
\centering
\begin{tabular}{|c|c|c|c|c|c|c|}
\hline
&  &  &  &  &  &\\
m & n & $R^{(m,n)_t}$  & $R^{(m,n)}_{\varphi} $  & $R^{(m,n)}_r$  &$L_{0} + \Bar{L}_{0}$ & $L_{0} - \Bar{L}_{0}$\\
&  &  &  &  &  &\\
\hline
&  &  &  &  &  &\\
1 & 0 & $\sqrt{\frac{M + 1}{\pi}}\frac{r}{2(1 + r^2)^\frac{M+1}{2}}$ & $ - \sqrt{\frac{M + 1}{\pi}}\frac{r}{2(1 + r^2)^\frac{M+1}{2}}$ & $ \sqrt{\frac{M + 1}{\pi}} \frac{i}{2 (1 + r^2)^{(M + 3)/2}}$ & $M + 1$ & $1$ \\
&  &  &  &  &  &\\
2 & 0 & $\sqrt{\frac{(M+2)(M+1)}{\pi}}\frac{r^2}{2(1 + r^2)^\frac{M+2}{2}}$ & $ - \sqrt{\frac{(M+2)(M+1)}{\pi}} \frac{r^2}{2(1 + r^2)^\frac{M+2}{2}}$ & $\sqrt{\frac{(M+2)(M+1)}{\pi}} \frac{ir}{2(1+ r^2)^{(M+4)/2}}$ & $M+2$ & $2$\\
&  &  &  &  &  &\\
3 & 0 & $\sqrt{\frac{(M+3)(M+2)(M+1)}{2 \pi}}\frac{r^3}{2(1 + r^2)^\frac{M+3}{2}}$ & $ - \sqrt{\frac{(M+3)(M+2)(M+1)}{2 \pi}} \frac{r^3}{2(1 + r^3)^\frac{M+3}{2}}$ & $\sqrt{\frac{(M+3)(M+2)(M+1)}{2 \pi}} \frac{i r^2}{2 (1+r^2)^{(M+5)/2}}$ & $M + 3$ & $3$ \\
&  &  &  &  &  &\\
0 & 1 & $\sqrt{\frac{M+1}{\pi M}}\frac{Mr^2 - 2}{(1 + r^2)^\frac{M+2}{2}}$ & $\sqrt{\frac{M+1}{\pi M}} \frac{- Mr^2}{(1 + r^2)^\frac{M+2}{2}}$ & $- \sqrt{\frac{M+1}{\pi M}} \frac{M+2}{2}$ $\frac{i r}{(1 + r^2)^{\frac{M+4}{2}}}$& $M + 2$ & $0$ \\
&  &  &  &  &  &\\
-1 & 1 & $\frac{1}{2 \sqrt{2 \pi M}} \frac{r(6 + 4M - M (1 + M) r^2)}{(1 + r^2)^\frac{M+3}{2}} $ & $\frac{1}{2 \sqrt{2 \pi M}} \frac{r(2 + M (1 + M) r^2)}{(1 + r^2)^\frac{M+3}{2}} $ & $\frac{1}{2 \sqrt{2 \pi M}} \frac{i \big( (M + 1)(M + 4)r^2 - 2 \big)}{(1 + r^2)^{\frac{M + 5}{2}}}$& $M + 3$ & $-1$ \\
&  &  &  &  &  &\\
-2 & 1 & $\frac{\sqrt{M + 2}}{2\sqrt{6 \pi M}}\frac{r^2(6 (2 + M) - M (1 + M) r^2)}{(1 + r^2)^\frac{M+4}{2}} $ & $\frac{\sqrt{M + 2}}{2\sqrt{6 \pi M}}\frac{r^2(6 + M (1 + M) r^2)}{(1 + r^2)^\frac{M+4}{2}} $ & $\frac{\sqrt{M + 2}}{2\sqrt{6 \pi M}} \frac{i r \big(6 - (M + 1)(M + 6) r^2 \big)}{(1 + r^2)^{\frac{M + 6}{2}}}$ & $M + 4$ & $-2$\\
&  &  &  &  &  &\\
1 & 1 & $- \frac{(M+1)}{\sqrt{\pi M}} \frac{r(Mr^2 - 3)}{2(1 + r^2)^\frac{M+3}{2}}$ & $\frac{(M+1)}{\sqrt{\pi M}} \frac{r(Mr^2 - 1)}{2(1 + r^2)^\frac{M+3}{2}}$ & $- \frac{(M+1)}{\sqrt{\pi M}} \frac{i ((M+2)r^2 - 1)}{2(1+r^2)^{(M+5)/2}}$ & $M + 3$ & $1$\\
&  &  &  &  &  &\\
2 & 1 & $\sqrt{\frac{M+2}{2\pi M}} \Big(\frac{M+1}{2}\Big) \frac{r^2(Mr^2 - 4)}{(1 + r^2)^\frac{M+4}{2}}$ & $- \sqrt{\frac{M+2}{2\pi M}} \Big(\frac{M+1}{2}\Big) \frac{r^2(Mr^2 - 2)}{(1 + r^2)^\frac{M+4}{2}}$ & $ - \sqrt{\frac{M+2}{2\pi M}} \Big(\frac{M+1}{2}\Big) \frac{i r \big((M+2)r^2 - 2 \big)}{(1+r^2)^{\frac{M+6}{2}}}$& $M + 4$ & $2$\\
&  &  &  &  &  &\\
1 & 2 & $\frac{M + 2}{4 \sqrt{\pi M}} \frac{r(10 - 2 (5 + 4 M) r^2 + M (1 + M) r^4)}{(1 + r^2)^\frac{M+5}{2}}$ & $- \frac{M + 2}{4 \sqrt{\pi M}} \frac{r(2 - 2 (1 + 2 M) r^2 + M (1 + M) r^4)}{(1 + r^2)^\frac{M+5}{2}}$ & $\frac{M + 2}{4 \sqrt{\pi M}} \frac{i \big( 2 - 2(2M + 7)r^2 + (M+1)(M+4)r^4 \big)}{(1+r^2)^{\frac{M + 7}{2}}}$& $M + 5$ & $1$ \\
&  &  &  &  &  &\\
\hline
\end{tabular}}
\caption{Explicit wave functions of single-particle states for various $(m,n)$, together with their quantum numbers under $L_0 \pm \bar L_0$.}
\label{low lying states}
\end{table}

\subsection*{Mapping bulk excitations to CFT states} \label{mapbulkcft}

Single-particle states of vector excitations are created by the 
action of the oscillators on the $AdS_3$ global vacuum 
\begin{eqnarray}
|\psi_{m, n } \rangle = a_{m, n}^\dagger |0\rangle.
\end{eqnarray}
Since $AdS_3$ admits $SL(2, R) \times SL(2, R)$ isometries, we can classify the states in terms of the eigenvalues of the Cartan generators  $L_0, \bar L_0$. The Killing vectors of these isometries are given by 
\begin{eqnarray} 
\label{iso gen in uvrho1} L_0 &=& i \partial_v, \\
\label{iso gen in uvrho2} L_{-1} &=& i \exp{(-i v)} \Bigg( \frac{\cosh{2 \rho}}{\sinh{2 \rho}} \partial_v - \frac{1}{\sinh{2 \rho}} \partial_u + \frac{i}{2} \partial_\rho  \Bigg), \\
\label{iso gen in uvrho3} L_1 &=& i \exp{(i v)} \Bigg( \frac{\cosh{2 \rho}}{\sinh{2 \rho}} \partial_v - \frac{1}{\sinh{2 \rho}} \partial_u - \frac{i}{2} \partial_\rho  \Bigg),
\end{eqnarray}
where $r = \sinh{\rho}$, $v = t - \varphi$, and $u = t + \varphi$. These generate the left-moving isometries of $AdS_3$, while the right-moving generators follow from $u \leftrightarrow v$. Note that we have interchanged what we call $L_a$ and $\bar L_a$ from that in \cite{Chowdhury:2024fpd} for our convenience. On vector fields, these isometries act via the Lie derivative corresponding to the Killing vector $\xi^\mu$.
\begin{eqnarray}
{\cal L}_\xi A_\mu = \xi^\nu \partial_\nu A_\mu + A_\nu \partial_\mu \xi^\nu,
\end{eqnarray}
and $\xi^\mu$ can be read out from (\ref{iso gen in uvrho1} - \ref{iso gen in uvrho3}). A similar equation exists for the generators $ {\cal L}_{\bar \xi}$, where $\bar \xi_\mu$ are the Killing vectors for the right-moving $SL(2, R) $. We will also label the Lie derivative by $\{ -1, 0, 1\}$ corresponding to the Killing vectors (\ref{iso gen in uvrho1} - \ref{iso gen in uvrho3}). From the action of ${\cal L}_0, \bar {\cal L}_0$ it is clear that the wave functions $A^{(m, n)}_\pm $ in (\ref{aminusm}), (\ref{aplusm}) and $A_r^{(m, n )}$ are such that 
\begin{eqnarray} \label{waveqn}
( {\cal L}_0 + \bar {\cal L}_0) A_\mu^{(m, n ) }  &=& \Omega_{(m, n ) } A_\mu^{(m, n ) }  = 
(2n +|m|  +M ) A_\mu^{(m, n ) } , \\ \nonumber
({\cal L}_0 - \bar {\cal L}_0) A_\mu^{(m, n ) } &=& m  A_\mu^{(m, n ) }, \\  \nonumber
{\rm with} & & n = 0, m = 1, 2, \cdots, \\ \nonumber
 {\rm and \;for }  && n = 1, 2, 3, \cdots , \qquad m \in \mathbb{Z}.
\end{eqnarray}
This is because these Lie derivatives act as derivatives in $t, \varphi$ respectively. Therefore  the states in the bulk $|\psi_{m, n }\rangle$ created by the corresponding oscillators have quantum numbers
\begin{eqnarray} \label{statqn}
L_0  |\psi_{m, n }\rangle  &=&  \frac{1}{2} (2n +|m|  +m  +M )  |\psi_{m, n }\rangle,  \\ \nonumber
\bar L_0  |\psi_{m, n }\rangle  &=& \frac{1}{2} (2n +|m|  -m  +M ) |\psi_{m, n }\rangle.
\end{eqnarray}
To obtain the map of these states to that in the $CFT$, let us first identify the state corresponding to the primary in the CFT. This must have the lowest energy and smallest possible angular momentum. It is clear from (\ref{waveqn}), this must be the state $|\psi_{1, 0} \rangle$. The corresponding vector-valued wave functions can be read out from table \ref{low lying states} and 
are given by
\begin{eqnarray} \label{Amu 10}
A_t^{(1,0)} &=& \sqrt{\frac{M+1}{\pi}} \frac{r}{2 (1 + r^2)^{\frac{M + 1}{2}}} e^{-i (M + 1)t} e^{i \varphi}, \\
{A_\varphi}^{(1,0)} &=& - \sqrt{\frac{M+1}{\pi}} \frac{r}{2 (1 + r^2)^{\frac{M + 1}{2}}} e^{-i (M + 1)t} e^{i \varphi}, \\
{A_r}^{(1,0)} &=& \sqrt{\frac{M+1}{\pi}} \frac{i}{2 (1 + r^2)^{\frac{M+3}{2}}} e^{-i (M + 1)t} e^{i \varphi}.
\end{eqnarray}
The behaviour of this state under the action of the lowering operators can be explicitly computed, and results in 
\begin{eqnarray}
{\cal L}_1 A_\mu^{(1,0)} = 0, \qquad \bar {\cal L}_1 A_\mu^{(1,0)} = 0, \qquad \mu = t, r, \varphi.
\end{eqnarray}
Since the isometries of $AdS_3$ map to the global $SL(2, R)\times SL(2, R)$ symmetries of the CFT, we can identify this state as the primary. From (\ref{statqn}), we can read out the quantum numbers of the primary in the CFT and identify
\begin{eqnarray} \label{cftsingleid}
|\psi_{1, 0} \rangle \leftrightarrow \Big| 1 + \frac{M}{2} , \frac{M}{2} \Big\rangle,
\end{eqnarray}
where the labels $|h, \bar h \rangle$ are the left-moving and right-moving weights of the CFT primary. Note that, as expected, the CFT primary dual to the vector has spin $1$ \footnote{ As we have mentioned earlier, we have chosen to work with $M>0$. Had we taken $M<0$, we would have obtained the lowest energy state to be dual to the primary $\Big| 1, 1+ \frac{|M|}{2}  \Big\rangle$. This can be seen by going through the same analysis with $M\rightarrow -M$ }.

Now that we have identified the quantum numbers of the 
lowest lying state, and the isometries of the $AdS_3$ with that of the CFT, the states $|\psi_{m, n } \rangle$ for arbitrary $(m, n )$ must be related to the primary by the action of the raising operators $(L_{-1})^{n_1} L_{-1}^{n_2} $ on $|\psi_{1, 0 }\rangle$. 
Let us verify that this is indeed the case for a few low-lying states. Again, by explicit computation, it can be shown that 
\begin{eqnarray}
{\cal L}_{-1} A_{\mu}^{(1,0)} = \sqrt{M+2} A_\mu^{(2, 0)} , \qquad \mu = t, r, \varphi.
\end{eqnarray}
The wave functions $A_\mu^{(2, 0)}$ are listed in table \ref{low lying states}. Therefore, we have 
\begin{eqnarray}
L_{-1} |\psi_{1, 0 } \rangle  = \sqrt{M+2} |\psi_{2, 0 } \rangle, 
\end{eqnarray}
which corresponds to the relation 
\begin{eqnarray}
L_{-1}\Big| 1 + \frac{M}{2} , \frac{M}{2} \Big\rangle =  \sqrt{M+2} \Big| 2 + \frac{M}{2} , \frac{M}{2} \Big\rangle. 
\end{eqnarray}
It is evident that $\sqrt{M+2}  = \sqrt{2h} $ with $h = \frac{M}{2} +1$ is precisely the additional factor that should occur if the states are unit normalized by the $SL(2, R)$ norm \footnote{ The norm of the level 1 descendant of a primary in the CFT is determined by the $SL(2, R)$ algebra and is given by $\langle h, \bar h |L_{-1} L_{1} |h, \bar h \rangle = 2h $}. Similarly, we find the relation 
\begin{eqnarray} \label{righex}
\bar{{\cal L}}_{-1} A_\mu^{(1,0)} = - \sqrt{M} A_\mu^{(0,1)}, \qquad \mu = t,r,\varphi,
\end{eqnarray}
which corresponds to the CFT relation 
\begin{eqnarray}
\bar L_{-1}\Big| 1 + \frac{M}{2} , \frac{M}{2} \Big\rangle =  -\sqrt{M} \Big| 1 + \frac{M}{2} ,  1+ \frac{M}{2} \Big\rangle. 
\end{eqnarray}
Note again the factor $\sqrt{M} = \sqrt{ 2\bar h} $ with $\bar h = \frac{M}{2}$ in (\ref{righex}) is consistent with that
in the  CFT. There are other relations one can find, for example, 
\begin{eqnarray}
{\cal L}_{-1} A_\mu^{(m,0)} = \sqrt{m(M + m + 1)} A_\mu^{(m+1, 0)} \;  ,  \qquad 
\bar{{\cal L} }_{-1} A_\mu^{(0,1)} = \sqrt{2} \sqrt{M + 1} \: A_\mu^{(-1,1)}, \nonumber \\
\end{eqnarray}
which corresponds to the CFT relations
\begin{eqnarray}
L_{-1}\Big| m + \frac{M}{2} , \frac{M}{2} \Big\rangle &=&  \sqrt{m(M + m + 1)} \Big| m+1  + \frac{M}{2} ,  1+ \frac{M}{2} \Big\rangle , 
\\
\nonumber
\bar{L}_{-1} \Big| 1 + \frac{M}{2} ,  1+ \frac{M}{2} \Big\rangle &=&  \sqrt{2} \sqrt{M + 1} 
\Big| 1 + \frac{M}{2} ,  2+ \frac{M}{2} \Big\rangle. 
\end{eqnarray}
These observations allow us to obtain the following dictionary of states between the single-particle bulk excitations and 
descendants of the primary of spin $1$ and conformal dimension $M +1$.
\begin{eqnarray} \label{dictdescn1}
|\psi_{m, n } \rangle = a_{m, n}^\dagger|0\rangle_{\rm bulk} \longleftrightarrow 
\Big| 1 + \frac{M}{2} +n_1 , \frac{M}{2}  +n_2 \Big\rangle, 
\\ \nonumber
{\rm with} \qquad  2n +  |m| = n_1 +n_2 +1, \qquad m = n_1 - n_2 +1, \\ \nonumber
{\rm for }\; n = 0 , \;  m = 1, 2, 3,  \cdots, \qquad {\rm and \; for\; }
 n = 1, 2, 3, \cdots, \qquad m \in \mathbb{Z}.
\end{eqnarray}
The $SL(2, R) \times SL(2, R)$ descendants in the CFT  are related to the primary  by 
\begin{eqnarray} \label{dictdescn}
\Big| 1 + \frac{M}{2} +n_1 , \frac{M}{2}  +n_2 \Big\rangle
= {\cal N}_{M, n_1, n_2} L_{-1}^{(n_1)} \bar L_{-1}^{(n_2)}  \Big| 1 + \frac{M}{2}  , \frac{M}{2} \Big\rangle,
\end{eqnarray}
where  the factor  ${\cal N}_{M, n_1, n_2}$ ensures that all states are unit normalized. 

\section{Corrections to minimal area} \label{correctminarea}

The single-particle excitations in the bulk carry stress tensor density. The expectation value of the stress tensor in such states produces a back reaction that deforms the $AdS_3$ geometry. 
In this section, we obtain the metric of the back-reacted geometry to the leading order in $G_N$ by solving the Einstein equations perturbatively in $G_N$. We then evaluate the leading corrections to the minimal area and obtain the corrections to the Ryu-Takayanagi entanglement entropy. In section \ref{bulkent}, we will show that these corrections, together with those from bulk entanglement entropy, precisely reproduce the result (\ref{maincftres}) obtained from the CFT.

\subsection{Back reacted geometry of single-particle states} \label{backreactsec}

We consider the following  single-particle excited states in $AdS_3$ 
\begin{eqnarray}
|\psi_{m, n} \rangle = a_{m, n }^\dagger |0\rangle_{{\rm Bulk}}.
\end{eqnarray}
The energy density induced by the excited state back reacts at the leading order in $G_N$, and the $AdS_3$ geometry is deformed. In this section, we evaluate the corrections to the geometry at the leading order in $G_N$ for single-particle excitations  $|\psi_{m, 0} \rangle, m =1, 2, \cdots $. From the $AdS/CFT$ dictionary obtained in the section \ref{mapbulkcft}, we see that these states are dual to holomorphic excitations of the primary
\begin{eqnarray}
a_{m , 0}^\dagger | 0\rangle_{\rm Bulk} \longleftrightarrow ( L_{-1})^l \Big| 1+ \frac{M}{2}, \frac{M}{2} \rangle, 
\qquad l = m -1 , \qquad m = 1, 2, \cdots.
\end{eqnarray}
In section \ref{shiftarea}, we will evaluate the Ryu-Takayanagi entanglement entropy in the deformed geometry, which will result in a leading correction in $G_N$ to the single interval entanglement entropy. 

We begin with evaluating the stress tensor, which is given by
\begin{eqnarray}
	T_{\mu \nu} = - \frac{2}{\sqrt{-g}} \frac{\delta S}{\delta g^{\mu \nu}}.
\end{eqnarray}
This results in
\begin{eqnarray} \label{stressmassivcs}
	T_{\mu \nu} = : 2 M (A_{\mu} A_{\nu} - \frac{1}{2} g_{\mu \nu} A_{\alpha} A^{\alpha}):.
\end{eqnarray}
Note that it is only the mass term that contributes to the stress tensor since the Chern-Simons term is topological. We have also normal-ordered the stress tensor to ensure that its expectation value on the vacuum state vanishes. This implies that we have ignored the UV effects, which are state-independent; these state-independent contributions cancel, which we consider the difference between the entanglement entropy of the excited state and that of the vacuum. 

We evaluate the expectation value $\langle \psi_{m, 0 }| T_{\mu\nu} |\psi_{m, 0 } \rangle$ by substituting the mode expansion of the vector fields in (\ref{aminusr}), (\ref{aplusr}) and using the oscillator algebra (\ref{oscalgeb}). Once the expectation value of the stress tensor is obtained, we solve the Einstein equations perturbatively to obtain the leading-order corrections to the geometry. 

We first carry out the analysis for the state $ |\psi_{1, 0} \rangle$  to demonstrate the details and then present the results for the tower of states $ |\psi_{m, 0} \rangle, m >1$. 

\subsubsection*{The primary: $| \psi_{1,0} \rangle$} \label{backgeo gs}

The non-zero components of the expectation value of the stress-energy tensor on the state $| \psi_{1,0} \rangle$ are given by 
\begin{eqnarray} \label{expecttt}
	\langle \psi_{1,0} | T_{tt} | \psi_{1,0} \rangle &=& \frac{M(M+1)}{\pi} \frac{1}{(1+r^2)^M}, \\ \nonumber
	\langle \psi_{1,0} | T_{\varphi \varphi} | \psi_{1,0} \rangle &=& \frac{M(M+1)}{\pi} \frac{r^4}{(1+r^2)^{M+2}}, \\ \nonumber
	\langle \psi_{1,0} | T_{t \varphi} | \psi_{1,0} \rangle &=& - \frac{M(M+1)}{\pi} \frac{r^2}{(1+r^2)^{M+1}}.	
\end{eqnarray}
We have further verified that the components satisfy the conservation law
\begin{eqnarray}
	\nabla^\mu \langle \psi_{1,0} | T_{\mu \nu} | \psi_{1,0} \rangle = 0.
\end{eqnarray}
The components of the stress tensor have radial dependence; the fact that the component expectation value of the component $T_{t\varphi}$ is non-zero indicates that the metric must have angular momentum. This leads us to consider the ansatz for the metric
\begin{eqnarray} \label{metric ansatz 1,0}
	ds^2 &=& -(r^2 + H_1^2(r)) dt^2 + \frac{dr^2}{(r^2 + H_2^2(r))} + r^2 d\phi^2 + H_3(r) dt d\varphi, \\ \nonumber
	H_1(r) &=& 1 + G_N a_1 (r), \qquad H_2(r) = 1 + G_N a_2 (r), \qquad H_3(r) = G_N a_3 (r).
\end{eqnarray}
Substituting this ansatz in Einstein's equation 
\begin{eqnarray} \label{einstein equation}
	R_{\mu \nu} - \frac{1}{2} g_{\mu \nu} R - g_{\mu \nu} = 8 \pi G_N \langle \psi | T_{\mu \nu} | \psi \rangle, 
\end{eqnarray}
leads us to the following differential equations for $a_1 (r)$, $a_2 (r)$ and $a_3(r)$ at order $(G_N)$,
\begin{eqnarray} \label{soln for 1,0}
	& &a_2'(r) = - 8 M (M + 1) \frac{r}{(1 + r^2)^{M+1}}, \\ \nonumber
	& &(1 + r^2) a_1'(r) - 2 r a_1 (r) + 2 r a_2 (r) = 0, \\ \nonumber
	& &r a_3''(r) - a_3'(r) = 32 M (M + 1) \frac{r^3}{(1 + r^2)^{M + 2}}.
\end{eqnarray}
These equations result from the $tt$, $rr$ and $t \phi$ components of Einstein's equations, respectively. All the other Einstein's equations are trivially or non-trivially satisfied. The solutions for these equations are
\begin{eqnarray} \label{slonbackgs}
	& &a_2 (r) = \frac{4 (M + 1)}{(1 + r^2)^M} + c_1, \\ \nonumber
	& &a_1 (r) = \frac{4}{(1 + r^2)^M} + c_1 + (1 + r^2) c_2, \\ \nonumber
	& &a_3 (r) = \frac{8}{(1 + r^2)^M} + \frac{r^2}{2} d_1 + d_2,
\end{eqnarray}
where  $c_1$, $c_2$, $d_1$ and $d_2$ are the constants of
integration. Setting the constants  $c_2 = d_1 = 0$ ensures that the metric asymptotes to $AdS_3$ for $r \rightarrow \infty$. The constants $c_1, d_2$ are fixed by demanding that the boundary stress tensor evaluated from the bulk using the Fefferman-Graham coordinates agrees with the expectation value of the stress tensor evaluated in the CFT for the primary $\big|1+ \frac{M}{2}, \frac{M}{2} \big\rangle $. In the work of scalars by \cite{Belin:2018juv}, one can rely on the fact that the back-reacted solution at order $G_N$ coincides with the conical defect and use its mass to fix the integration constant \footnote{See discussion below equation (3.18) of \cite{Belin:2018juv}.}. Here too, the constant $c_1$ can be fixed using this procedure, and certainly it agrees with our approach of using the Fefferman-Graham coordinates. We use the Fefferman-Graham approach since the solution has angular momentum, and therefore, we need to generalize the identification of the leading order solution to rotating conical defects. The Fefferman-Graham approach is convenient and easy; an alternative approach would be to identify the leading-order solution to rotating conical defects.

The Fefferman-Graham expansion of the metric in (\ref{metric ansatz 1,0}) is given in appendix \ref{appenfg}. From (\ref{holttval}), we obtain the following result for the boundary stress tensor
\begin{eqnarray} \label{Ttt FG gs}
	\langle \psi_{1,0} | T_{tt} | \psi_{1,0} \rangle_{FG} = \frac{1}{4 G_N} \Big( - \frac{1}{2} - G_N c_1 \Big). 
\end{eqnarray}
For the CFT on the cylinder, the stress $T_{tt}$ component of the stress tensor admits the expansion 
\begin{eqnarray} \label{FG}
	T_{tt} (t, \varphi) = \sum_{n = -\infty}^{\infty} \big( L_n e^{i n (t + \varphi)} + \bar{L}_n e^{i n (t - \varphi)} \big) - \frac{c + \bar{c}}{24}.
\end{eqnarray}
Evaluating the expectation value of the CFT stress tensor on the primary results in 
\begin{eqnarray} \label{in cft}
	\frac{ \big\langle 1 + \frac{M}{2}, \frac{M}{2}  | T_{tt} (t, \varphi) | 1+ \frac{M}{2} , \frac{M}{2} \big \rangle}
	{\big\langle 1+ \frac{M}{2} , \frac{M}{2}  | 1+ \frac{M}{2} , \frac{M}{2}  \big\rangle } 	&=& - \frac{c}{12} + (M + 1),
\end{eqnarray}
where we have used $c=\bar c$ in the last line.  We can relate the central charge to Newton's constant using the Brown-Henneaux \cite{Brown:1986nw} formula
\begin{eqnarray} \label{BH formula}
	\frac{1}{G_N} = \frac{2 c}{3}.
\end{eqnarray}
Then requiring \eqref{Ttt FG gs} and \eqref{in cft} to agree and using \eqref{BH formula}, we obtain
\begin{eqnarray} \label{constant for 1,0}
	c_1 = - 4 (M + 1).
\end{eqnarray}
Though we would not need the constant $d_2$, we can determine it using the Fefferman-Graham expansion of the metric to read out the $t\varphi$ component of the boundary stress tensor. From (\ref{for d2 gs FG}), we obtain
\begin{eqnarray} \label{holtpm}
\langle \psi_{1,0} | T_{t\varphi} | \psi_{1,0} \rangle_{FG}  =  \frac{d_2}{4}. 
\end{eqnarray}
In the CFT, we examine the corresponding component of the stress tensor, which is given by the expansion 
\begin{eqnarray}
T_{tt} (t, \varphi) = \sum_{n = -\infty}^{\infty} \big( L_n e^{i n (t + \varphi)} - \bar{L}_n e^{i n (t - \varphi)} \big) .
\end{eqnarray}
Here we have used $c = \bar c$.  This component of the stress tensor reads out the angular momentum quantum number of the CFT state
\begin{eqnarray}\label{cftttm}
\frac{ \big\langle 1 + \frac{M}{2}, \frac{M}{2}  | T_{t\varphi } (t, \varphi) | 1+ \frac{M}{2} , \frac{M}{2} \big \rangle}
{\big\langle 1+ \frac{M}{2} , \frac{M}{2}  | 1+ \frac{M}{2} , \frac{M}{2}  \big\rangle } &=& 1.
\end{eqnarray}
Requiring the holographic stress tensor to agree with that evaluated from the CFT, from (\ref{holtpm}) and (\ref{cftttm}), we obtain 
\begin{equation}
d_2 = 4.
\end{equation}
It is important to note that though the stress tensor vanishes for $M=0$ as seen in (\ref{expecttt}),  both $c_1$ and $d_2$ do not vanish. 

\subsubsection*{The tower of states $|\psi_{m, 0 } \rangle$ }

We proceed to the tower of states  $|\psi_{m, 0 } \rangle, \, m>1$ which are obtained by the action of the left-moving $SL(2, R)$ raising operator on the primary state with lowest energy $|\psi_{1, 0 } \rangle$. Proceeding as before, the non-trivial components of the stress tensor expectation value are given by 
\begin{eqnarray}
	\langle \psi_{m,0} | T_{tt} | \psi_{m,0} \rangle &=& \frac{Mm}{\pi} \binom{M+m}{m} \frac{r^{2(m-1)}}{(1+r^2)^{M-m+1}}, \\ \nonumber
	\langle \psi_{m,0} | T_{\phi \phi} | \psi_{m,0} \rangle &=& \frac{Mm}{\pi}\binom{M+m}{m} \frac{r^{2(m+1)}}{(1+r^2)^{M+m+1}}, \\ \nonumber
	\langle \psi_{m,0} | T_{t \phi} | \psi_{m,0} \rangle &=& - \frac{Mm}{\pi}\binom{M+m}{m} \frac{r^{2m}}{(1+r^2)^{M+m}}.	
\end{eqnarray}
Again, the components of the stress tensor satisfy the conservation law
\begin{eqnarray}
	\nabla^\mu \langle \psi_{m,0} | T_{\mu \nu} | \psi_{m,0} \rangle = 0.
\end{eqnarray}
The ansatz for the back-reacted metric is as given in (\ref{metric ansatz 1,0}). To distinguish this case, we will parametrize the 
corrections by $b_1, b_2, b_3$. 
\begin{eqnarray} \label{metric ansatz m,0}
	ds^2 &=& -(r^2 + \mathcal{H}_1^2(r)) dt^2 + \frac{dr^2}{(r^2 + \mathcal{H}_2^2(r))} + r^2 d\phi^2 + \mathcal{H}_3(r) dt d\phi, \\ \nonumber
	\mathcal{H}_1(r) &=& 1 + G_N b_1 (r), \qquad \mathcal{H}_2(r) = 1 + G_N b_2 (r), \qquad \mathcal{H}_3(r) = G_N b_3 (r).
\end{eqnarray}
To solve the Einstein equations in the leading order in $G_N$, the functions  $b_1, b_2, b_3$ satisfy the following linear differential equations 
\begin{eqnarray} \label{toweree}
	& &b_2'(r) = - 8 M m \binom{M + m}{m} \frac{r^{2m - 1}}{(1 + r^2)^{M + m}}, \\ \nonumber
	& &(1 + r^2) b_1'(r) - 2 r b_1 (r) + 2 r b_2 (r) = 0, \\ \nonumber
	& &r b_3''(r) - b_3(r) = 32 M m \binom{M + m}{m} \frac{r^{2 m + 1}}{(1 + r^2)^{M + m +1}}.
\end{eqnarray}
We will see subsequently that to obtain the corrections to the minimal area, we would need only the form of $b_2(r)$ explicitly, which is given by 
\begin{eqnarray} \label{rr comp sol}
	b_2 (r) &=& k_1 - 4 r^{2m} \frac{\Gamma(M + m + 1)}{\Gamma(M) \Gamma(m + 1)} {}_2 F_1 \Big( m, M + m, 1 + m, -r^2 \Big) ,  \\ \nonumber
	& & = k_1 - 4 (m +M)  +  r^{-2M} \left[ 
	 \frac{ 4 \Gamma( 1+m +M) }{\Gamma(m) \Gamma( 1+M) }  + O( r^{-2} ) \right].
\end{eqnarray}
As done in section  \ref{backgeo gs}, we will fix the value of $k_1$ by requiring that the boundary stress tensor evaluated using the Fefferman-Graham coordinates agrees with the expectation value of the CFT stress tensor on the state  $(L_{-1})^l \big|1+ \frac{M}{2}, \frac{M}{2} \rangle, \; l = m -1$. Here too, we can adopt this approach; however using the results from the appendix  \ref{appenfg}, we obtain (\ref{towerholfg})
\begin{eqnarray} \label{Ttt FG m,0}
	\langle \psi_{m,0} | T_{tt} | \psi_{m, 0} \rangle_{FG} = \frac{1}{4 G} \bigg( - \frac{1}{2} - \big(k_1 - 4(M + m) \big) G_N \bigg).
\end{eqnarray}
Given the stress-tensor of CFT on a cylinder \eqref{FG}, the expectation value of the stress tensor is given by 
\begin{eqnarray} \label{Ttt exp m,0}
\frac{ \big\langle 1 + \frac{M}{2}, \frac{M}{2}  | T_{tt} (t, \varphi) | 1+ \frac{M}{2} , \frac{M}{2} \big \rangle}
	{\big\langle 1+ \frac{M}{2} , \frac{M}{2}  | 1+ \frac{M}{2} , \frac{M}{2}  \big\rangle }  =
   - \frac{c + \Bar{c}}{24} + M + 1 + l = - \frac{c}{12} + M + m. 
\end{eqnarray}
Then requiring \eqref{Ttt FG m,0} and \eqref{Ttt exp m,0} to agree and using \eqref{BH formula}, we obtain
\begin{eqnarray} \label{constant m,0}
	k_1 = 0.
\end{eqnarray}

\subsection{Shift in minimal area} \label{shiftarea}

\begin{figure}{}
\begin{center}
\begin{tikzpicture}[rotate=90,transform shape]
%\begin{tikzpicture}
\coordinate (A) at (0,-0.5);
\node[xshift= 0.3 cm,rotate =-90] at (A) {\color{brown! 120}{$r_m$}};
\coordinate (B) at (-1.4,-0.9);
\node [rotate=-90]at (B) {I};
\coordinate (C) at (1.4,-0.9);
\node[rotate=-90] at (C) {II};
\coordinate (D) at (0,-1.3);
\node[rotate=-90] at (D) {\color{violet}$\gamma_A$};
\coordinate (E) at (0,-2.2);
\node[rotate=-90] at (E) {\color{violet}$\Sigma_A$};
\coordinate (F) at (0,1.4);
\node[rotate=-90] at (F) {\color{violet}$\Sigma_{\bar{A}}$};
\coordinate (G) at (0,3.3);
\node[rotate=-90] at (G) {\color{violet}$\bar{A}$};
\coordinate (H) at (0,-3.3);
\node[rotate=-90] at (H) {\color{violet}$A$};
\draw[color=gray,thick] (0,0) circle [radius=3];
\filldraw [black] (0,-1) circle (0.8 pt) node{};
\filldraw [brown!120] (2.15, -2.15) circle (0.8 pt) node[rotate=-90,below][anchor=west]{$\varphi=2\pi x$};
\filldraw [brown!120] (-2.15, -2.15) circle (0.8 pt) node[rotate=-90,below][anchor=west]{$\varphi=0$};
\filldraw [black] (0,0) circle (0.8 pt) node{};
\draw[ dashed, thick, gray!100] (0,0)--(0,-1);
\draw[red] (2.15,-2.15) arc (37:88:2.94);
\draw[blue] (-2.15,-2.15) arc (148:88:2.48);
\end{tikzpicture}
\end{center}
\caption{The $t=0$ slice of $AdS_3$. We consider the entanglement between system $A$ and $\bar{A}$ on the boundary. The minimal surface $\gamma_A$ is the geodesic in the bulk connecting the endpoints of $A$. $\gamma_A$ splits the bulk into right(left) wedge denoted by $\Sigma_A(\Sigma_{\bar{A}})$. $\gamma_A$ consists of:  Branch I, where $\varphi^\prime(r) < 0$ and Branch II with $\varphi^\prime(r) > 0$. When excitations break the isometry in $\varphi$, we need to evaluate the minimal area for the above branches separately.}
\label{fig: Rindler}
\end{figure}
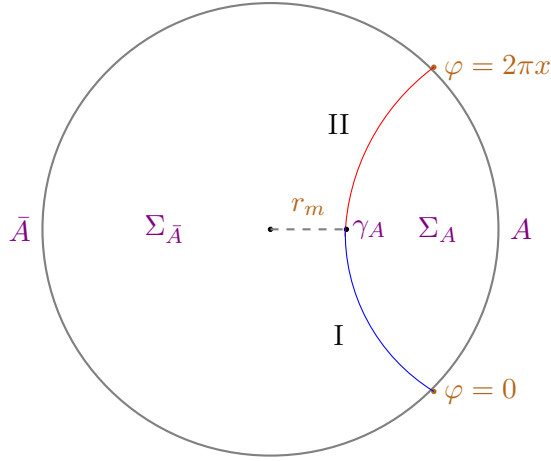

The first term in the FLM formula in (\ref{flm}) is the minimal area, which is the length of the minimal geodesic between the two endpoints of the interval on the boundary, as shown as $\gamma_A$ in figure \ref{fig: Rindler}. As we have seen in the previous section, the single-particle excitations deform the geometry at order $G_N$; this implies that the minimal area is corrected. Since we are interested in the difference in the entanglement entropy between the ground state and the single-particle excited state, it is sufficient to consider the correction to the minimal length. Thus, to the leading order in $G_N$, the change in the minimal length is given by 
 \begin{eqnarray}
	\delta A = A [g_0 + \delta g] - A [g_0].
\end{eqnarray}
Here $A[g]$ is the length of the geodesic $\gamma_A$ in the background metric $g$. $g_0$ refers to the  metric for global $AdS_3$ and $g_0 + \delta g$ is the back-reacted metric.

The geodesic $\gamma_A$ at $t=0$ in the global $AdS_3$ metric is given by minimizing the length 
\begin{eqnarray} \label{minarea}
	A [g_0] = 2 \int_{r_m}^\infty dr \sqrt{\frac{1}{1 + r^2} + r^2 (\varphi'(r))^2} \; ,
\end{eqnarray}
where $r_m$ is the turning point on the geodesic. This results in the  differential equation
\begin{eqnarray}
	\varphi'(r) = - \frac{r_m}{r \sqrt{(1 + r^2)(r^2 - r_m^2)}} \qquad &:& \text{Branch} \,1 \\ \nonumber
	\varphi'(r) = \frac{r_m}{r \sqrt{(1 + r^2)(r^2 - r_m^2)}} \qquad &:& \text{Branch}\,  2.
\end{eqnarray}
From the solution of these equations, we can relate $r_m$ to the angle between the endpoints of the interval on the boundary. This relation is given by 
%We choose the initial angle  $\varphi = 0$, this results in the following equation for $\varphi$
%\begin{eqnarray}
%	\varphi = \frac{\pi}{2} - \arctan \Big( \frac{r^2 + \sqrt{(1 + r^2)(r^2 - r_m^2)}}{r} \Big) \qquad &:& \text{Branch}\; 1 \\ \nonumber
%	\varphi = \frac{\pi}{2} - \arctan \Big( \frac{r^2 - \sqrt{(1 + r^2)(r^2 - r_m^2)}}{r} \Big) \qquad &:& \text{Branch}\; 2 \\ \nonumber
%\end{eqnarray}
%Let us define
\begin{eqnarray} \label{int length}
	\pi x = \arctan \frac{1}{r_m} = \arccot r_m  \equiv \frac{\theta}{2}.
\end{eqnarray}
Note that 2 branches contribute equally to the length, which results in the factor of $2$ in (\ref{minarea}). Since we are interested in the correction to the leading order in $G_N$, the change in area is obtained by substituting the shifted metric and integrating it along the geodesic. It is easy to see from the ansatz of the form (\ref{metric ansatz 1,0}) or (\ref{metric ansatz m,0}), the corrections to the minimal area depend only on the corrections to the $g_{rr}$ component of the metric. Let the coefficient of $g_{rr}$ be given by
\begin{eqnarray}
	g_{rr} = \frac{1}{1 + r^2 + J_2 ( r )}. 
\end{eqnarray}
On evaluating the shift in minimal area, we obtain
\begin{eqnarray} \label{change in area formula}
	\delta A = - \frac{1}{2}\int_{r_m}^\infty dr J_2 ( r ) \frac{(r^2 - r_m^2)^{\frac{1}{2}}}{r (1 + r^2)^{\frac{3}{2}}}.
\end{eqnarray}

\subsection*{Area shift for the state $| \psi_{1,0} \rangle$}

We evaluate the shift in area for the back-reacted geometry due to the state $| \psi_{1,0} \rangle$. Reading out the perturbed metric from \eqref{metric ansatz 1,0} with \eqref{slonbackgs}, \eqref{constant for 1,0} and applying the expression \eqref{change in area formula}, results in the integral
\begin{eqnarray}
	\delta A_{| \psi_{1,0} \rangle} = 2 G_N \int_{r_m}^\infty dr \Bigg[ 4 (M + 1) - \frac{4 (M + 1)}{(1 + r^2)^M}  \Bigg] \frac{(r^2 - r_m^2)^{\frac{1}{2}}}{r (1 + r^2)^{\frac{3}{2}}}. 
\end{eqnarray}
The integral evaluates to
\begin{eqnarray}
	\frac{\delta A_{| \psi_{1,0} \rangle}}{4 G_N}& = &2 (1+M) (1 - r_{m} \cot^{-1}(r_{m})) \\ \nonumber
	&&  - 
	\frac{\sqrt{\pi} \Gamma(M+2) } { 2\Gamma(M + \frac{5}{2}) \; r_{m}^{2(M+1)} } \; {}_2 {F}_1(1+M, M + 3/2, M + 5/2, -1/r_{m}^2) .
\end{eqnarray}
Substituting $r_m$ in terms of interval length $\pi x$ from \eqref{int length}, and keeping the leading order term we obtain
\begin{eqnarray} \label{corrected area 1,0}
	\frac{\delta A_{| \psi_{1,0} \rangle}}{4 G_N} = 2 (1+M) (1 - \pi x \cot^{-1}(\pi x)) - \frac{\Gamma(M+2) \Gamma(\frac{3}{2})}{\Gamma(M + \frac{5}{2})} (\pi x)^{2(M+1)} + \cdots.
\end{eqnarray}
Comparing this result to the answer from the CFT in (\ref{gscft}) and identifying $h = 1+ \frac{M}{2}, \bar h = \frac{M}{2}$, we see that the leading term agrees, while the sub-leading term differs. In fact, the dependence of this term on the CFT is of the form $x^{4(M+1)}$, while the shift in area goes as $x^{2(M+1)}$. We will see in section \ref{bulkent} that the second term in the FLM formula (\ref{flm}) precisely has contributions which cancel the sub-leading term in (\ref{corrected area 1,0}) and reproduce the result from the CFT. This pattern of cancellation was seen earlier for primaries, and descendants dual to a scalar \cite{Belin:2018juv,Chowdhury:2024fpd}.

\subsection*{Area shift for the states $| \psi_{m,0} \rangle$}

The back-reacted metric corresponding to the states $| \psi_{m,0} \rangle$ is given in \eqref{metric ansatz m,0}, with correction to the metric coefficient $g_{rr}$ given in \eqref{rr comp sol} and using \eqref{constant m,0}. Substituting this correction in the expression for the shifted area \eqref{change in area formula}, we obtain
\begin{eqnarray}
	\delta A_{| \psi_{m,0} \rangle} =  - 2 G_N \int_{r_m}^\infty dr \; 4 r^{2m} \frac{\Gamma(M + m + 1)}{\Gamma(M) \Gamma(1+m)} {}_2 F_1 (m, M + m, 1 + m, -r^2) \frac{(r^2 - r_m^2)^{\frac{1}{2}}}{r (1 + r^2)^{\frac{3}{2}}}.
	\nonumber
	\\
\end{eqnarray}
The above integral can be evaluated to the leading orders in short distance using the expansion for the hypergeometric function given in the second line of (\ref{rr comp sol}). Substituting $r_m$ with $\pi x$ from \eqref{int length}, and keeping the leading order terms in the non-analytic expansion of $\pi x$, we obtain
\begin{eqnarray} \label{corrareades}
	\frac{\delta A}{4 G_N} = 2 (M + m ) (1 - \pi x \cot(\pi x) ) - \frac{\Gamma(3/2) \Gamma(M+2)}{\Gamma(M + \frac{5}{2})} \Bigg( \frac{\Gamma(M + m+1)}{\Gamma(M+2) (m-1)!} \Bigg) (\pi x)^{2(M + 1)} + \cdots \nonumber.
	\\
\end{eqnarray}
This perturbative excitation in the bulk is dual to the boundary state  $L_{-1}^l \big| 1 + \frac{M}{2}, \frac{M}{2} \big\rangle$ with $l = m-1$. As expected, this result reduced to (\ref{corrected area 1,0}) for $m =1$. Comparing (\ref{corrareades})  with the corresponding CFT result in (\ref{maincftres}), we see that the leading term agrees, but again the sub-leading non-analytical term differs. We will subsequently see that the contribution from the 
bulk entropy $S_{\rm bulk}$ will be such that the result from the CFT is precisely reproduced.

\section{Bulk entanglement entropy} \label{bulkent}

In this section, we describe the computation of bulk entanglement entropy $S_{\rm bulk}^{EE} (\Sigma _A) $ corresponding to the single interval of length $\frac{\theta}{2}= \pi x$ on the boundary in order $G_N^0$.  This is the second term in the Faulkner, Lewkowycz and Maldacena proposal in (\ref{flm}). At this order, the contribution from this term is given by the Von-Neumann entropy of the reduced density matrix obtained by tracing over excitations of bulk fields outside the Ryu-Takayanagi minimal surface. Performing such a trace is non-trivial in curved spacetime. To do this, we will adopt the same approach introduced by \cite{Belin:2018juv} and further developed in \cite{Chowdhury:2024fpd}. This uses the map, which takes the Ryu-Takayanagi surface to the Rindler horizon in BTZ, then tracing over the exterior of the Ryu-Takayanagi surface corresponds to tracing over the left half of the Rindler BTZ. This map, which was originally introduced by \cite{Casini:2011kv}, is given by 
%coordinate $(t,r,\varphi)$ to AdS-Rindler coordinate $(\tau,\rho,x)$.
\begin{align} 
r=&\sqrt{\rho^2 \sinh^2 x + \left( \sqrt{\rho^2-1} \cosh \eta \cosh \tau + \rho \cosh x \sinh \eta\right)^2} \nonumber\, ,\\ 
t=& \arctan\left( \frac{\sinh \tau \sqrt{\rho^2-1}}{\rho \cosh x \cosh \eta +\sqrt{\rho^2-1} \cosh\tau \sinh \eta}\right)  \nonumber \, , \\ 
\varphi -\frac{\theta}{2}=& \arctan\left( \frac{\rho \sinh x}{\sqrt{\rho^2-1} \cosh \tau \cosh \eta +\rho \cosh x \sinh \eta}\right) \,. \label{coordchange}
\end{align} 
Here $r, t, \varphi$ are global $AdS_3$ coordinates and $\rho, \tau, x$ are that of the Rindler BTZ with metric 
given in (\ref{btzrinda}). 

This transformation maps the Ryu-Takayanagi surface $\gamma_A$  in global $AdS_3$ to the Rindler horizon $\rho^2=1$ at $t=0$ on the following identification.
\begin{align}\label{eq: map eta theta}
\cosh \eta= \sqrt{r_m^2+1}= \frac{1}{\sin \frac{\theta}{2}}.
\end{align}
This can be directly seen by observing that the Rindler horizon maps to the surface 
\begin{align}
r^2= \sinh^2 x+\cosh^2 x \sinh^2 \eta,~~\tan(\varphi-\frac{\theta}{2})=\frac{\sinh x}{\cosh x \sinh \eta},
\end{align}
using (\ref{coordchange}). These equations are the parametric solutions to the differential equation of the minimal surface
\begin{align}\label{eq: eq minimal length}
\left(\frac{d \varphi}{d r}\right)^2= \frac{r_m^2}{r^2(1+r^2)(r^2-r_m^2)}.
\end{align}
%using the equation of the Rindler horizon which is given by
%\begin{align}
%r^2= \sinh^2 x+\cosh^2 x \sinh^2 \eta,~~\tan(\varphi-\frac{\theta}{2})=\frac{\sinh x}{\cosh x \sinh \eta}.
%\end{align}
The equation for the minimal surface is given  by
%The perturbed metric breaks the angular symmetry hence the sign of $\tan (\varphi-\frac{\theta}{2}) $ in terms of $r$ will be ambiguous upto a sign. This is also reflected on the first order equation of minimal surface derived from \eqref{eq: eq minimal length}.
%\begin{subequations}
\begin{align}
& \tan(\varphi-\frac{\theta}{2})=-\frac{\sqrt{r^2-r_m^2}}{r_m \sqrt{1+r^2}},~~\text{Branch-I},\\ \nonumber
& \tan(\varphi-\frac{\theta}{2})=-\frac{\sqrt{r^2-r_m^2}}{r_m \sqrt{1+r^2}},~~\text{Branch-II}.
\end{align}
%\end{subequations}
In Branch I, the angle increases as $r$ decreases towards $r_m$ from infinity and in Branch II, the angle continues to increase as $r$ increases from $r_m$ to infinity. It is useful to note that the identification in (\ref{eq: map eta theta}) of the boost parameter, with the angle in the short distance or large boost limit, reduces to 
\begin{eqnarray} \label{largeeta}
\lim_{\eta\rightarrow \infty} \cosh\eta = \frac{e^{\eta}}{2} = \frac{2}{\theta} = \frac{1}{\pi x}.
\end{eqnarray}
Here we have kept the leading terms in the limit and used the relation (\ref{int length}). 

The Rindler BTZ coordinates in (\ref{coordchange}) do not cover the entire global $AdS_3$. This is because the range of $\varphi$ is restricted once a particular branch of $\arctan$ is chosen. Just as in the flat space case, one of the branches is the right Rindler wedge, and one is the left. The right Rindler wedge corresponds to the region $\Sigma_A$, while the left corresponds to $\Sigma_{\bar{A}}$. Single-particle excitation in global $AdS_3$, once mapped into the Rindler BTZ, will have support in both these 
wedges. This discussion implies the global vacuum in the $AdS_3$  can be written as a thermo-field double state in Rindler left and right vacua \footnote{The global vacuum is a sum over super selection sectors labelled by the perpendicular component of the electric field at the horizon. We will discuss this sum and the contribution to the entanglement entropy due to this sum over super selection sectors in the section. For the present, we will work in a definite super selection sector. }
\begin{align}\label{eq: therofield double}
|0\rangle= \sum_n e^{-\frac{2 \pi E_n}{2}} | n^* \rangle_L |n \rangle_R.
\end{align}
$L, R$ denote the left and right Rindler Hilbert spaces, and inverse temperature of the thermo-field double is $\beta = 2\pi$, which can be obtained from the Rindler BTZ geometry.  We formally label the states with $n$ and their energies as $E_n$. $| n^* \rangle_L$ is the CPT conjugate of the $|n \rangle_R$. 

Single-particle excitations due to the massive Chern-Simons field in Rindler BTZ can be obtained by quantizing the field in this geometry. This has been carried out in appendix \ref{appenc}. 
The result of this exercise is the following. Let us define 
\begin{eqnarray}
A_+ = A_\tau + A_x, \qquad A_-  = A_\tau - A_x.
\end{eqnarray}
Here, it is understood that these are vectors in the BTZ background. We can solve the Chern-Simons equations and expand the vector in terms of modes, which results in 
\begin{eqnarray} \label{rinamexp}
A_-  = \sum_{I \in L, R} \int_{\omega >0} \frac{d\omega dk}{4\pi^2} \left[ 
 b_{\omega, k , I } A_-^{(\omega, k ) I } e^{-i \omega \tau + i k x}
+  b_{\omega, k, I }^\dagger   A_-^{(\omega, k ) I * } e^{i \omega \tau - i k x} \right],
\end{eqnarray}
where the modes are  given by 
{\small \begin{eqnarray}\label{aminusexpmain}
A_-^{(\omega, k)\, I }&=& N_{(\omega, k )} 
 \rho^{-M} \Big(1-  \frac{1}{\rho^2}\Big)^{- \frac{i \omega}{2} } 
 {}_2 F_1\Big( \frac{1}{2} ( M - i ( k + \omega)) , \frac{1}{2} ( M + i ( k - \omega) ) , M , \frac{1}{\rho^2} \Big), \nonumber \\
 {\rm where}\; &&   I \in \{L, R\},  \qquad \qquad {\rm and} \\ \nonumber
N_{\omega, k } &=& \frac{ \Big|\Gamma( 1+ i \omega) \Gamma\big[ \frac{1}{2} ( M - i ( k -\omega) ) \big] \Gamma\big[ \frac{1}{2} ( M+2  +i ( k +\omega) ) \big] \Big| } { \pi \csch(\pi \omega) \Gamma(M) \sqrt{M \omega}}. 
\end{eqnarray}}
Similarly, the expansion of the component $A_+$ is given by 
\begin{eqnarray} \label{rinapexpam}
A_+  = \sum_{I \in L, R} \int_{\omega >0} \frac{d\omega dk}{4\pi^2} \left[
  b_{\omega, k , I } A_+^{(\omega, k ) I } e^{-i \omega \tau + i k x}
+  b_{\omega, k I }^\dagger   A_+^{(\omega, k ) I * } e^{i \omega \tau - i k x} \right],
\end{eqnarray}
with
{\small \begin{eqnarray}
A_+^{(\omega, k)\, I}&=& C_+^{(\omega, k )} 
\rho^{-(M+2) } \Big(1-  \frac{1}{\rho^2} \Big)^{- \frac{i \omega}{2} } 
{}_2 F_1\Big( \frac{1}{2} ( M +2  - i ( k + \omega)) , \frac{1}{2} ( M +2 + i ( k - \omega) ) , M+2 , \frac{1}{\rho^2} \Big),
\nonumber \\
C_+^{(\omega, k )}  &=&   -\frac{M^2  + ( k - \omega)^2 }{ 4 M ( 1+M) } N_{\omega, k }.
\end{eqnarray} }
We can find the radial component $A_\rho$ by using the 2nd  Chern-Simons equation in (\ref{cseqrin}). In the appendix \ref{appenc}, we have shown that quantization of these modes results in the oscillator algebra
\begin{eqnarray} \label{rincomrel}
[b_{\omega, k, I} , b_{\omega', k', J}^\dagger] = (2\pi)^2 \delta ( \omega -\omega') \delta(k-k') \delta_{IJ}.
\end{eqnarray}
These oscillators create single-particle excitations in  BTZ.  
 
As we mentioned in the introduction, the massive Chern-Simons field satisfies the equations 
\begin{eqnarray}  \label{diffbc}
\nabla^i  F_{i \tau  }  =  M^2 A_\tau  , \qquad \nabla^i A_i  =  - \nabla^\tau A_\tau. 
\end{eqnarray}
This implies that field configurations or initial conditions must satisfy the conditions
\begin{eqnarray}\label{normbc}
n^\mu F_{\mu \tau  }^R|_{\gamma} =  n^\mu F_{\mu \tau  }^L|_{\gamma}, 
\qquad  n^\mu A_\mu^R|_{\gamma} =  n^\mu A_\mu^L|_{\gamma}. 
\end{eqnarray}
Here $n^\rho$ is the unit  normal to a constant $\rho$ surface given by 
\begin{eqnarray}
n^\mu = \big( 0 , \sqrt{\rho^2 -1}, 0  \big). 
\end{eqnarray}
The boundary conditions in (\ref{normbc}) result from using the simple Gaussian pill box argument on the equations (\ref{diffbc}). We will choose the surface $\gamma$ to be close to the horizon at $\rho = 1+ \epsilon$ with $0<\epsilon<<1$. In the appendix (\ref{appenc}), we show that for non zero $\omega$ we can ensure that the LHS and RHS of (\ref{normbc}) can be made to vanish on $\gamma$, that is 
\begin{equation} \label{pmc}
n^\mu F_{\mu \tau  }^R|_{\gamma} =  n^\mu F_{\mu \tau  }^L|_{\gamma} = 0, \qquad 
\qquad  n^\mu A_\mu^R|_{\gamma} =  n^\mu A_\mu^L|_{\gamma}  =0.
\end{equation}
This results in the quantization of the frequencies
\begin{eqnarray}
\omega_n = \frac{2\pi n } { |\log ( 2\epsilon) |}, \qquad n = 1, 2, \cdots .
\end{eqnarray}
We see that since $|\log( 2\epsilon) |$  is large, the frequencies are closely spaced and therefore for all modes other than those with strictly $\omega = 0$, we can use the continuum approximation as written in equations (\ref{aminusexpmain}), (\ref{rinapexpam}). 
We call these non-zero frequencies bulk modes.
 
The $\omega = 0$ mode near the horizon is different since the indicial roots of the radial equation for the wave functions of the vectors coincide. This leads to the fact that it would not be
possible to satisfy the more rigid boundary conditions (\ref{pmc}). However, in the zero-mode sector, we would be able to satisfy the boundary conditions (\ref{normbc}). In section \ref{edgesection}, we will show that the quantization of the $\omega = 0$ sector results in super selection sectors and edge modes at $\gamma$.   Though there is a contribution to entanglement entropy from these edge modes, it vanishes in the $\epsilon \rightarrow 0 $ limit. The complete Hilbert space in both the left and the right wedges consists of the tensor product of bulk modes as well as the zero-mode sector. In the rest of this section, we will work within the Hilbert space of the bulk modes and deal with the edge modes in the section \ref{edgesection}.

As the $AdS_3$ vacuum is a thermo-field double in Rindler  space  given in  (\ref{eq: therofield double}), the trace over the left Rindler wedge $\Sigma_{\bar{A}}$ leads to the thermal state
\begin{align}\label{eq: rho0}
\rho_0= {\rm Tr}_{\Sigma_{\bar{A}}} |0 \rangle \langle 0 |= e^{-2 \pi H_R}.
\end{align}
$H_R$ is the Hamiltonian of the single-particle excitations in the right wedge
\begin{align}
H_R= \sum_{\omega,k} \omega b^\dagger_{\omega,k,R} b_{\omega,k,R}.
\end{align}
The  properties of  the thermo-field double allow us to relate  the action of creation(annihilation) operators of the left Rindler wedge to that of the right
\begin{align}\label{eq: relation operator left right}
& b_{\omega,k,L}|0 \rangle=e^{-\pi \omega} b^\dagger_{\omega,-k,R}|0 \rangle,\qquad 
b^\dagger_{\omega,k,L}|0 \rangle=e^{\pi \omega} b_{\omega,-k,R}|0 \rangle.
\end{align}
This indicates the left and right wedges are entangled with each other non-trivially. Since the vector field can be expanded as modes in either global $AdS_3$ or in Rindler space,  the creation (annihilation) operators in the global and Rindler coordinates are related by \bbgv coefficients as follows,
\begin{align}\label{eq: definition bogo}
a_{m,n}=\sum_{I,\omega,k} \left( \alpha_{m,n;\omega,k, I} b_{\omega,k,I}+\beta_{m,n;\omega,k, I} b_{\omega,k,I}^\dagger  \right).
\end{align}
Here we use the notation 
\begin{align}  \label{notatsum}
\sum_{\om}\equiv \int_0^\infty \frac{d \om}{2 \pi},\qquad \qquad \sum_{k}\equiv \int_{-\infty}^\infty \frac{d k}{2 \pi}.
\end{align}
From the fact that the global vacuum is annihilated by the oscillators  $a_{m,n}$ using (\ref{eq: definition bogo}), we see that Bogoliubov coefficients must satisfy the relations
\begin{align}\label{eq: relation bogo left right}
\alpha_{\omega,k,L}=- e^{\pi \omega} \beta^*_{\omega,-k,R},\qquad\qquad \beta^*_{\omega,k,L}=- e^{-\pi \omega} \alpha_{\omega,-k,R}. 
\end{align}
Using the commutation relation $\left[a^\dagger_{m,n},a_{m',n'}\right]=\delta_{m,m'}\delta_{n,n'} $ we can derive a relation which relates the \bbgv coefficients in left and right wedge. 
\begin{align}\label{eq: normalization left right}
\sum_{I,\omega,k} \left(\alpha_{m,n;\omega,k} \alpha^*_{m',n';\omega,k,I}-\beta^*_{m,n;\omega,k} \beta_{m',n';\omega,k,I} \right)=\delta_{m,m'} \delta_{n,n'}.
\end{align}
Using \eqref{eq: relation bogo left right} in \eqref{eq: normalization left right}, we obtain a constraint between \bbgv coefficients in the right wedge only.
\begin{align}\label{eq: normalization right right}
\sum_{\omega,k} \left[ \alpha_{m,n;\omega,k,R}\alpha^*_{m',n';\omega,k,R}(1-e^{-2\pi \omega})-\beta_{m,n;\omega,k,R}\beta^*_{m',n';\omega,k,R}(1-e^{2\pi \omega})\right]=\delta_{m,m'} \delta_{n,n'}.
\end{align}
This means that we can write any single-particle state of the global $AdS_3$ solely in the basis of operators in the right wedge. Indeed, one can see that upon substituting \eqref{eq: relation bogo left right} in \eqref{eq: definition bogo}.
\begin{align}\label{eq: state in right mode}
	\nn |\psi_{m,n}\rangle  = a^\dagger_{m,n}|0 \rangle & = \sum_{\omega,k} \left[ (1-e^{-2 \pi \omega}) \alpha^*_{m,n;\omega,k,R} b^\dagger_{\omega,k,R}+(1-e^{2 \pi \omega}) \beta_{m,n;\omega,k} b_{\omega,,k,R}\right]|0 \rangle\\
	&\equiv \sum_{\om,k}\mathcal{A}_{m,n;\om,k;R}|0\rangle,
\end{align}
where we define
\begin{align}
\mathcal{A}_{m,n;\om,k;R}= (1-e^{-2 \pi \omega}) \alpha^*_{m,n;\omega,k,R} b^\dagger_{\omega,k,R}+(1-e^{2 \pi \omega}) \beta_{m,n;\omega,k} b_{\omega,,k,R}.
\end{align}
This relation simplifies the process of integrating out the left wedge, as we will see subsequently. We will now suppress the index $R$ as a subscript in the operators and \bbgv coefficients with the understanding that we always mean the operators on the right Rindler wedge. 

Next, we state the general formula for the bulk entanglement entropy of an arbitrary excited state with quantum numbers $m$ and $n$. The reduced density matrix of the right wedge $\Sigma_A$ is given by tracing out the left sector.
\begin{align}\label{eq: bulk reduced density matrix }
\nn \rho_{\rm bulk}= &\text{Tr}_{\Sigma_{\bar{A}}} \rho\\
 = &\Big( \sum_{\om',k'}\mathcal{A}_{m,n;\om',k';R} \Big) ~\rho_0 
 \Big( \sum_{\om,k}\mathcal{A}_{m,n;\om,k;R}^\dagger \Big) .
\end{align}
The expression of $\rho_0$ is given in \eqref{eq: rho0}. The computation of the nth \ren entropy can be performed by taking $n$-copies of $\rho_{\rm bulk}$. 

We can obtain the Von-Neuman entropy using the R\'{e}nyi entropy and taking  $n\rightarrow 1$ analytically at the end.
\begin{align}\label{eq: formula bulk entanglement entropy}
S_{n: {\rm bulk}}(\Sigma_A)=\frac{1}{1-n} \log \frac{\text{Tr} (\rho_{\rm bulk})^n}{\text{Tr} \rho_0^n},\qquad\qquad \,~ S_{\rm bulk}(\Sigma_A)= \lim_{n\rightarrow 1} S_{n: {\rm bulk }}(\Sigma_A).
\end{align} 
The division by $ \text{Tr} \rho_0^n $ ensures that the contribution to the bulk entanglement entropy due to the vacuum is subtracted. From the functional form of $\mathcal{A}_{m,n,R}$ in \eqref{eq: state in right mode} we see that the evaluation of $\text{Tr} (\rho_{\rm bulk})^n$ needs the \bbgv  coefficients and the two-point function in Rindler coordinate
\begin{align}\label{eq: 2pt correlator Rindler}
\text{Tr} (\rho_0 b^\dagger_{\omega,k} b_{\omega',k'})= \frac{4\pi^2 \delta(\omega-\omega')\delta(k-k')}{e^{2\pi \om}-1}.
\end{align}
In appendix  \ref{append}, we have evaluated the Bogoliubov coefficients for the states $|\psi_{m, 0} \rangle,  \; m = 1, 2, 3, \cdots$. We observe that these coefficients admit a short interval  expansion, note that the information of the length of the interval is contained in $\eta$  , which determines the transformation from global $AdS_3$ to BTZ. This enables us to evaluate the entanglement entropy (\ref{eq: formula bulk entanglement entropy}) as a perturbative expansion in the interval size. This perturbative expansion has been discussed in detail in \cite{Belin:2018juv} and generalized in  \cite{Chowdhury:2024fpd}. Here we just state the result. The leading contribution to the bulk entanglement is given by 
\begin{align}\label{eq: bulk EE 1st order formula}
S^{(1)}_{bulk}(\Sigma_A)= 2 \pi \sum_{\om,k} \omega \left( |\alpha_{m,n; \omega, k }|^2+ |\beta_{m,n; \omega, k }|^2 \right).
\end{align}
The sum over $\omega, k$ denotes an integral as in (\ref{notatsum}); here, we have suppressed the $I $ subscripts on the Bogoliubov coefficients as defined in (\ref{eq: definition bogo}), since it will be understood that we will refer only to the right Rindler wedge. As shown in the appendix \ref{append}, the Bogoliubov coefficients are proportional to $(\sin \pi  x)^{M+1}$, therefore these corrections are of the order $(\pi x )^{(M+1)} $. The second-order corrections are given by the expression 
\begin{align} 
\nonumber
&&S^{(2)} (\Sigma_A) =
 \frac{1}{2} \sum_{\stackrel{\omega_1, \omega_2}{k_1, k_2} }  \Bigg[
 2\pi  (\omega_1 - \omega_2) \frac{ ( 1-e^{-2\pi \omega_1} ) (1- e^{2\pi \omega_2} ) }{ 1- e^{2\pi ( \omega_2 - \omega_1)}}
  \Big|\alpha_1^*  \alpha_2 + \beta_1^* \beta_2 \Big|^2
  \\
 && +2\pi ( \omega_1 + \omega_2) \frac{ (1-e^{2\pi \omega_1})  (1-e^{2\pi \omega_2})}{ 1- e^{ 2\pi (\omega_1 + \omega_2) } }
 2 \Big\{ | \alpha_1^*|^2 |\beta_2|^2 + \alpha_1^* \beta_2^* \alpha_2 \beta_1 \Big\}
 \Bigg]. \label{2ndorderbulkf}
\end{align}
Here $\alpha_1 = \alpha_{m, n ; \omega_1, k_1}, \alpha_2 =
\alpha_{m, n;\omega_2, k_2} $ and similarly for $\beta_1, \beta_2$. 
The last term in the 2nd line can be shown to be real by
interchanging the dummy variables $1\leftrightarrow2$. This compact form of the second-order correction was written down in \cite{Chowdhury:2024fpd}. Note these corrections are of the order $(\pi x )^{2(M+1) }$.

\subsection{Bulk entanglement for $|\psi_{1, 0}\rangle$}

Let us now use the expressions of corrections to the bulk entanglement entropy for the state  $|\psi_{1, 0}\rangle$. This is the lowest excited state for the massive Chern-Simons field and, as discussed earlier, corresponds to the state $|1 + \frac{M}{2}, \frac{M}{2} \rangle$. The Bogoliubov coefficients have been evaluated in (\ref{eq: alpha m=1 n=0 large eta}), (\ref{eq: beta m=1 n=0 large eta}). The  boost parameter $\eta$  in the transformation (\ref{coordchange}) is related to the interval length as $\lim_{\eta\rightarrow\infty}(\cosh \eta)^{-1} = \pi x $ from (\ref{largeeta}). It is sufficient to retain only the leading
dependence on the interval length in the Bogoliubov coefficients to obtain the leading non-analytical contributions for the bulk
entanglement entropy. The \bbgv coefficients given by 
{\small \begin{eqnarray}
&&\lim_{\eta\rightarrow \infty}  \alpha_{1, 0;\omega, k } = - \lim_{\eta\rightarrow\infty} 
\beta_{1, 0, \omega, k }  \\ \nonumber
&& = \frac{ 2^M }{\Gamma( M+2) } \sqrt{ \frac{ M ( M +1) }{ \pi \omega}}  
(\cosh \eta)^{-(M+1)}\left|\Gamma(1+ i \om) \Gamma\left(\frac{M}{2}+\frac{i(k-\omega)}{2}\right) \Gamma\left(\frac{M}{2}+1+\frac{i(k+\omega)}{2}\right)\right|. 
\end{eqnarray} }
It is interesting to note that the Gamma functions that contain the left-moving $k+\omega$ and the right-moving $k - \omega$  combination of energy and momentum shifts according to the combination of left-moving and right-moving weights of the primary.
The Bogoliubov coefficients are proportional to $(\pi x)^{M+1}$.

\subsubsection*{First order contribution}
Substituting the Bogoliubov coefficients in the expression for the first-order correction to the bulk entanglement entropy in \eqref{eq: bulk EE 1st order formula}, we obtain the integral 
{\small \begin{align}\label{eq: integration 1st order bulk} 
S^{(1)}_{\rm bulk}(\Sigma_A) &= \frac{ 2^{2M+2} M ( M +1) }{ ( \Gamma ( M +2) )^2 } 
(\cosh \eta)^{-2(M+1)}  \\ \nonumber
& \times  \int_0^\infty  \frac{d \om}{2\pi}
\int_{-\infty}^\infty\frac{d k}{2\pi} \left| \Gamma(1+ i \om) \Gamma\left(\frac{M}{2}+\frac{i(k-\omega)}{2}\right) \Gamma\left(\frac{M}{2}+1+\frac{i(k+\omega)}{2}\right)\right|^2.
\end{align} }
These integrals can be done analytically as done in \cite{Colin-Ellerin:2024npf}. The integration over $k$  can be performed using the Barnes'-Beta integral and equation (5.13.3) of \cite{NIST:DLMF}. 
{\small \begin{align}\label{eq: integration k 1st order bulk}
\notag & \int_{-\infty}^\infty \frac{d k}{2\pi} \left|\Gamma\left(\frac{M}{2}+\frac{i(k-\omega)}{2}\right) \Gamma\left(\frac{M}{2}+1+\frac{i(k+\omega)}{2}\right)\right|^2\\
& = 2 \frac{\Gamma(M) \Gamma(M+2)}{\Gamma(2M+2)}\Big|\Gamma(M+1+i \om)\Big|^2.
\end{align}}
Then the integral over $\omega$ can be performed using equations (6.413) of \cite{gradshteyn2007}. 
\begin{align}\label{eq: integration om 1st order bulk}
& \notag \int_0^\infty \frac{d \om}{2\pi} \Big|\Gamma(1+i \om)\Gamma(M+1+i \om)\Big|^2\\
& = \frac{1}{2 \pi } \frac{\sqrt{\pi}}{2} \frac{\Gamma(M+1)\Gamma\big(M+\frac{3}{2} \big)\Gamma\big(\frac{3}{2}\big)\Gamma(M+2)}{\Gamma\big(M+\frac{5}{2} \big)}.
\end{align}
Combining \eqref{eq: integration k 1st order bulk} and \eqref{eq: integration om 1st order bulk}, we evaluate \eqref{eq: integration 1st order bulk} as
\begin{align} \label{finfirstord}
S^{(1)}_{\rm bulk}(\Sigma_A)_{|\psi_{1, 0 } \rangle }
= (\cosh \eta)^{-2(M+1)} \frac{ \Gamma\big( \frac{3}{2} \big) \Gamma(M+2)}{\Gamma\big(M+\frac{5}{2} \big)}.
\end{align}
On substituting the relation (\ref{largeeta}), we see that this term at the leading order precisely cancels the correction to the area term in (\ref{corrected area 1,0}) when we substitute for the bulk entanglement entropy in the FLM formula (\ref{flm}).

\subsubsection*{Second order contribution}

Let us proceed to evaluate the second-order term in the bulk entanglement entropy. Since at the leading order $|\alpha_{1, 0; \omega k }|  =  \beta_{1, 0; \omega k } | $, the expression in (\ref{2ndorderbulkf}) simplifies to 
{\small 
\begin{eqnarray}
S^{(1)}_{\rm bulk} ( \Sigma_A) = 
4 \pi \sum_{\stackrel{\omega_1, \omega_2}{k_1, k_2} } |\alpha_1|^2 |\alpha_2|^2 \Bigg[
 (\omega_1 - \omega_2) \frac{ ( 1-e^{-2\pi \omega_1} ) (1- e^{2\pi \omega_2} ) }{ 1- e^{2\pi ( \omega_2 - \omega_1)}}+ ( \omega_1 + \omega_2) \frac{ (1-e^{2\pi \omega_1})  (1-e^{2\pi \omega_2})}{ 1- e^{ 2\pi (\omega_1 + \omega_2) }}\bigg]. \nonumber \\
\end{eqnarray}
}
Substituting the expressions for the Bogoliubov coefficients, we obtain 
{\small 
\begin{eqnarray} \label{eq: integration bulk 2nd order}
S^{(2)} (\Sigma_A)\notag  &=& 4 \pi  \left(  \frac{ M ( M +1) 2^{2M} }{ \pi \Gamma( M + 2) }  \right)^2
 (\cosh \eta)^{-4(M+1)}  \\ \nonumber
& & \times  \notag \int \frac{ d\om_1 d\om_2 dk_1 dk_2}{(2\pi)^4}\Bigg[
 (\omega_1 - \omega_2) \frac{ ( 1-e^{-2\pi \omega_1} ) (1- e^{2\pi \omega_2} ) }{ 1- e^{2\pi ( \omega_2 - \omega_1)}}+ ( \omega_1 + \omega_2) \frac{ (1-e^{2\pi \omega_1})  (1-e^{2\pi \omega_2})}{ 1- e^{ 2\pi (\omega_1 + \omega_2) }}\bigg] \\
 & &\times \prod_{i=1,2} \frac{1}{\om_i}\left| \Gamma(1+ i \om_i) \Gamma\left(\frac{M}{2}+\frac{i(k_i-\omega_i)}{2}\right) \Gamma\left(\frac{M}{2}+1+\frac{i(k_i+\omega_i)}{2}\right)\right|^2.
\end{eqnarray}
}
Here, it is understood that the integrals in $\omega$ run from $0$ to $\infty$. The integrals over $k_1, k_2$ can be done using \cite{NIST:DLMF}. 
\begin{align}\label{eq: integration k 2nd order bulk}
& \notag \int_{-\infty}^\infty \frac{dk_i}{2\pi}  \left|\Gamma\left(\frac{M}{2}+\frac{i(k_i-\omega_i)}{2}\right) \Gamma\left(\frac{M}{2}+1+\frac{i(k_i+\omega_i)}{2}\right)\right|^2\\
&  =2\frac{\Gamma(M+2)\Gamma(M)}{\Gamma(2M+2)} \Big|\Gamma(M+1+ i \om_i) \Big|^2.
\end{align}
Now consider the integral over $\omega_1$, which is given by 
{\small \begin{eqnarray}
I(\omega_1)  &= &
\int_0^\infty \frac{d \om_1}{2 \pi}\Bigg\{  \bigg[ (\omega_1 - \omega_2) \frac{ ( 1-e^{-2\pi \omega_1} ) (1- e^{2\pi \omega_2} ) }{ 1- e^{2\pi ( \omega_2 - \omega_1)}}+ ( \omega_1 + \omega_2) \frac{ (1-e^{2\pi \omega_1})(1-e^{2\pi \omega_2})}{ 1- e^{ 2\pi (\omega_1 + \omega_2) }}\bigg]  \nonumber \\
&& \qquad\qquad \times 
\frac{1}{\om_1} \Big|\Gamma[M+1+ i \om_1]\Gamma[1+i \om_1] \Big|^2 \Bigg\}.
\end{eqnarray}
}
We can rewrite all the exponentials in terms of Gamma functions, using the reflection formula
{\small 
\begin{eqnarray} \label{eq: integration om1 2nd order bulk}
I &= &
 \frac{2 i \pi}{\Gamma(i \om_2) \Gamma(1- i \om_2)]}\int_0^\infty \frac{d \om_1}{2 \pi} \left \lbrace 
 \Big|\Gamma(1-i(\om_1+\om_2))\Big|^2+\Big|\Gamma\big(1-i(\om_1-\om_2)\big)\Big|^2 \right \rbrace 
 \Big|\Gamma(M+1+ i \om_1) \Big|^2 \nonumber \\
&=& \notag \frac{2 i \pi}{\Gamma(i \om_2)  \Gamma(1- i \om_2) }\int_{-\infty}^\infty \frac{d \om_1}{2 \pi}
 \Big|\Gamma\big(1+i(\om_1+\om_2 )\big)\Big|^2 
 \Big|\Gamma(M+1+ i \om_1]) \Big|^2 \\
&=& 2 i \pi \frac{\Gamma(2M+2)}{\Gamma(2M+4)}\frac{\Big|\Gamma(M+2+ i \om_2)\Big|^2}{\Gamma(i \om_2)\Gamma(1- i \om_2)}.
\end{eqnarray}
}
Lastly, we perform the $\om_2$ integral using \cite{gradshteyn2007}.
{\small \begin{eqnarray} \label{eq: integration om2 2nd order bulk}
J &=& 
\int_0^\infty \frac{d \om_2}{2 \pi} \frac{\Big|\Gamma(M+1+ i \om_2) \Gamma(M+2+ i \om_2)\Big|^2}{\Gamma(i \om_2)\Gamma(1-i \om_2)}\frac{\Big|\Gamma(1+ i \om_2)\Big|^2}{\om_2} \nonumber \\ 
 &=&  i \frac{1}{2 \pi}\frac{\sqrt{\pi}}{2}\frac{\Gamma(M+1)\Gamma(M+2)\Gamma\big(M+\frac{3}{2} \big)\Gamma\big(M+\frac{5}{2} \big) \Gamma ( 2M+3)}{\Gamma\big(2M+3+\frac{1}{2} \big) }.
\end{eqnarray}
}
Combining \eqref{eq: integration k 2nd order bulk}, \eqref{eq: integration om1 2nd order bulk} and \eqref{eq: integration om2 2nd order bulk} we evaluate \eqref{eq: integration bulk 2nd order} as
{\small \begin{eqnarray} \label{finsecondor}
\notag S^{(2)} (\Sigma_A) |_{|\psi_{1, 0} \rangle }&=&
4 \pi  \left(  \frac{ M ( M +1) 2^{2M} }{ \pi \Gamma( M + 2) }  \right)^2 (\cosh \eta)^{-4(M+1)} \\ \nonumber
&& \times \left(2\frac{\Gamma(M+2)\Gamma(M)}{\Gamma(2M+2)}\right)^2  \times 2 i \pi \frac{\Gamma(2M+2)}{\Gamma(2M+4)} \\ \nonumber
&& \times  i \frac{1}{2 \pi}\frac{\sqrt{\pi}}{2}\frac{\Gamma(M+1)\Gamma(M+2)\Gamma\big (M+\frac{3}{2} \big)\Gamma\big( M+\frac{5}{2} \big) \Gamma( 2M+3) }{\Gamma\big( 2M+3+\frac{1}{2} \big) }
\\
&=&   -(\cosh \eta)^{-4(M+1)} \frac{ \Gamma\big( \frac{3}{2} \big)  \Gamma(2M+3) }{\Gamma\big(2M+\frac{7}{2} \big) }.
\end{eqnarray}}
Let us now combine the corrections to the minimal area in (\ref{corrected area 1,0}), the first-order contribution (\ref{finfirstord}) and the second-order contribution  (\ref{finsecondor}) to evaluate all the terms in the FLM formula (\ref{flm})
\begin{eqnarray} \label{resultflmgs}
S_{\rm FLM} ( \rho_A)|_{|\psi_{1, 0 }\rangle}
  &=&  2 (1+M) (1 - \pi x \cot^{-1}(\pi x))    \\ \nonumber
   &&  -(\pi x )^{-4(M+1)} \frac{ \Gamma\big( \frac{3}{2} \big)  \Gamma(2M+3)}{\Gamma\big( 2M+\frac{7}{2} \big) } +\cdots. 
\end{eqnarray}
Here, we have used $\lim_{\eta\rightarrow\infty} (\cosh\eta)^{-1}  = \pi x $ in the short interval limit (\ref{largeeta}) in the terms that depend non-analytically on the interval. As mentioned earlier, in this limit, the correction to the minimal surface is precisely cancelled by the first-order correction to the bulk entanglement entropy. We are now in a position to compare this result to that evaluated in the CFT (\ref{gscft}). From the discussions  around equation (\ref{cftsingleid}) the state $|\psi_{ 0, 1}\rangle $ is dual to the primary with weights $h = 1+ \frac{M}{2}, \bar h = \frac{M}{2} $. Substituting these values in the expression from the CFT(\ref{gscft}), we see that it reduces to the result in (\ref{resultflmgs}), thus verifying the FLM formula for the lowest energy excitation of vectors in the bulk. 

As mentioned in the introduction, taking the massless limit in (\ref{resultflmgs}), the expression reduces to the leading terms evaluated in (\ref{u1cftres})  for the entanglement entropy of the $U(1)$  current in large $c$ CFT. 

\subsection{Bulk entanglement for $|\psi_{m, 0}\rangle$}

We can proceed with evaluating the bulk entanglement entropy for the tower of states  $|\psi_{m, 0}\rangle$, where only $m \geq 1$ values are allowed. From (\ref{dictdescn}) we see that these states are dual to descendants $ (L_{-1})^l | 1 + \frac{M}{2} , \frac{M}{2} \rangle$ with $l = m -1$. We need the Bogoliubov coefficients which relate the oscillators in $a_{m, 0 }$ in the global $AdS_3$ vacuum to those in the BTZ, which have been evaluated in the (\ref{ratiohigherb}). Just as was observed for states dual to the descendants of the scalar primary in \cite{Chowdhury:2024fpd}, the Bogoliubov coefficients of these holomorphic descendants in the short interval limit are proportional to those of the state $|\psi_{0, 1} \rangle$ in the large $\eta$ or short interval limit, i.e.,
\begin{eqnarray} \label{mainratio}
\lim_{\eta\rightarrow\infty}
\frac{  \alpha_{m ,0;\omega,k} }{\alpha_{1,0;\omega,k} } 
=\sqrt{\frac{\Gamma(M+m+1) }{\Gamma(m)\Gamma(M+2) } }.
\end{eqnarray}
It is therefore easy to evaluate the contribution to bulk entanglement from these states. To obtain the correction to the leading order bulk entanglement, we just need to multiply the result in (\ref{finfirstord}) by the square of the above ratio. This results in 
\begin{eqnarray} \label{forderm}
S^{(1)}_{\rm bulk}(\Sigma_A)_{|\psi_{m, 0 } \rangle }
= (\cosh \eta)^{-2(M+1)} \frac{ \Gamma\big( \frac{3}{2} \big) \Gamma(M+2) }{\Gamma\big( M+5/2 \big) } \times 
\frac{\Gamma(M+m+1)}{\Gamma(m)\Gamma(M+2) }.
\end{eqnarray}
Observe that again, this contribution is equal and opposite to the correction to the minimal area. For the second-order bulk entanglement entropy, we need to multiply the result in (\ref{finsecondor}) by the 4th power of the ratio in (\ref{mainratio}). This yields
\begin{eqnarray}\label{sorderm}
S^{(2)}_{\rm bulk}(\Sigma_A)_{|\psi_{m, 0 } \rangle }
=    -(\cosh \eta)^{-4(M+1) } \frac{ \Gamma( \frac{3}{2})  \Gamma(2M+3) }{\Gamma\big(2M+\frac{7}{2} \big) }
\times  \left(  \frac{\Gamma(M+m+1) }{\Gamma(m)\Gamma(M+2) } \right)^2.
\end{eqnarray}
We can now combine the correction to the minimal area (\ref{corrareades}) and the bulk entanglement, both the first-order term (\ref{forderm}) and the second-order term (\ref{sorderm}), to evaluate the terms in the FLM formula (\ref{flm}). We obtain 
\begin{eqnarray} \label{desflmres}
S_{\rm FLM} ( \rho_A)|_{|\psi_{m, 0 }\rangle} &=& 
 2 (M + m ) (1 - \pi x \cot(\pi x) )  \\ \nonumber
 && - 
(\pi x )^{-4(M+1) } \frac{ \Gamma\big( \frac{3}{2}\big)  \Gamma(2M+3) }{\Gamma\big(2M+\frac{7}{2} \big) }
\times  \left(  \frac{\Gamma(M+m+1) }{\Gamma(m)\Gamma(M+2)} \right)^2  + \cdots.
\end{eqnarray}
Now comparing this result to that in the CFT (\ref{maincftres}) for a descendant state at level $l=m-1$  with dimensions $h= 1 + \frac{M}{2}, \bar h = \frac{M}{2}$, we see that they precisely agree. Interestingly, the precise multiplicative factor that occurs at the first non-trivial correction in the short-distance expansion in CFT is reproduced in the bulk. Note that the bulk computation involves several steps, which kept track of the normalization of the wave functions both in the global $AdS_3$ and Rindler BTZ, which contributed to the ratio of the Bogoliubov coefficients in (\ref{mainratio}). 

There is a possible contribution to the entanglement entropy 
since the global vacuum is a sum over the superselection sectors.
This contribution may spoil the nice agreement of the result using the FLM formula in (\ref{resultflmgs}) and (\ref{desflmres}) with the CFT. In the next section, we study this issue in detail and show that the contribution of entanglement due to the superselection sectors vanishes, and indeed, the agreement of the FLM formula to the CFT is complete

\section{Edge modes of massive Chern-Simons fields}
\label{edgesection}

In this section, we focus on the $\omega=0$ sector in the Rindler-BTZ coordinates. The solutions to the Chern-Simons equations in this sector are given by,
\begin{eqnarray} \label{fullzeromode}
A_-^{(0, k)}  &=&  C_-^{(0, k )} \rho^{-M}  {}_2 F_1
 \Big( \frac{1}{2} ( M - i k ) , \frac{1}{2} ( M + i k ), M \frac{1}{\rho^2} \Big) , \\ \nonumber
  A_+ ^{(0, k )} &=&  C_+^{(0, k )} \rho^{-(M+2) }  {}_2 F_1
 \Big( \frac{1}{2} ( M +2  - i k ) , \frac{1}{2} ( M +2+  i k ), M \frac{1}{\rho^2} \Big), 
\end{eqnarray}
with the ratio,
\begin{eqnarray}
\frac{C_+^{(0, k ) } }{C_-^{(0, k ) }}  = - \frac{M^2 + k^2}{ 4 M ( 1+M ) }.
\end{eqnarray}
As we did for the bulk modes, we will determine the normalization $C_-^{(0, k ) }$ by requiring that the canonical commutation relations are satisfied. Before we proceed, let us examine the behaviour of these modes near the horizon at  $\rho = 1+\epsilon$. 
\begin{eqnarray} \label{expansionsh}
A_{-}^{(0, k)} |_{\rho =  1+ \epsilon }  &=&  -   C_-^{(0, k )} \frac{  \Gamma(  M) }{ \Gamma(\frac{M + i k}{2} ) 
 \Gamma(\frac{M - i k}{2} ) } \log ( 2 \epsilon  ) + O(\epsilon \log (\epsilon), \epsilon^0), \\ \nonumber
 A_+^{(0, k ) } |_{\rho =  1 + \epsilon }  &=&  C_-^{(0, k )} \frac{  \Gamma(  M) }{ \Gamma(\frac{M + i k}{2} ) 
 \Gamma(\frac{M - i k}{2} ) } \log ( 2 \epsilon  ) + O(\epsilon \log (\epsilon), \epsilon^0) , \\ \nonumber
 A_x^{(0, k ) }|_{\rho = 1+ \epsilon } &=&     C_-^{(0, k )} \frac{  \Gamma(  M) }{ \Gamma(\frac{M + i k}{2} ) 
 \Gamma(\frac{M - i k}{2} ) }   \log ( 2 \epsilon ) + O(\epsilon \log (\epsilon), \epsilon^0) , \\ \nonumber
A_\rho^{(0, k ) }|_{\rho = 1+ \epsilon}  &=& 
  \frac{C_-^{(0, k )}}{\epsilon} i k     \frac{  \Gamma(  M) }{ 4  \Gamma(\frac{M + i k +2}{2} ) 
 \Gamma(\frac{M - i k+2 }{2} ) } 
+ O (  \epsilon, \epsilon \log \epsilon  ) , \\ \nonumber
A_\tau^{(0, k ) }|_{\rho = 1+ \epsilon}  &=&  C_-^{(0, k ) } \frac{ M \Gamma(M) }
{2  \Gamma(\frac{M + 2 + i k}{2} ) \Gamma(\frac{M +2 - i k}{2} ) }   + O (  \epsilon, \epsilon \log \epsilon  ) . 
\end{eqnarray}
It is clear from these expansions that we can't satisfy the more restricted boundary  conditions (\ref{pmc}), since 
$A_\rho$ diverges. For non-zero $\omega$, we could choose $\omega$ to be discrete so that $A_{\rho}$ vanishes as in (\ref{discfreq}). Therefore, we will aim to satisfy the boundary conditions (\ref{normbc}) as operator relations in the zero-mode sector of the Hilbert space. 

\subsection*{Quantization of the zero mode sector}

Let us first quantize the zero-mode sector. We carry out the analysis in the right Rindler wedge. To do this, we need to examine the electric flux across an infinitesimal area element on the surface $\gamma$ and the radial Wilson line passing through a point on this surface, which are canonically conjugate variables. Using the last Chern-Simons equations in (\ref{cseqrin}), we see that the electric flux is proportional to $A_x$. While the radial Wilson line is given by,
\begin{eqnarray} \label{definwilson}
W( \tau, x) = \int_{1+\epsilon}^\infty d\rho A_\rho( \tau, \rho, x).
\end{eqnarray}
From the  commutation  relation in (\ref{dracbrakbtz}), we see that 
\begin{eqnarray} \label{wilsoncom}
[A_x( \tau, \rho = 1+ \epsilon, x), W( \tau, x') ] = \frac{i}{2} \delta( x-x'). 
\end{eqnarray}
We will use this commutation relation to quantize the edge modes. This will also fix the normalization $C_-^{(0, k ) }$ in (\ref{fullzeromode}). 

The behaviour of the field $A_x$ at the  brick wall $\rho = 1+ \epsilon$  in 
(\ref{expansionsh}), determines its mode expansion to be 
\begin{eqnarray} \label{nearhorax}
&&A_ x (\tau, \rho = 1+\epsilon, x  )  \\ \nonumber
&&=   \log( 2\epsilon) 
 \int \frac{dk}{2\pi }
 \frac{  \Gamma(  M) }{ \Gamma(\frac{M + i k}{2} ) 
 \Gamma(\frac{M - i k}{2} ) }  
   \left[ 
 C_-^{(0, k )}b_{0, k}  e^{i k x}  +   C_-^{(0, k ) *} b^\dagger_{0, k} 
 e^{ - i k x}   \right]. 
\end{eqnarray}
We have kept the leading order in the expansion of the wave function, which is sufficient for our purpose. To evaluate the Wilson line, it is useful to write down the wave function of the radial component of the Chern-Simons vector
{\small \begin{eqnarray}
A_{\rho}^{(0, k ) } =  \frac{C_-^{0,k} k  \rho ^{-M-1}  }{k-i M} 
 \left[\, _2F_1\big(\frac{1}{2} (M-i k),\frac{1}{2} (M+i k );M;\frac{1}{\rho ^2}\big)-\, _2F_1\big(\frac{1}{2} (M+2 -ik ),\frac{1}{2} (M + i k );M;\frac{1}{\rho ^2}\big)\right]. \nonumber \\
\end{eqnarray} }
Using this expression we can evaluate the Wilson line (\ref{definwilson}) for a definite momentum mode $k$ 
{\small  \begin{eqnarray}
W^{(0, k ) } &&  =  - \frac{C_-^{0,k} k  \rho ^{-M}  }{M( k-i M) } 
\left[\, _3F_2\big( \frac{1}{2} ( M-ik)  ,\frac{1}{2} ( M+i k ) ,  \frac{M}{2};M, \frac{M}{2}  + 1;\frac{1}{\rho ^2}\big) \right. \nonumber \\ 
& & \qquad\qquad\qquad\qquad  \left.  -\, _2F_1
\big(\frac{1}{2} ( M+2 - i k ) ,\frac{1}{2} ( M+ i k ) ;M, \frac{M}{2}  + 1 ;\frac{1}{\rho ^2}\big)\right] \Bigg|_{\rho = 1+ \epsilon}^\infty.
\end{eqnarray} }
Expanding this at the endpoints, we obtain 
\begin{eqnarray}
W^{(0, k ) } =   \frac{ i k \Gamma (M) }{ 4 \Gamma \big(\frac{ M+2 - i k }{2} \big) 
\Gamma \big(\frac{ M+2 +i k }{2} \big) } \log ( 2\epsilon)  + O( \epsilon^0, \epsilon \log(\epsilon) ). 
\end{eqnarray}
This allows us to expand the Wilson line in terms of Fourier modes as 
\begin{eqnarray}
&&W( \tau, x')  \\ \nonumber
&& = 
 \log(2 \epsilon) 
 \int \frac{dk}{2\pi }
 \frac{   \Gamma(  M) }{  4  \Gamma(\frac{M + i k+2 }{2} ) 
 \Gamma(\frac{M - i k+2 }{2} ) }  
   \left[ 
i k C_-^{(0, k ) } b_{0, k}  e^{i k x}  - i k  C_-^{(0, k ) *}  b^\dagger_{0, k} 
 e^{ - i k x}   \right]. 
\end{eqnarray}
We can now promote the Wilson line and the field $A_x$ in (\ref{nearhorax}) to be operators. Substituting this expansion in the commutation relations (\ref{wilsoncom}) and demanding the canonical commutation relations 
\begin{eqnarray} \label{righbcomm}
[b_{0, k, R}, b_{0, k, R }^\dagger] = 2\pi \delta( k - k'). 
\end{eqnarray}
We obtain the normalization
\begin{eqnarray} \label{zeromodnorm}
|C_{-}^{(0, k ) } |^2 = \frac{1}{|k|  \big[  \Gamma(M) \log( 2\epsilon  )\big]^2}    \Big|
  \Gamma\big(\frac{M + i k}{2} \big)  \Gamma\big(\frac{M  +2 + i k}{2} \big)  \Big|^2.
 %\Gamma(\frac{M - i k}{2} )  \Gamma(\frac{M  +2 - i k}{2} ). 
\end{eqnarray}
In (\ref{righbcomm}), we have reinstated the subscript $R$ to denote that we are in the right Rindler wedge. 
  
\subsection*{Edge states and boundary conditions}
  
It is convenient to define position and momentum operators by considering the linear combination 
\begin{eqnarray} \label{posmomop}
\hat a_{k, R } = i \frac{ b_{0, k,  R}-  b_{0, -k, R }^\dagger}{\sqrt{2} }  , \qquad \hat q_{k, R} = \frac{  b_{0, -k, R} + b_{0, k, R}^\dagger}{\sqrt{2}}.
\end{eqnarray}
As expected, these operators obey the commutation relations of position and momentum operators
\begin{eqnarray}
[\hat a_{k, R}, \hat q_{k', R}]  = 2\pi  i \delta( k -k'),
\end{eqnarray}
with the reality property,
\begin{eqnarray}
\hat a_{k, R} ^\dagger = \hat a_{-k, R}, \qquad  \hat q^\dagger_{k, R} = \hat q_{-k, R}. 
\end{eqnarray}
We will show subsequently that the modular Hamiltonian is diagonal in the momentum operators. These operators $a_{k, R}$ can be used to create normalized momentum eigenstates, which are defined by 
\footnote{$\sum_k = \int_{-\infty}^\infty \frac{dk}{2\pi }$. }
\begin{eqnarray}
\big|\{\varepsilon\}  \big\rangle_R = e^{ i \sum_k  \varepsilon_k  \hat a_{k, R} } |0\rangle_R , 
\end{eqnarray} 
Note that vacuum obeys $q_{k, R} |0\rangle_R  = 0 $. Therefore we have 
\begin{eqnarray}
\hat q_{k, R}  \big|\{\varepsilon\}  \big\rangle_R  = \varepsilon_k  \big|\{\varepsilon\}  \big\rangle_R. 
\end{eqnarray}
The parameters obey the reality condition  $\varepsilon_k^* = \varepsilon_{-k}$. The momentum eigenstates are normalized 
\begin{eqnarray}
{}_R\big\langle\{\varepsilon' \} \big| \{\varepsilon\}\big \rangle_R =  \prod_k  \delta ( \varepsilon'_k - \varepsilon_k). 
\end{eqnarray}
Here, for convenience, we restrict ourselves to the Hilbert space of the zero modes, but this Hilbert space occurs in a tensor product with the non-zero modes. We can repeat the same analysis in the left Rindler wedge. For this, we need to reverse the orientation of the Wilson line in (\ref{definwilson}), since the normal points inwards. This results in the commutation relation
\begin{eqnarray}
[b_{0, k, L}, b_{0, k, L  }^\dagger] = - 2\pi \delta( k - k'). 
\end{eqnarray}
We can go through the same construction of the position and momentum operators as in 
(\ref{posmomop}), which leads to 
\begin{eqnarray}
[\hat a_{k, L}, \hat q_{k', L}]  =-  2\pi  i \delta( k -k').
\end{eqnarray}
We then define the momentum eigenstate as 
\begin{eqnarray}
\big| \{\varepsilon \} \big\rangle_L  = e^{-  i \sum_k \varepsilon_k \hat a_{-k, L} } \big|0\big\rangle_L.
\end{eqnarray}
We use the momentum eigenstates to construct the  edge states as follows 
\begin{eqnarray}
\big| \{\varepsilon \} \big\rangle_R  \big| \{\varepsilon \} \big\rangle_L  = 
e^{   i \sum_k \varepsilon_k \hat a_{k, R}  -  i \sum_k \varepsilon_k \hat a_{-k, L} }|0\rangle_R \; |0\rangle_L.
\end{eqnarray}
It is these states that satisfy the condition 
\begin{eqnarray} \label{statebc1}
( \hat q_{k, R} - \hat q_{-k, L} ) \big| \{\varepsilon \} \big\rangle_R  \big| \{\varepsilon \} \big\rangle_L  = 0. 
\end{eqnarray}
Furthermore, it is easy to show using oscillator algebra that the overlap 
\begin{eqnarray}
{}_R \big\langle   \varepsilon'_{k}  \big|  \; {}_L \big\langle  \epsilon'_{k}  \big|       \big( \hat a_{k, L } - \hat a_{-k, R} \big) 
\big| \{\varepsilon \} \big\rangle_R  \big| \{\varepsilon \} \big\rangle_L  
&=&   i \partial_{\varepsilon_k}  \delta ( \varepsilon'_{k }- \varepsilon_k) \left( 
\prod_{k , k\neq k'} \delta( \varepsilon_k) -  \prod_{k , k\neq k'} \delta( \varepsilon_k)  \right) , \nonumber \\
& =& 0, 
\end{eqnarray}
where ${}_L\langle  \varepsilon_k  |=  {}_L\langle 0 |e^{- i \varepsilon_{k}^* \hat a_{-k} }$. 

Here, we have used the fact that the expectation value of the position operator on momentum eigenstates results in the derivative of the delta function. Note that if we take the overlap with states with non-zero $\varepsilon_{k'}, k' \neq k $, we again obtain a vanishing result. These results are due to the left-right symmetry of the physical states. So we say that the following state is null. 
\begin{eqnarray} \label{statebc2}
( \hat a_{k, L } - \hat a_{-k, R} )  \big| \{\varepsilon \} \big\rangle_R  \big| \{\varepsilon \} \big\rangle_L\approx 0.
\end{eqnarray}
The equations  (\ref{statebc1}) and (\ref{statebc2}) ensure that boundary conditions (\ref{normbc}) hold on the state $\big| \{\varepsilon \} \big\rangle_R  \big| \{\varepsilon \} \big\rangle_L$ since the electric field and the field $A_\rho$ can be expanded in terms of the operators $\hat q_k, \hat a_k$. 

\subsection*{Modular Hamiltonian and probability distribution of super selection sectors}

Each such state $\big| \{\varepsilon \} \big\rangle_R  \big| \{\varepsilon \} \big\rangle_L$  results in one particular field configuration ${\varepsilon}$. This is a given super selection sector to construct the edge state we need to sum over the super selection sectors with the probability as follows \footnote{ $\int{\cal D} {\varepsilon} = \int \prod_k d \varepsilon_k $ }
\begin{eqnarray} \label{defedgesup}
|E\rangle = \frac{1}{\sqrt{Z_{\rm edge}} } \int {\cal D} {\varepsilon}  {\cal D} {\varepsilon^* }  \sqrt{p ( {\varepsilon}, {\varepsilon^*} ) }\big| \{\varepsilon \} \big\rangle_R  \big| \{\varepsilon \} \big\rangle_L.
\end{eqnarray}
The probability with which the superselection sectors are distributed is determined by the modular Hamiltonian, and $\sqrt{Z_{\rm edge}}$ ensures that the state is normalized. For the zero-mode sector, the modular Hamiltonian is defined by 
\begin{eqnarray}
K^{\rm edge} = 2\pi \int_{\rho = 1+\epsilon}^\infty d \rho \sqrt{g_{\Sigma} } \xi^\mu \eta^\nu T_{\mu\nu}|_{\omega =0},
\end{eqnarray}
where 
\begin{eqnarray} \label{quantmodular}
\sqrt{ g_{\Sigma}} = \frac{\rho}{\sqrt{\rho^2-1}} , \qquad
\eta^\mu = \Big( \frac{1}{\sqrt{\rho^2 - 1}}, 0 , 0 \Big) , \qquad \xi^\mu = ( 1, 0 , 0 ), 
\end{eqnarray}
${g_{\Sigma}}$ is the determinant of the metric on the $\tau =0$ slice, $\eta^\mu$ is the unit normal to $\Sigma$, and $\xi^\mu$ is the time-like Killing vector corresponding to boosts in the Rindler BTZ coordinates. The stress tensor is that of the massive Chern-Simons theory, which is given in  (\ref{stressmassivcs}). Substituting the expressions from (\ref{quantmodular}) we obtain 
\begin{eqnarray}
K^{\rm edge}  = 2\pi \int _{\rho = 1+\epsilon}^\infty d\rho dx  \frac{\rho}{\rho^2-1  } T_{\tau\tau},
\end{eqnarray}
where the stress tensor of the massive Chern-Simons theory is given by 
\begin{eqnarray}
T_{\tau\tau} = 2M :\Big( A_\tau A_\tau - \frac{g_{\tau\tau }}{2} ( A_\mu A^\mu ) \Big):.
\end{eqnarray}
Now, using the Chern-Simons equations of motion together with the fact that the $\omega =0$ modes are time independent, we obtain
\begin{eqnarray}
K^{\rm edge}  = 2\pi \int _{\rho = 1+\epsilon}^\infty d\rho dx \Big[ - \partial_\rho ( A_\tau A_x)  + \partial_x( A_\tau A_\rho )  \Big].
\end{eqnarray} 
It is easy to see that by momentum conservation, the 2nd term vanishes, and we are left with 
\begin{eqnarray} \label{simpmodham}
K^{\rm edge}  =  2\pi \lim_{\rho = 1+ \epsilon}  \int_{-\infty}^\infty  dx A_\tau A_x.
\end{eqnarray}
The expansion of $A_\tau$ near the horizon can be written from (\ref{expansionsh}) and  is given by
\begin{eqnarray} \label{atauzex}
&& A_ \tau(\tau, \rho = 1+\epsilon, x  )  \\ \nonumber
& & = \int \frac{dk}{2\pi }
 \frac{ M  \Gamma(  M) }{  2 \Gamma(\frac{M + 2+  i k}{2} ) 
 \Gamma(\frac{M+2  - i k}{2} ) }  
   \left[ 
 C_-^{(0, k )}b_{0, k, R}  e^{i k x}  +   C_-^{(0, k ) *} b^\dagger_{0, k, R} 
 e^{ - i k x}   \right]. 
\end{eqnarray}
Here, we have looked at the expansion in the right Rindler wedge, and we will see that due to the property (\ref{statebc1}), it does not matter in which wedge we evaluate the modular Hamiltonian. The important aspect of the expansion of $A_\tau$ is that it does not have $\log(2\epsilon)$ dependence. This leads to the suppression of the modular Hamiltonian, which results in the fact that the edge modes do not contribute to the entanglement entropy in the limit $\epsilon\rightarrow 0$.

We can substitute the expansions (\ref{nearhorax}) and (\ref{atauzex}) 
into the expressions for the modular Hamiltonian (\ref{simpmodham}) to obtain
\begin{eqnarray}
K^{\rm edge}
 & =&  2 \pi \int \frac{dk}{2\pi }   |C_-^{(0, k ) } |^2   \frac{   \log( 2\epsilon  )  \Gamma(  M) }{ \Gamma(\frac{M + i k}{2} ) 
 \Gamma(\frac{M - i k}{2} ) }    \frac{ M  \Gamma(  M) }{  2 \Gamma(\frac{M + 2+  i k}{2} ) 
 \Gamma(\frac{M+2  - i k}{2} ) }  \\ \nonumber
 & & \times \left( 
 b_{0,  k, R} b_{0 , -k, R} +    b_{0,  k, R}  b^\dagger_{0,  k , R }  + b^\dagger_{0 , k , R} b_{0 , k , R}+ b^\dagger_{0,  k, R} 
 b^\dagger_{0,  -k , R}
 \right). 
\end{eqnarray}
Using the normalization found in (\ref{zeromodnorm}) results in 
\begin{eqnarray}
K^{\rm edge}  &=&  2 \pi \int \frac{dk}{2\pi }   \frac{M}{ 2 |k|  |\log( \epsilon)| }   
\left( 
 b_{0,  k, R} b_{0 , -k, R} +    b_{0,  k, R }b^\dagger_{0,  k, R}  + b^\dagger_{0,  k, R} b_{0,  k, R} + b^\dagger_{0,  k, R} b^\dagger_{0,  -k, R }
 \right).  \nonumber \\
\end{eqnarray}
We write this expression in terms of the momentum operators defined in (\ref{posmomop}) to obtain 
\begin{eqnarray} \label{modhaminq}
K^{\rm edge}   &=&  2 \pi \int \frac{dk}{2\pi }   \frac{M }{  |k|  |\log( 2 \epsilon)| }    \left( \hat q_{k, R} \hat q_{-k, R} \right).
\end{eqnarray}
On evaluating the expectation value of the modular Hamiltonian on the edge state, we get 
\begin{eqnarray}
{}_R \big\langle \{\varepsilon \} \; \big| {}_R \big\langle \{\varepsilon \} \big|
 K^{\rm edge} \big| \{\varepsilon \} \big\rangle_R  \big| \{\varepsilon \} \big\rangle_L =
 2 \pi \int \frac{dk}{2\pi }   \frac{M |\varepsilon_k |^2}{  |k|  \log(  2 \epsilon) }.  
\end{eqnarray}
The structure of the modular Hamiltonian in (\ref{modhaminq}) and its expectation value on the edge stage is similar to that found by \cite{Colin-Ellerin:2024npf} for the massless vector field in higher dimensions. Here, since the stress tensor is proportional to the mass, we obtain the factor $M$ as an overall factor. Furthermore, note that the entanglement Hamiltonian is diagonal in the momentum eigenstates, hence the probability distribution is given by 
\begin{eqnarray}
p( \{\varepsilon\}, \{\varepsilon^*\} ) =  e^{- 2 \pi \int \frac{dk}{2\pi }   \frac{M | \varepsilon_k|^2  }{  |k|  \log(  2 \epsilon) }
}.
\end{eqnarray}
%
%We are now ready to evaluate the density matrix and the corrections to entanglement entropy. 
%The modular  Hamiltonian is diagonal in the momentum eigen states, therefore we can the 
%probability distribution as 
%\begin{eqnarray}
%p( \{\varepsilon\}, \{\varepsilon^*\} ) =  2 \pi \int \frac{dk}{2\pi }   \frac{M}{  |k|  \log(  2 \varepsilon) }  
%\left( \varepsilon_k  \varepsilon_{-k} \right) 
%\end{eqnarray}
Therefore, the reduced density matrix of the vacuum restricted to the Hilbert space of the edge modes obtained by tracing out the left sector is given by
\begin{eqnarray}
\rho_{\rm edge} &=&   {\rm Tr}_{L}  |E\rangle \langle E| \\ \nonumber
&=& \frac{1}{ Z_{\rm edge}  }  \int  {\cal D}\epsilon  {\cal D} \varepsilon^*  \;
e^{  -2 \pi \int \frac{dk}{2\pi }   \frac{M | \varepsilon_k |^2  }{  |k|  \log(  2 \epsilon) } 
 }  
\big| \{\varepsilon \} \big\rangle \big \langle  \{\varepsilon \} \big|,
\end{eqnarray}
where we have suppressed the $R$ label. The normalization  $Z_{\rm edge}$ ensures that the density matrix is normalized. The full reduced density matrix of the vacuum is the tensor product of the non-zero mode along with the edge modes. 

\subsection*{Edge state contribution to the vacuum subtracted entanglement}

Let us evaluate the change in the density matrix of the edge modes due to the excitation 
$a^\dagger_{1, 0}|0\rangle = |\psi_{1,0} \rangle$ on the global vacuum.  A similar analysis will carry through for all other states. As mentioned earlier, we need the Bogoliubov coefficients, which relate the creation operators in the Global $AdS_3$ vacuum to the Rindler-BTZ vacuum. We have already treated the modifications due to the Bogoliubov coefficients of the bulk modes. Here we focus on the edge modes or the $\omega =0$ sector.
\begin{eqnarray} \label{globalzerocon}
a_{1, 0 } = a_{1, 0; {\rm bulk} } +  
 \sum_{I, k } \left(  \hat\alpha_{1, 0 ;  k, I } \hat a_{k, I} +  \hat \beta_{1, 0; k, I } \hat q_{k, I} \right) , \\ \nonumber
a_{1,0}^\dagger = a_{1, 0\, {\rm bulk}  }^\dagger + 
\sum_{I, k } \left(  \hat\alpha_{1, 0 ; - k, I }^* \hat a_{k, I} +  \hat\beta_{1, 0; -k, I }^* \hat q_{-k, I} \right).
\end{eqnarray}
The sum over $I$ runs over $L, R$, and $\hat \alpha, \hat \beta$ refer to the Bogoliubov coefficients in the zero-mode sector. $a_{1, 0; {\rm bulk} }$ refers to the bulk contribution of the oscillator written as an expansion in bulk Rindler modes (\ref{eq: definition bogo}). We know that the global $AdS_3$ vacuum is annihilated by $a_{1, 0 } $.  The evaluation of the Bogoliubov coefficients for the bulk modes in Rindler-BTZ ensures that these modes annihilate this vacuum due to the relations (\ref{eq: relation operator left right}) and (\ref{eq: relation bogo left right}),  that is $a_{1, 0; {\rm bulk} } $ annihilates the $AdS_3$ vacuum. This also has to hold for the edge mode sector, which results in the equation
\begin{eqnarray}\label{edgeannihilate}
 \sum_{I, k } \left(  \hat\alpha_{1, 0 ;  k, I } \hat a_{k, I} +  \hat\beta_{1, 0; k I } \hat q_{k, I} \right)|E\rangle =0.
\end{eqnarray}
From equation (\ref{statebc1}) and the fact that we need the state to be null as in (\ref{statebc2}), we obtain the relations 
\begin{eqnarray}
\hat\alpha_{m, n ; k , R }  = - \hat\alpha_{m, n ; -k, L}, \qquad  \hat \beta_{m, n ; k , R  } = -\hat\beta_{m, n ;  -k, L },
\end{eqnarray}
which ensures that the (\ref{edgeannihilate}) holds. 
  
We can evaluate the correction on the vacuum states of the zero modes
\begin{eqnarray}
( a_{1,0}^\dagger -a_{1, 0\, {\rm bulk}  }^\dagger) |0\rangle &=&  \sum_k  \hat \alpha^*_{1, 0 ; - k R} ( \hat a_{k, R}  - \hat a_{-k, L})   |E\rangle, \\ \nonumber
&\equiv& \delta\big(  |E\rangle\big).
\end{eqnarray}

Observe that only the Bogoliubov coefficient $\hat \alpha$ contributes to this correction due to the property $(\ref{statebc1})$ of the edge state. Let us proceed with expanding the RHS
\begin{eqnarray}
\delta \big( |E\rangle \big) &=& 
\int {\cal D} \varepsilon {\cal D} \varepsilon^* 
e^{-  \pi \int \frac{dk'}{2\pi }   \frac{M  |\varepsilon_{k'}|^2 }{  |k'|  \log(  2 \epsilon) }  }
\sum_k  \hat\alpha^*_{1, 0 ; - k R} ( \hat a_{k, R}  - \hat a_{-k, L}) 
\big| \{\varepsilon \} \big\rangle_R  \big| \{\varepsilon \} \big\rangle_L, \nonumber \\
&=&  - i \sum_k  \hat \alpha^*_{1,0; - k, R}  \int {\cal D} \varepsilon {\cal D} \varepsilon^* 
 \Big[
e^{-  \pi \int \frac{dk'}{2\pi }   \frac{M   |\varepsilon_{k'}|^2 }{  |k'| \log(  2 \epsilon) }  }\Big]
 \partial_{\varepsilon_k} \big| \{\varepsilon \} \big\rangle_R  \big| \{\varepsilon \} \big\rangle_L.
\end{eqnarray}
We  integrate by parts and obtain
\begin{eqnarray}
\delta \big( |E\rangle \big) &=&  i 
  \sum_k  \hat \alpha^*_{1, 0 ; - k,  R}  \int {\cal D} \varepsilon {\cal D} \varepsilon^* 
 \partial_{\varepsilon_k} \Big[
e^{-  \pi \int \frac{dk'}{2\pi }   \frac{M   |\varepsilon_{k'}|^2 }{  |k'| \log(  2 \epsilon)}  }\Big]
 \big| \{\varepsilon \} \big\rangle_R  \big| \{\varepsilon \} \big\rangle_L.
\end{eqnarray}
Here, it is understood that the derivative acts as
\begin{eqnarray}
\partial_{\varepsilon_k} \varepsilon_{k'} = \delta( k - k'). 
\end{eqnarray}
This change in the edge state results in the following correction of the reduced density matrix, which is given by 
{\small \begin{eqnarray}
\delta \rho_{\rm edge} = \frac{1}{Z_{\rm edge} }  \sum_k  |\hat \alpha_{1, 0 ; - k,R}  |^2
 \int {\cal D}  \varepsilon{\cal D} \varepsilon^* 
 \partial_{\varepsilon_k} \Big[
e^{-  \pi \int \frac{dk'}{2\pi }   \frac{M   |\varepsilon_{k'}|^2 }{  |k'| |\log(  2 \epsilon)| }  }\Big] 
\big| \{\varepsilon \} \big\rangle \big \langle  \{\varepsilon \} \big|  \partial_{\varepsilon_k ^*} \Big[
e^{-  \pi \int \frac{dk'}{2\pi }   \frac{M   |\varepsilon_{k'}|^2 }{  |k'| |\log(  2 \epsilon)| }  }\Big].  \nonumber \\
\end{eqnarray}}
The correction to the entanglement entropy is given by \footnote{We have restricted ourselves to the zero mode sector for a less cumbersome notation. The zeroth-order density matrix has a part that involves the bulk modes, which cancels against the normalization in the evaluation of the trace.}
\begin{eqnarray} \label{finaledge}
S ( \delta \rho_{\rm edge} ) |_{|\psi_{1, 0} \rangle } &=&-  {\rm Tr}( \delta\rho_{\rm edge} \rho_{{\rm edge}}) , 
\\ \nonumber
&=&- \int \frac{dk}{2\pi}   |\alpha_{1, 0 ; k,R  }  |^2 \frac{\pi M}{|k \log( 2\epsilon) | }. 
\end{eqnarray}
Substituting the value of the Bogoliubov coefficient in the zero mode sector derived in (\ref{finalbgza}), we obtain
\begin{eqnarray}
S ( \delta \rho_{\rm edge} ) |_{|\psi_{1, 0}\rangle} &=&  \frac{M (M+1)  2^{2(M+1)}
   (\cosh \eta)^{-2( M+1)} }{ 2 |\log 2\epsilon| \big( \Gamma\left(M+2\right) \big)^2}  
%  2^{2(M+1)}
%   (\cosh \eta)^{-2( M+1)}
%\int \frac{dk}{2\pi} 
%   2^{2(M+1)}
%   (\cosh \eta)^{-2( M+1)}
   \nonumber \\
     & &  \qquad  \times \int  \frac{dk}{2\pi} 
     \left|  \Gamma\left(\frac{M}{2}+\frac{i k}{2}\right)  \Gamma\left(\frac{M}{2}+1+\frac{i k }{2}\right) 
     \right|^2.
\end{eqnarray}
The integral over $k$ is convergent, which implies that the contribution to the vacuum subtracted entanglement entropy due to the edge modes is suppressed as the brick wall cut off is taken to zero, $\epsilon \rightarrow 0$. The reason this occurs is that the modular Hamiltonian or the probability distribution of the edge modes is suppressed by $|\log(2\epsilon) |^{-1}$. 

On proceeding to evaluate the 2nd order contribution to vacuum subtracted entanglement in the short distance expansion, this feature of the probability distribution, together with the fact that at 2nd order we have a quartic power of the Bogoliubov coefficient, ensures that this contribution too is suppressed in the  $\epsilon \rightarrow 0$ limit. 

Though we have not repeated the evaluation of the contribution of the edge modes for the descendants, it is reasonably easy to see that the suppression of this contribution will occur in the descendants due to the same features seen for the primary. This concludes the complete verification of the FLM formula for the case of the massive vectors in $AdS_3$. 

\section{Conclusions} \label{conclus}
 
We have quantized the massive Chern-Simons fields both in the global $AdS_3$ and Rindler BTZ. We have also evaluated the Bogoliubov coefficients relating the creation(annihilation) operators of these fields in global $AdS_3$ to those in the Rindler-BTZ. Using the results of this quantization, we extended the approach of \cite{Belin:2018juv,Chowdhury:2024fpd} to obtain the entanglement entropy of perturbative vector excitations in $AdS_3$.  We considered the massive Chern-Simons field, which is dual to a state of spin $1$ and conformal dimensions $M+1$. We showed that the quantum corrections to the entanglement entropy of a single interval given by the FLM formula precisely agree with those evaluated in large $c$ CFT. 

On taking the massless limit, we showed that the leading and sub-leading contributions to the entanglement entropy using the FLM formula coincide with the exact expression obtained using methods of topological Chern-Simons theory in the bulk \cite{Belin:2019mlt}. 
We should point out that those methods relied on using only edge modes on the `entanglement cut' or the Ryu-Takayanagi surface to evaluate $S_{\rm bulk}$. Our approach is different; we turn on the mass, which breaks the gauge symmetry as well as the topological nature of the Chern-Simons theory. This renders the check similar to that developed in \cite{Belin:2018juv,Chowdhury:2024fpd,Colin-Ellerin:2024npf,Colin-Ellerin:2025dgq}. In this approach, we showed that the edge modes do not contribute to the vacuum-subtracted entanglement entropy.

We perform the calculation with non-zero mass and, in the end, set the mass to zero. The fact that the leading terms agree with those of the methods in \cite{Belin:2019mlt} of dealing with the massless Chern-Simons theory is satisfying. We also mention that the analysis in this paper was done with $M>0$; we can repeat the analysis with $M<0$ and obtain the same conclusions. In that case, the lowest energy state of the massive Chern-Simons field will be dual to the primary of dimension $| \ha |M|, 1+ \ha |M| \rangle$. 
This agreement at zero mass is interesting and worth investigating further. As mentioned in the introduction, the methods of \cite{Belin:2019mlt} involved topological Chern-Simons theory, and the contribution of the entanglement was only due to edge modes of the `entanglement cut'. In our study, there was no contribution from the edge modes, and the naive assumption is that, setting $M\rightarrow 0$, the origin of entanglement continues to arise from the bulk modes. Observe that in the setting $M=0$, the edge mode contribution in (\ref{finaledge}) vanishes. Perhaps the bulk-boundary duality of topological Chern-Simons can be used to understand this phenomenon. That is, there is a duality relating topological Chern-Simons theory to chiral $U(1)$ currents on any boundary, not just on the $AdS$ boundary. It was this feature that was exploited in  \cite{Belin:2019mlt} to arrive at the result for the vacuum-subtracted entanglement. 
It is interesting to investigate this issue further. We suspect the same behaviour for gravitons and higher spin fields in $AdS_3$ since they are all topological, their degree of freedom lies on the boundary of the manifold. 

In fact, this approach of turning on the mass, to break gauge symmetry and evaluate entanglement, is interesting. It is worth re-examining the calculations of the entanglement entropy of the Maxwell field as well as the graviton across a spherical surface in $d=4$ done in \cite{Soni:2016ogt,David:2022jfd} with this approach and check if, on taking the massless limit, the coefficient of the logarithmic term coincides with or without the inclusion of the edge modes. 

Finally, in \cite{Datta:2011za,Datta:2012gc}, the wave equation of massive higher spin bosons, including gravitons and fermions, has been solved in the BTZ geometry. As done in this paper, it was the method of solving the equations developed in \cite{Datta:2011za} that helped us quantize the massive Chern-Simons fields both in global $AdS_3$ and BTZ. 
These methods can be generalised to higher spin fields, especially that of the graviton, which is interesting to study. In \cite{Colin-Ellerin:2025dgq}, a `quantum extremal gauge' was chosen to define the area operator of the graviton in higher dimensions. Since, in three dimensions, it is easy to turn on a mass for the graviton so that gauge symmetries 
as well as its topological nature can be broken, it will be interesting to check how all the details for the verification of the FLM formula go through. 

\appendix 
\section{ Global $AdS_3$ and Rindler  BTZ }

The metric of $AdS_3$ in global coordinates is given by 
\begin{eqnarray}
ds^2 = -(1+ r^2) dt^2 + \frac{dr^2}{ 1+ r^2} + r^2 d\varphi^2.
\end{eqnarray}
Note that the radius of $AdS_3$ is set to unity. The non-zero values of the Christoffel symbols  in this background are given by 
\begin{eqnarray} \label{chrisglobal}
&& \Gamma^{t}_{tr} =  \frac{r}{ 1+r^2},  \\ \nonumber
&& \Gamma^{r}_{tt} = r( 1+r^2), \qquad \Gamma^r_{rr} = - \frac{r}{ 1+r^2 } , \qquad
\Gamma^{r}_{\varphi\varphi} = - r( 1+r^2),  \\ \nonumber
&& \Gamma^\varphi_{r\varphi} = \frac{1}{r}.
\end{eqnarray}
The Riemann curvature in global $AdS_3$ is given by
\begin{eqnarray}
R_{\mu\nu\rho\sigma} = g_{\mu\sigma} g_{\nu\rho} - g_{\mu \rho}g_{\nu\sigma}.
\end{eqnarray}
The metric of Rindler-BTZ is given by 
\begin{eqnarray} \label{btzrinda}
ds^2 = - ( \rho^2 -1) d\tau^2 + \frac{d\rho^2}{\rho^2 -1} + \rho^2 dx^2,
\end{eqnarray}
with $x, \tau$ taking values from $(-\infty$, $\infty)$, and $\rho$ taking values from $1$ to $\infty$. $\tau$ is periodic in the imaginary direction  and is identified as $\tau \sim \tau + 2\pi i$. The non-zero values of the Christoffel connections in this background are given by 
\begin{eqnarray} \label{chrisbtz}
&& \Gamma^{\tau}_{\tau \rho} = \frac{\rho}{ \rho^2 -1}, \\ \nonumber
&& \Gamma^\rho_{\tau\tau} = \rho( \rho^2 -1),  \qquad 
 \Gamma^\rho_{\rho\rho} = - \frac{\rho}{\rho^2 -1}, \qquad  \Gamma^\rho_{xx} = - \rho( \rho^2 -1), 
 \\ \nonumber
 && \Gamma^x_{\rho x} = \frac{1}{\rho}.
\end{eqnarray}
The Riemann curvature of the Rindler-BTZ space also satisfies the condition
\begin{eqnarray}
R_{\mu\nu\rho\sigma} = g_{\mu\sigma} g_{\nu\rho} - g_{\mu \rho}g_{\nu\sigma}.
\end{eqnarray}

\section{Vector modes in $AdS_3$ and their quantization} \label{appenb}

In this appendix, we provide the details of obtaining the solutions to the massive Chern-Simons equations of motion (\ref{cherneq}) in $AdS_3$ as well as their quantisation. The massive Chern-Simons equations are given by 
\begin{eqnarray} \label{cherneqa}
\frac{g_{tt} }{ \sqrt{-g} }\Big( \partial_r A_{\varphi} - \partial_{\varphi} A_r \Big) = - M A_t, 
\\ \nonumber
\frac{g_{rr}}{\sqrt{-g} } \Big( \partial_{\varphi} A_t - \partial_t A_{\varphi} \Big) = - M A_r, 
\\ \nonumber
\frac{g_{\varphi\varphi} }{\sqrt{-g} } \Big( \partial_t A_r - \partial_r A_t \Big) = 
- M A_{\varphi }.
\end{eqnarray}
In (\ref{2ndorder}), we have shown that using the Chern-Simons equations, we can obtain the decoupled second-order equations for the components $A_-$ and $A_+$. 
\begin{eqnarray} \label{2ndordera}
( \Box  -( M + 1)^2 +1 ) A_+ =0, \\ \nonumber
( \Box  -( M - 1)^2 +1 ) A_- =0.
\end{eqnarray}
Using the fact that the directions $\varphi$ and $t$ are isometries, and then solving the radial equation, we can obtain  the following modes  
{\small 
\begin{eqnarray}\label{aminus}
A_ -^{(m, n)} &=&  R_-^{(m, n)} e^{- i \Omega_{m \,n} t} e^{ i m \varphi},\\ \nonumber
R_{-}^{(m, n)} &=& C_-^{(m,n)} r^{|m| } (1+ r^2)^{ \frac{\Omega_{m, n }}{2} } 
{}_2 F_1\Big( \frac{1}{2} ( 2 + \Omega_{(m, n)} + |m|  -M ), 
\frac{1}{2} ( \Omega_{(m,  n)} + |m|  +M ), 1+ |m|, -r^2 \Big), 
\end{eqnarray}  }
where 
\begin{eqnarray}
\Omega_{(m, n)} = 2 n + |m| +M ,  & & \qquad n =0, \qquad m = 1, 2, 3,  \cdots \\ \nonumber
& & \qquad n =1,2, \cdots   \qquad m \in \mathbb{Z}.  
\end{eqnarray}
The superscript on $A_-$ refers to the fact that we are focused on one particular mode. These wavefunctions are smooth at the origin, and normalizable at infinity for $2M+1>0$. We will see subsequently that the condition for $n=0$, only positive integral values of $m$ are allowed, arises from the requirement that the vector components solve the first-order Chern-Simons equations. Similarly, the modes for $A_+$ are given by 
{\small 
\begin{eqnarray}\label{aplus}
A_+^{(m ,n)} &=&  R_+^{(m, n)} e^{- i \Omega_{(m ,n)} t} e^{ i m \varphi}, \\ \nonumber
R_{+}^{(m, n)} &=& C_+^{(m, n)} r^{|m| }  (1+ r^2)^{ \frac{\Omega_{(m, n) }}{2} } 
{}_2 F_1\Big( \frac{1}{2} (   \Omega_{(m, n)} + |m|  -M ), 
\frac{1}{2} ( 2+ \Omega_{(m, n)} + |m| +M ), 1+ |m|, -r^2 \Big), 
\end{eqnarray} }
where 
\begin{eqnarray}
\Omega_{m\, n} = 2 n + |m| +M ,  & &\qquad n =1, 2, 3,\cdots,  \qquad m \in \mathbb{Z}.
\end{eqnarray}
Note that $n=0$ is not allowed for the modes of $A_+$. Also, observe that the modes of $A_-$ are related to those of  $A_+$ by the shift $M\rightarrow M+2,$ which is expected, since the 
equations in (\ref{2ndordera}) are related by this shift of $M$.    The constants $C_+^{(m, n)}$ will be related to the constants $C_-^{(m, n)}$  in (\ref{aminus}) by demanding that the Chern-Simons equations are satisfied. It is important to point out that the
complex conjugate of these solutions in (\ref{aminus}) and (\ref{aplus}) will also solve the 2nd order equations (\ref{2ndordera}). 

Using the 2nd Chern-Simons equations in  (\ref{cherneqa}), we can find the component $A_r$ for each mode of $A_\pm$, and then it can be checked whether these three components consistently solve the remaining two equations in (\ref{cherneqa}). An easy way to check for this consistency is to examine the behaviour at $r\rightarrow 0$. 

\subsection*{$n=0$}

Let us do this consistency check for the $n=0$ modes. We will find that the first-order Chern-Simons equations are satisfied if the condition $m>1$ is imposed. For $n=0$ modes only $A_-^{m, 0}$ is non-zero and $A_+^{m,0}$ vanishes. Using this, we obtain 
\begin{eqnarray}
A_{t}^{ (m,  0)  }  &=&   \frac{A_-^{( m, 0 )} }{2}, \\ \nonumber
A_{\varphi}^{( m,  0)  } &=&  - \frac{A_-^{( m,  0)  } }{2}, \\ \nonumber
A_{r}^{(m, 0 )}  &=&  \frac{i}{M} \frac{1}{ r ( 1+ r^2) } ( M - m + |m| )  \frac{A_-^{( m , 0 ) } }{2} .
\end{eqnarray}
To obtain $A_r$, we have used the 2nd Chern-Simons equation in (\ref{cherneqa}). It is clear that if $A_r$ needs to be regular at $r=0$, we need $|m|\geq 1$. We can check the consistency of this solution by substituting these components in the first Chern-Simons equation for $A_t$ in (\ref{cherneqa}). 
%Let us study this  for $m\leq -1$. Then we have 
%\begin{equation}
%A_{r}^{m , 0}  =  \frac{i}{M} \frac{1}{ r ( 1+ r^2) } ( M -  2 m )  \frac{A_-^{m \, 0 } }{2} , \qquad m<0
%\end{equation}
Now at $r\rightarrow 0$, the LHS of the Chern-Simons equation is given by 
\begin{equation}\label{lim1}
\lim_{r\rightarrow 0}  -\frac{ ( 1+r^2) }{r} \left(-  \frac{1}{2} \partial_r  A_-^{(m, 0 ) } - i m  A_r^{( m, 0)} \right) 
=-  \frac{ |m| - m }{2 M} r^{|m| - 2}  C_-^{( m, 0)  } + O (r^{|m|} ). 
\end{equation}
The  behaviour of the RHS of the  Chern-Simons equation (\ref{cherneqa}) at the origin is given by 
\begin{equation}\label{lim2}
\lim_{r\rightarrow 0 } - \frac{M}{2}  A_-^{( m , 0)  }  = - \frac{M}{2}   C_-^{(m, 0) }  r^{|m|} .
\end{equation}
Comparing the behaviour in (\ref{lim1}), ( \ref{lim2}), we see that non-trivial solutions can exist only for positive $m$ and therefore  $m\geq 1$. This results in 
\begin{eqnarray}
A_r^{( m, 0)  } = \frac{i }{ 2 r ( 1+ r^2) } A_-^{( m, 0) } , \qquad m \geq 1.
\end{eqnarray}
It is easy to verify that the remaining two Chern-Simons equations are satisfied for $m\geq 1$ due to the properties of the hypergeometric functions, which ensure that  for $m\geq 1$ 
\begin{eqnarray}
\frac{1+r^2}{r} \left(- \frac{1}{2} \partial_r A_- ^{ ( m , 0 ) }  - i m A_r^{( m, 0 ) } \right) = \frac{M}{2}  A_-^{( m,  0) }
, \qquad  m\geq 1 \\ \nonumber
r\Big[ - i ( M + m ) A_r^{( m, 0)  } - \frac{1}{2} \partial_r  A_- ^{( m,  0 ) } \Big]  = \frac{M}{2} A_-^{( m,  0) }, \qquad m \geq 1.
\end{eqnarray}
Here, the first equation ensures the Chern-Simons equation for $A_t$ in (\ref{cherneqa}) is satisfied, while the 2nd equation ensures that the Chern-Simons equation for $A_{\varphi}$ is satisfied. Note that the overall constant $C_-^{( m \, 0)  }$  for $n=0$  is not determined by this analysis.  We will choose the constant so that when these modes are promoted to be operators, the creation and annihilation operators commute and are canonically normalized. 

\subsection*{$n\geq 1, m\neq 0$}

For $n\geq 1$, both the modes in $A_-$ as well as $A_+$ are non-zero, therefore given $(n, m )$ we would need both these modes to obtain all three components  $A_\mu$. Let us first discuss the case of $m \neq 0$. Using the Chern-Simons equation for $A_r$, we obtain 
\begin{eqnarray}
A_r^{(m, n ) } = - \frac{i }{ 2M ( 1+r^2) r} \left[
( m + \Omega_{(m, n ) } ) A_+^{(m, n) }  + ( m - \Omega_{(m, n ) } A_-^{(m, n ) } \right].
\end{eqnarray}
We can now look at the equation for the component $A_t$, which is given by 
\begin{eqnarray}
\frac{1+r^2}{r} 
\left[ \frac{1}{2} \partial_r \big( A_+^{(m, n)} - A_-^{(m, n )} \big) -i m A_r^{(m, n) } \right] 
= \frac{M }{2} \big( A_+^{(m, n)} +A_-^{(m, n )} \big). 
\end{eqnarray}
For $m\neq0$, it is clear that the LHS of the above equation behaves as $r^{|m| -2}$. However, the RHS behave as $r^{|m|}$, therefore we must have the coefficient of the power $r^{|m| -2}$ to vanish. The leads to the following condition on the coefficients $C_\pm^{(m, n )}$
\begin{eqnarray} \label{goodratio}
\frac{C_+^{(m, n) } }{ C_-^{(m, n ) } } 
= \frac{M + |m| - {\rm sgn}(m) \Omega_{(m, n ) } }{ M - |m| - {\rm sgn} ( m ) \Omega_{(m, n )} }, 
\qquad n \geq 1, m \neq 0. 
\end{eqnarray}

Here ${\rm sgn}$ refers to the signum function. Once this condition is satisfied, it is easy to verify that the Chern-Simons equation, which determines $A_{\varphi}$, is satisfied. The fact that the ratio of $C_+$ to $C_-$ is determined implies that the modes 
$A_+, A_-$ and $ A_r$  are determined up to one overall constant, which we will normalize demanding the canonical commutation relation. 

\subsection*{$n\geq 1, m=0$}
Again, using the Chern-Simons equation for $A_r$, we obtain for this case
\begin{eqnarray} \label{armzero}
A_r^{(0, n )}  = - \frac{i}{ 2 M r( 1+ r^2) } \Omega_{0, n} \big( A_+^{(0, n)} - A_-^{(0, n ) } \big). 
\end{eqnarray}
From demanding that $A_r$ is smooth in  the $r\rightarrow 0$ limit, we obtain the condition
\begin{eqnarray}\label{eqcoeff}
C_+^{(0, n ) } = C_-^{(0, n )}.
\end{eqnarray}
Once this condition is used for the multiplicative constants, we can see that the wavefunctions satisfy the equations
\begin{eqnarray}
&&\frac{1+r^2}{r} \partial_r \big( A_+^{(0, n ) } - A_-^{(0, n ) } \big) = M \big( A_+^{(0, n ) } + A_-^{(0, n ) } \big) , \\ \nonumber
&&r \left[ 
\frac{ \Omega_{(0, n)}^2 }{ M r ( 1+ r^2) }   \big( A_+^{(0, n ) } - A_-^{(0, n ) } \big) 
-\partial_r    \big( A_+^{(0, n ) } +  A_-^{(0, n ) } \big)  \right] 
= M   \big( A_+^{(0, n ) } - A_-^{(0, n ) } \big).
\end{eqnarray}
These equations hold due to the identities satisfied by the hypergeometric function. The first equation ensures that the Chern-Simons equation for $A_t$ is satisfied, while the second equation ensures that the Chern-Simons equation of motion for $A_{\varphi}$ is satisfied. Again, we see that due to the condition (\ref{eqcoeff}) and the equation (\ref{armzero}), all three components of the vector for these modes are determined to an overall constant. 
 
\subsection{Modes in terms of Jacobi polynomials}
 
It is convenient to cast the modes in $A_+^{(m, n ) }$ and $A_-^{(m, n ) }$ in terms of Jacobi polynomials. Note that they are solutions to the scalar Laplacian in global $AdS_3$. Therefore, these solutions can also be written in terms of Jacobi polynomials \cite{Hamilton:2006az}\footnote{We can derive this using equations (15.3.3),(15.3.4),(15.4.6) and (22.1.6) in \cite{abramowitz1965handbook}.}. This makes it easy to use the orthonormality properties of Jacobi polynomials and, therefore, to obtain the commutation relations between creation and annihilation operators. The  radial functions $R_\pm ^{(m, n) } $ in (\ref{aminus}) and (\ref{aplus}) can be written as 
\begin{eqnarray}\label{jacob}
R_-^{(m, n) } &=& C_-^{(m, n ) } (-1)^n  \frac{n! |m| ! }{ (n+|m|)! }
( \cos \sigma)^{ M} (\sin \sigma)^{|m| } P_n^{(M-1, |m|)} ( -\cos 2\sigma) ,  \\ \nonumber
R_+^{(m, n) }&=& C_+^{(m, n ) } (-1)^{n-1}  \frac{ (n-1)! |m| !  }{ (n-1+|m| )!  }
( \cos \sigma)^{ M+2} (\sin \sigma)^{|m|} P_{n-1}^{(M+1, |m| )} ( -\cos 2\sigma) ,
\end{eqnarray}
where 
\begin{equation}
r= \tan \sigma, \qquad 0\leq \sigma <\frac{\pi}{2} .
\end{equation}
The pre-factors involving the gamma functions in (\ref{jacob}) ensure that the behaviour at $r\rightarrow 0$ obeyed by the wave functions in terms of hypergeometric functions in  (\ref{aminus}), (\ref{aplus}) is preserved. Jacobi polynomials enjoy the orthogonality property
{\small \begin{eqnarray} \label{orthjacob}
\int_{-1}^1 P_n^{( \alpha, \beta) } ( x) P_m^{(\alpha, \beta) }(x) 
( 1-x)^\alpha ( 1+x)^\beta dx = \frac{ 2^{\alpha + \beta +1}}{ 2n + \alpha + \beta + 1} 
\frac{ \Gamma( n+ \alpha + 1)  \Gamma( n+ \beta + 1) }{ n!  \Gamma( n+ \alpha + \beta  + 1) } 
\delta_{n, m }. \nonumber \\
\end{eqnarray} }
For subsequent analysis, we define
\begin{eqnarray}
N^{\alpha, \beta}_{n} =  \frac{ 2^{\alpha + \beta +1}}{ 2n + \alpha + \beta + 1} 
\frac{ \Gamma( n+ \alpha + 1)  \Gamma( n+ \beta + 1) }{ n!  \Gamma( n+ \alpha + \beta  + 1) } .
\end{eqnarray}
 
\subsection*{Expansion of  vectors $A_+, A_-$  in modes}

We will restrict our attention to the vector components $A_+$ and $A_-$. Using (\ref{cherneqa}), we can write the expansion of $A_r$. 
From the solutions discussed earlier, we see that their expansion can be written as 
\begin{eqnarray} \label{aminusra}
A_- &= &\sum_{n=1 m =-\infty }^\infty \left(  R_{-}^{(m, n )} a_{m, n} e^{ - i\Omega_{(m, n ) } t }
e^{i m \varphi} 
+  R_{-}^{(m, n ) *} a_{m, n}^\dagger e^{  i\Omega_{(m, n )}  t } e^{-i m \varphi} \right)   \\ \nonumber
&&  +\sum_{m = 1}^\infty 
\left(  R_{-}^{(m, 0 )} a_{m, n} e^{ - i\Omega_{(m, 0 ) } t }
e^{i m \varphi} 
+  R_{-}^{(m, 0 ) *} a_{m, n}^\dagger e^{  i\Omega_{(m, 0 ) } t } e^{-i m \varphi} \right). 
\end{eqnarray}
The expansion for the component $A_+$ is given by 
\begin{eqnarray} \label{aplusr}
A_+ &= &\sum_{n=1 m =-\infty }^\infty \left(  R_{+}^{(m, n )} a_{m, n} e^{ - i\Omega_{(m, n ) } t }
e^{i m \varphi} 
+  R_{+}^{(m, n )*} a_{m, n}^\dagger e^{  i\Omega_{(m, n ) } t } e^{-i m \varphi} \right).
\end{eqnarray}
In both the expansions (\ref{aminusra}) and (\ref{aplusr}), we use the expressions (\ref{jacob}) in which the radial modes are given in terms of Jacobi Polynomials. To ensure that this expansion is a solution of the first-order Chern-Simons equations, the coefficients $C_+^{(m, n)}$ are determined in terms of $C_-^{(m, n) }$ by the ratio given in (\ref{goodratio}). Then promoting the Dirac brackets in (\ref{amamdb}), (\ref{apapdb}),  (\ref{apamdb}) to the canonical commutation relations and requiring 
\begin{eqnarray}
[a_{m, n }, a_{m', n' }^\dagger] = \delta_{m, m'} \delta_{n, n'},
\end{eqnarray}
we can fix the constants  $C_-^{(m, n) }$. 

\subsection{Quantization of the vector modes}

To proceed with the quantization of the vectors, we first write the creation(annihilation) operators in terms of $A_+, A_-$. We will treat three  cases $n=0, m\geq 1$,  $n\geq 1, m \neq 0$ and $n\geq 1 , m =0$  separately. 

\subsection*{$n=0, m\geq 1$ }
These modes contain only the $A _-$. Using the orthogonality of the Jacobi polynomials  (\ref{orthjacob}), we can write 
{\small \begin{eqnarray} \label{defa}
C_-^{(m, 0)} a_{m, 0}  e^{-i \Omega_{(m, 0)} t} &=& \frac{2^{M-1 + m }}{2\pi N^{(M-1, m )}_0} 
\int_{-1}^1dx \int_0^{2\pi } d\varphi  A_- (\cos \sigma)^{M-2} (\sin \sigma)^m P_0^{(M-1, m )}(x) e^{-im \varphi},
\nonumber 
\\
x  &=&  - \cos 2 \sigma. 
\end{eqnarray} }
A similar expression exists for $a_{m, 0}^\dagger$, which is obtained by taking the Hermitian conjugate of (\ref{defa}). 
{\small \begin{eqnarray} \label{defadagger}
C_-^{(m, 0) *} a_{m, 0}^\dagger   e^{i \Omega_{(m, 0)} t} = \frac{2^{M-1 + m }}{2\pi N^{(M-1, m )}_0} 
\int_{-1}^1dx \int_0^{2\pi } d\varphi  A_- (\cos \sigma)^{M-2} (\sin \sigma)^m P_0^{(M-1, m )}(x) e^{im \varphi}.
\nonumber 
\\
\end{eqnarray} }
In  (\ref{defa}), (\ref{defadagger}),  $\cos \sigma$ and $\sin \sigma$ needs to be written in terms of the variable $x$. 
We promote the Dirac bracket between the classical fields to a commutation relation. It is useful to write the delta function in (\ref{amamdb}) in terms of the variable $x$.  
\begin{eqnarray} \label{commamam}
[A_-(t, x, \varphi) , A_-( t, x', \varphi') ] = - i \frac{2}{M} ( 1-x) \delta ( x-x') \partial_\varphi \delta ( \varphi - \varphi'). 
\end{eqnarray}
Using  (\ref{defa}), ( \ref{defadagger}) and the above commutation relation, we find
\begin{eqnarray}
C_-^{(m, 0)} C_-^{(m, 0) *} [  a_{m, 0}, a_{m, 0}^\dagger] = \frac{1}{\pi} \frac{ \Gamma( M + m +1) }{ (m-1) ! \Gamma( M+1)  }.
\end{eqnarray}
We have used the orthogonality relations (\ref{orthjacob}) and the commutation relations  (\ref{commamam}) to obtain this result.  To obtain the standard commutation relation, we can choose the  constant to be 
\begin{eqnarray} \label{neqzero}
C_-^{(m, 0)} = \sqrt{  \frac{1}{\pi} \frac{ \Gamma( M + m +1) }{ (m-1) ! \Gamma( M+1)  }  }, 
\quad\qquad n=0, m \geq 1. 
\end{eqnarray}
For this paper, it is sufficient to fix the normalization constant of these modes, but we will proceed with the determination of the constants $C_+^{(m, n )}, C_-^{(m, n ) } $ for all the other modes for completeness. 

\subsection*{$n\geq 1, m\neq 0$ }

To write down an expression for $a_{m, n }$ with $n\geq 1, m \neq 0$, we need the mode expansion of both $A_- $ and $A_+$. Using  the orthogonality  relations  (\ref{orthjacob}), we obtain 
{\small \begin{eqnarray}
& &C_-^{( m, n  )} a_{m, n } e^{-i \Omega_{(m, n ) } t}  
+ C_-^{(- m , n )*} a_{-m , n }^\dagger e^{ i \Omega_{(m, n ) } t} = \\ \nonumber
& & \qquad \frac{ (-1)^n  (n+ |m|)! }{ n ! |m| ! } 
\frac{2^{M-1 + |m|  }}{2\pi N^{(M-1, |m| )}_n} 
\int_{-1}^1dx \int_0^{2\pi } d\varphi  A_- (\cos \sigma)^{M-2} (\sin \sigma)^{|m|}  P_n^{(M-1, |m|  )}(x) e^{-im \varphi}, \\ \nonumber, 
&& \qquad\equiv X_-^{(m, n ) }. \\ \nonumber
&& \\ \nonumber
& &C_+^{( m , n )} a_{m, n } e^{-i \Omega_{(m, n ) } t}  
+ C_+^{(- m, n  )*} a_{-m , n } ^\dagger e^{ i \Omega_{(m, n ) } t} = \\ \nonumber
& & \qquad \frac{ (-1)^{(n-1)}( n-1  + |m|  ) ! }{(n-1) !  |m|  !} 
\frac{2^{M+1 + |m| }}{2\pi N^{(M+1, |m| )}_{n-1}} 
\int_{-1}^1dx \int_0^{2\pi } d\varphi  A_+ (\cos \sigma)^{M} (\sin \sigma)^{|m|}  P_{n-1}^{(M+1, |m|  )}(x) e^{-im \varphi} \\ \nonumber
&& \qquad \equiv X_+^{(m, n ) }.
\end{eqnarray} }
Solving for the annihilation operator, we obtain 
\begin{eqnarray} \label{solforan}
a_{m, n } e^{-i \Omega_{(m, n ) } t } =
\frac{C_+^{(-m, n )* }  X_-^{(m, n ) } - C_-^{(-m, n ) *} X_+^{(m, n ) }} { C_+^{(-m, n ) *} C_-^{(m, n) } - 
C_-^{(-m, n )* } C_+^{(m, n ) } }.
\end{eqnarray}
Before proceeding, it is useful to evaluate the commutation relation between the operators $X_-^{(m, n ) }, X_{+}^{(m, n ) }$.
Using the commutation relation in (\ref{commamam}) and the orthogonality of the Jacobi polynomials, we obtain 
{\small \begin{eqnarray}
[X_-^{(m, n ) }, X_-^{(m', n' ) } ]  = \frac{m}{\pi M}  \frac{ ( n +|m|) !}{n! (m!)^2 }
 \frac{ ( 2n + M + |m| )  \Gamma( n + M + |m| ) }{ \Gamma( n+M)  } \delta_{n, n'} \delta_{m, -m'}.
% \nonumber \\
\end{eqnarray} }
From the Dirac bracket (\ref{apapdb}), we obtain  the commutation relation 
\begin{eqnarray}
[A_+(t, x, \varphi) , A_+( t, x', \varphi') ] =  i \frac{2}{M} ( 1-x) \delta ( x-x') \partial_\varphi \delta ( \varphi - \varphi') .
\end{eqnarray}
This admits
{\small \begin{eqnarray}
[X_+^{(m, n ) }, X_+^{(m', n' ) } ]   = -  \frac{m}{\pi M}  \frac{ ( n-1 +|m|) !}{(n-1) ! (m!)^2 }
 \frac{ ( 2n + M + |m| )  \Gamma( n + M +1  + |m| ) }{ \Gamma( n+M +1 )  } \delta_{n, n'} \delta_{m, -m'}.
 \nonumber
 \\
\end{eqnarray} }
The commutator between $X_+^{(m, n ) }$ and $X_-^{(m, n ) }$ is trivial,
\begin{eqnarray}
[X_+^{(m, n ) }, X_-^{(m', n' ) } ] =0. 
\end{eqnarray}
Also, we have the relations
\begin{eqnarray}
( X_+^{(m, n )  })^\dagger = X_+^{(-m, n) } , \qquad 
(X_-^{(m, n )} )^\dagger = X_-^{(-m, n) }.
\end{eqnarray}
We can now proceed to evaluate the commutation relation between $a, a^\dagger$ using the expression for the annihilation operator in 
(\ref{solforan}) and its conjugate.  We obtain 
\begin{eqnarray} \label{nneqzeroaa}
C_{-}^{(m, n )} C_-^{(m, n )*} 
[a_{m, n}, a_{m, n }^\dagger ] &= & \frac{1}{\pi M} \frac{ (n+m)!}{n! (m!)^2} 
\frac{  ( n+m ) \Gamma( M + n +m +1) }{\Gamma(M +n ) } , \\ \nonumber
&& n\geq 1  m >0.
\end{eqnarray}
This relation satisfies the property that in the limit  $n=0, m >0$, we obtain the result (\ref{neqzero}). Demanding that the creation-annihilation operators are canonically normalised and making the choice that $C_-^{(m, n ) } $ are real, we obtain 
\begin{eqnarray} \label{cformgzero}
C_{-}^{(m, n )} =  \left[ \frac{1}{\pi M} \frac{ (n+m)!}{n! (m!)^2} 
\frac{  ( n+m ) \Gamma( M + n +m +1) }{\Gamma(M +n ) }  \right]^{\frac{1}{2}}, 
\qquad n\geq 1, m >0.
\end{eqnarray}
Similarly $m<0$, we obtain 
\begin{eqnarray}
C_{-}^{(m, n )} C_-^{(m, n )*} 
[a_{m, n}, a_{m, n }^\dagger ]  &=& \frac{1}{\pi M} \frac{ (n-m)!}{(n-1)! (|m|!)^2}
\frac{  \Gamma( M + n -m  ) (M+n) }{\Gamma(M +n ) }, \\ \nonumber
&&  n\geq 1, m<0.
\end{eqnarray}
From this, we find 
\begin{eqnarray} \label{cformlzero}
C_{-}^{(m, n )} &=& \left[  \frac{1}{\pi M} \frac{ (n-m)!}{(n-1)! (|m|!)^2}
\frac{  \Gamma( M + n -m  ) (M+n) }{\Gamma(M +n ) } \right]^{\frac{1}{2}}, 
\qquad n\geq 1, m<0.
\end{eqnarray}
As a consistency check, we have evaluated the commutation relation $[a_{m, n}, a_{-m, n } ]$ using 
(\ref{solforan}) and shown that it vanishes. 

\subsection*{$n\geq 1, m=0$ }

This case is more involved compared to others. From (\ref{eqcoeff}), we see that $C_+^{(0, n)} = C_-^{(0, n )}$ and therefore, we  would not be able to get $2$ independent equations 
by considering $A_+$ and $A_-$  as we did in  earlier. We therefore look at the expansion of $A_\varphi, A_r$. Using the expansions in (\ref{aminusra})  and  (\ref{aplusr}),  we find that 
{\small \begin{eqnarray} \label{arnzero}
\frac{1}{2\pi} \int_0^{2\pi } d\varphi A_{\varphi} 
&=& \frac{1}{2} \sum_{n=1}^\infty  \left[ (-1)^{n-1} (\cos \sigma)^M 
\left( \cos^2 \sigma P_{n-1}^{(M +1, 0 )} ( - \cos 2\sigma)  + 
P_{n}^{(M -1, 0 )} ( - \cos 2\sigma)  \right)  \right.  \nonumber \\ 
&& \qquad\qquad \times \left. \left( C_-^{(0, n ) } a_{0, n } e^{-i \Omega_{(0, n ) } t } 
+ C_-^{(0, n ) *} a_{0, n }^\dagger e^{ i \Omega_{(0, n )} t }  \right ) \right]. \nonumber \\
\end{eqnarray} }
Using recurrence relations of Jacobi polynomials, we find the identity
{\small \begin{eqnarray}
\cos^2 \sigma P_{n-1}^{(M +1, 0 )} ( - \cos 2\sigma)  + 
P_{n}^{(M -1, 0 )} ( - \cos 2\sigma)   = 
\frac{M}{2n + M } \left[
P_n^{(M, 0) } ( - \cos 2 \sigma) 
+ P_{n-1}^{(M, 0)} ( - \cos 2 \sigma) 
\right].  \nonumber\\
\end{eqnarray} }
Substituting this identity in the expression in (\ref{arnzero}), we obtain 
{\small \begin{eqnarray}
\frac{1}{2\pi} \int_0^{2\pi}  d\varphi A_{\varphi} 
&=& \frac{1}{2} \sum_{n=1}^\infty \left[ (-1)^{n-1} (\cos \sigma)^M 
\frac{M}{2n + M } \
\left(  
P_n^{(M, 0) } ( - \cos 2 \sigma) 
+ P_{n-1}^{(M, 0)} ( - \cos 2 \sigma) 
 \right)  \right.  \nonumber \\ 
&& \quad\qquad \times  \left. \left( C_-^{(0, n ) } a_{0, n } e^{-i \Omega_{(0, n ) } t } 
+ C_-^{(0, n ) *} a_{0, n }^\dagger e^{ i \Omega_{(0, n )} t }  \right ) \right]. \nonumber \\
\end{eqnarray} }
Then using  similar steps together with (\ref{armzero})  the $m=0$ sector of $A_r$ is given by 
{\small \begin{eqnarray}
\frac{1}{2\pi} \int_0^{2\pi}  d\varphi A_{r}  &=&
 \frac{i}{ 2 r ( 1+ r^2) } 
 \sum_{n=1}^\infty  \left[ (-1)^{n} (\cos \sigma)^M 
\left(  
P_n^{(M, 0) } ( - \cos 2 \sigma) 
+ P_{n-1}^{(M, 0)} ( - \cos 2 \sigma) 
 \right) \right.  \nonumber \\ 
&& \qquad \qquad \qquad  \left. \times \left( C_-^{(0, n ) } a_{0, n } e^{-i \Omega_{(0, n ) } t } 
- C_-^{(0, n ) *} a_{0, n }^\dagger e^{ i \Omega_{(0, n )} t }  \right )  \right]. \nonumber \\
\end{eqnarray} }
We can now use the orthogonality of the Jacobi polynomials to obtain
{\small \begin{eqnarray} \label{aphiosc}
&& \frac{1}{2\pi} \int_{-1}^1 dx \int_0^{2\pi } A_{\varphi}  (\cos\sigma )^M  P_{n-1}^{(M, 0) } ( -\cos  2\sigma)  
\\ \nonumber
& & \qquad =  (-1)^{(n-1) } 
\frac{M}{2n +M}  \frac{1}{ 2n + M - 1}   \left( C_-^{(0, n ) } a_{0, n } e^{-i \Omega_{(0, n ) } t } 
+ C_-^{(0, n ) *} a_{0, n }^\dagger e^{ i \Omega_{(0, n )} t }  \right )
\\ \nonumber 
&& \qquad + 
(-1)^{(n-2)} 
\frac{M}{2n -2 +M}  \frac{1}{ 2n + M - 1}  
 \left( C_-^{(0, n -1 ) } a_{0, n-1 } e^{-i \Omega_{(0, n -1) } t } 
+ C_-^{(0, n-1 ) *} a_{0, n-1 }^\dagger e^{ i \Omega_{(0, n-1 )} t }  \right ).
\end{eqnarray} }
For the field $A_r$, we obtain 
{\small \begin{eqnarray} \label{arosc}
&& \frac{1}{2\pi} \int_{-1}^1 dx \int_0^{2\pi } A_{r}  (\cos\sigma )^M   r ( 1+ r^2) 
P_{n}^{(M, 0) } ( -\cos 2 \sigma)  \\ \nonumber
&& \qquad =  i (-1)^n  \frac{1}{ 2n + M +  1}   
  \left( C_-^{(0, n ) } a_{0, n } e^{-i \Omega_{(0, n ) } t } 
-  C_-^{(0, n ) *} a_{0, n }^\dagger e^{ i \Omega_{(0, n )} t }  \right ) 
\\ \nonumber
&& \qquad + i (-1)^{(n+1) } \frac{1}{ 2n + M + 1} 
  \left( C_-^{(0, n+1 ) } a_{0, n+1 } e^{-i \Omega_{(0, n+1 ) } t } 
-  C_-^{(0, n+1 ) *} a_{0, n+1 }^\dagger e^{i \Omega_{(0, n +1)} t }  \right ). 
\end{eqnarray} } 
Promoting the Dirac bracket between $A_r $ and $A_{\varphi}$ in (\ref{araphdb}) to a commutation relation and using the change of variables from the coordinate $r$ to $x$, we obtain 
\begin{eqnarray} \label{araphiq}
r ( 1+ r^2) [A_\varphi  ( t, x, \varphi) , A_{r}( t, x', \varphi') ] =   i (1+x) \delta ( x-x') \delta ( \varphi - \varphi'). 
\end{eqnarray}
Finally, we need the recursion relation 
\begin{eqnarray} \label{2ndrecur}
\Big( n + \frac{M}{2} - \frac{1}{2} \Big)  ( 1+x) P_{n-1}^{(M, 0)} 
&=& \frac{n ( M +n ) }{ 2n +M} P_n^{(M, 0 ) } + \frac{ n ( M + n ) }{ 2n + M } P_{n-1}^{(M, 0 ) } \\ \nonumber
&& +   [ n] \rightarrow [ n -1],
\end{eqnarray}
where the last line is obtained by replacing $n$ by $n-1$ in the two terms that appear on the first line. Now, using all these inputs, we can evaluate the commutator of the equations in (\ref{aphiosc}) and (\ref{arosc}). Note that the non-trivial commutator, which contributes from the RHS of these equations, is between $a_{0, n }$ and $a_{0, n }^\dagger$. To evaluate the LHS, we need to use (\ref{araphiq}) and the recursion relation in (\ref{2ndrecur}) and the orthonormality relations of the Jacobi polynomials. This enables us to determine and fix the normalisation as 
\begin{eqnarray}
C_{-}^{(0, n ) } C_{-}^{(0, n ) * } [a_{0, n }, a_{0, n }^\dagger]  = 
\frac{1}{\pi M} n ( n +M) ,
\end{eqnarray}
which results in 
\begin{eqnarray} \label{cformezero}
C_{-}^{(0, n ) }   =  \left[  \frac{1}{\pi M} n ( n +M)  \right]^{\frac{1}{2}}.
\end{eqnarray}
Note that extrapolating to the $m \rightarrow 0$ limit of both the normalizations, which were determined for $m>0$ in (\ref{cformgzero}) and $m<0$ in ( \ref{cformlzero}), we obtain the above result. This serves as a simple consistency check of the normalizations. 

From the expressions in (\ref{neqzero}), (\ref{cformgzero}),   (\ref{cformlzero}) and (\ref{cformezero}), we can write the general expression for the normalisation which reproduces these cases.
{\small  \begin{eqnarray} \label{finalcminus}
C_{-}^{(m, n )}  &=&  \left\{ \frac{1}{\pi M} \frac{ (n+m)!}{n! (m!)^2} 
\frac{  \big( n+ \frac{1}{2} \big[ 1 + { \rm sgn} (m) \big] \, m \big) \Gamma( M + n +|m| )  \big( M +n + 
\frac{1}{2} \big[ 1 + { \rm sgn} (m) \big] \, m  \big)}{\Gamma(M +n ) }  \right\}^{\frac{1}{2}},  \nonumber
\\
& & n = 0, \qquad m = 1,  2, 3 \cdots, \\ \nonumber
& & n \geq 1, \qquad m \in \mathbb{Z} .
\end{eqnarray} }
This completes our analysis of the quantization of the massive Chern-Simons equations in $AdS_3$.

\section{Fefferman-Graham Expansion}\label{appenfg}

Here, we obtain the expectation value of the boundary stress tensor for the perturbed metric derived in section(\ref{backreactsec},  using the Fefferman-Graham expansion. Consider a 3-dimensional bulk spacetime $(z, x_i)$ which asymptotes to the $AdS_3$ vacuum. Let the boundary of this spacetime be at $z \rightarrow 0$. The metric in the Fefferman-Graham form is given as
\begin{eqnarray} \label{FG form metric}
	ds^2 = \frac{1}{z^2} \big( dz^2 + g_{ij} (x,z) dx^i dx^j \big),
\end{eqnarray}
where $g(x,z)$ is expanded as,
\begin{eqnarray} \label{g expand}
	g(x,z)_{ij} = g_{(0)} + z^2  g_{(2) ij} + \cdots .
\end{eqnarray}
Once the metric is in this form, we can read out the expectation value of the stress tensor in the CFT using the relation \cite{Balasubramanian:1999re,deHaro:2000vlm}
\begin{eqnarray}
	\langle T_{ij} \rangle_{FG} = \frac{1}{4 G_N} {g_{(2)}}_{ij}
\end{eqnarray}

\subsection{Holographic stress tensor for $| \psi_{1,0} \rangle$}

The back-reacted metric is given in \eqref{metric ansatz 1,0} with \eqref{slonbackgs}.
\begin{eqnarray}
	ds^2 &=& -(r^2 + H_1^2(r)) dt^2 + \frac{dr^2}{(r^2 + H_2^2(r))} + r^2 d\varphi^2 + H_3(r) dt d\varphi \\ \nonumber
	H_1(r) &=& 1 + G_N a_1 (r), \qquad H_2(r) = 1 + G_N a_2 (r), \qquad H_3(r) = G_N a_3 (r),
\end{eqnarray}
where
\begin{eqnarray}
	a_2 (r) = \frac{4 (M + 1)}{(1 + r^2)^M} + c_1, \quad 
	a_1 (r) = \frac{4}{(1 + r^2)^M} + c_1, \quad 
	a_3 (r) = \frac{8}{(1 + r^2)^M}  + d_2.
\end{eqnarray}
To obtain the leading order term in the asymptotic form of the metric at $r \rightarrow \infty$, we can ignore the first terms in the expression for $a_1 (r)$, $a_2 (r)$ and $a_3 (r)$. These are sub-leading and vanish at the boundary if  $M > 0$, which we assume is always the case. Hence, we work with $a_1(r) = a_2(r) = c_1$ and $a_3 (r) = d_2$ to obtain the holographic stress tensor. We apply the following coordinate transformation to bring the metric to the Fefferman-Graham form
\begin{eqnarray} \label{fgcoordt}
	t \rightarrow t, \qquad r \rightarrow \frac{1}{z + \alpha z^3 + O(z^5) }, \qquad \varphi \rightarrow \varphi. 
\end{eqnarray}
The metric transforms to 
\begin{eqnarray} \label{FG form metric for gs}
ds^2 &=& \Big[ \frac{-1}{z^2}  + (-1 + 2\alpha  -2 c_1 G_N) + O(z^2) \Big] dt^2 
+ \Big[ \frac{1}{z^2} + (-1 + 4 \alpha - 2 c_1 G_N +O(z^2)) \Big] dz^2  \nonumber \\
&& + \Big[  \frac{1}{z^2} -2 \alpha  + O(z^2)  \Big] d\varphi^2 
+ \Big[G_N d_2  + O(z^2) \Big] dtd\varphi.
%	\begin{split}
%		ds^2 = dt^2 \Bigg[ - 2 c_1 G_N  - \frac{1}{(z + \alpha z^3)^2} - 1 \Bigg] +  dz^2 \frac{(1 + 3 \alpha z^2)^2}{(z + \alpha z^3)^4} \Bigg[\frac{1}{\frac{1}{(z + \alpha z^3)^2} + 1} &- \frac{2 c_1 G_N}{\Big(\frac{1}{(z + \alpha z^3)^2} + 1 \Big)^2} \Bigg] \\
%		&+ \frac{d \varphi^2}{(z + \alpha z^3)^2} + G_N d_2 dt d\varphi
%	\end{split}
\end{eqnarray}
We can determine the coefficient $\alpha$ by requiring the $zz$ component of the metric to be $1/z^2$ to the order of $\mathcal{O} (z^0)$, resulting in
\begin{eqnarray}
	\alpha = \frac{1}{4} (1 + 2 c_1 G_N).
\end{eqnarray}
Substituting this in \eqref{FG form metric for gs} and collecting the coefficients of the $\mathcal{O} (z^0)$ in $tt$ component of the metric, we obtain
\begin{eqnarray} \label{holttval}
	\langle \psi_{1,0} | T_{tt} | \psi_{1, 0} \rangle_{FG} = \frac{1}{4 G} \bigg( - \frac{1}{2} - c_1 G_N \bigg).
\end{eqnarray}
Similarly, the holographic expectation value $\langle T_{t \varphi} \rangle$ is given by 
\begin{eqnarray} \label{for d2 gs FG}
	\langle \psi_{1, 0} | T_{t \varphi} | \psi_{1,0} \rangle_{FG} = \frac{1}{4 G_N} \times G_N d_2.
\end{eqnarray}

%The $t \varphi$ component of the stress tensor of the CFT is given as
%
%\begin{eqnarray}
%	T_{t \varphi} (t, \varphi) = L_0 - \bar{L}_0
%\end{eqnarray}
%
%The expectation value of the CFT stress tensor 
%
%\begin{eqnarray} \label{for d2 gs cft}
%	\frac{\langle h, \bar{h} | T_{t \varphi} (t, \varphi) | h, \bar{h} \rangle}{\langle h, \bar{h} | h, \bar{h} \rangle} = h - \bar{h} = 1
%\end{eqnarray}
%
%Then requiring \eqref{for d2 gs FG} and \eqref{for d2 gs cft} to agree, we obtain
%
%\begin{eqnarray}
%	d_2 = 4
%\end{eqnarray}

\subsection{Holographic stress tensor for $| \psi_{m,0} \rangle$}
The back-reacted metric is given in \eqref{metric ansatz m,0} with the metric components satisfying the differential equations in \eqref{toweree} 
\begin{eqnarray} \label{towermetric}
	ds^2 = [-(r^2 + 1) - 2 b_1 (r) G_N]dt^2 + \Big[ \frac{dr^2}{1 + r^2} - dr^2 \frac{2 b_2 (r) G_N}{(1 + r^2)^2}\Big] + r^2 d\varphi^2 + G_N b_3 (r) dt d\varphi.
\end{eqnarray}
The solution for $b_2(r)$ is given by 
\begin{eqnarray} \label{b2r}
	b_2 (r) &=& k_1 - 4 r^{2m} \frac{\Gamma(M + m + 1)}{\Gamma(M) \Gamma(1+m)} {}_2 F_1 (m, M + m, 1 + m, -r^2)
	\\ \nonumber
	&  =& k_1 - 4 (m +M)  +  r^{-2M} \left[ 
	 \frac{ 4 \Gamma( 1+m +M) }{\Gamma(m) \Gamma( 1+M) }  + O( r^{-2} ) \right].
\end{eqnarray}
For our analysis, it is sufficient to obtain the asymptotic behaviour of  $b_1 (r)$ at $r\rightarrow \infty$. The differential equation for $b_1 (r)$ is given by (\ref{toweree}). 
\begin{eqnarray} \label{eqforb1}
	(1 + r^2) b_1^{'} (r) - 2 r b_1 (r) + 2 r b_2 (r) = 0.
\end{eqnarray}
Substituting for $b_2 (r)$ from \eqref{b2r} and solving for the leading terms in $b_(r)$ as $r\rightarrow\infty$ we obtain
\begin{eqnarray}
b_1(r) &=&  \Big[ c -4 \frac{ \Gamma( 1-M) \Gamma( 1+m +M) }{ \Gamma( m)} \Big] r^2 
+ 
\Big[ k_1  - 4 (m+M) + c -4 \frac{ \Gamma( 1-M) \Gamma( 1+m +M) }{ \Gamma( m)} \Big]  \nonumber \\
&& +  4 \frac{\Gamma(1+m +M) }{ \Gamma(m) \Gamma( 2+M) } r^{-2M} + \cdots .
\end{eqnarray}
Here $c$ is the undetermined constant from the homogeneous solution of (\ref{eqforb1}). To ensure that the solution is asymptotically $AdS_3$, we need to choose 
\begin{eqnarray}
c= 4 \frac{ \Gamma( 1-M) \Gamma( 1+m +M) }{ \Gamma( m)}. 
\end{eqnarray}
Therefore we obtain 
\begin{eqnarray}\label{solb1r} 
b_1(r) = k_1 - 4(m+M) + 4 \frac{\Gamma(1+m +M) }{ \Gamma(m) \Gamma( 2+M) }  r^{-2M} + \cdots.
\end{eqnarray}
Now that we have obtained the solution for $b_1$ (\ref{solb1r}) and  $b_2$  in  (\ref{b2r}), we can use the coordinate transformation in (\ref{fgcoordt}) and bring the metric in (\ref{towermetric}) to the Fefferman-Graham form. In this case, this results in 
\begin{eqnarray}
\alpha = \frac{1}{4} \Big( 1 + 2 ( k_1 - 4(m+M) ) G_N \Big).
\end{eqnarray}
Proceeding to extract the holographic stress tensor, we obtain 
\begin{eqnarray} \label{towerholfg}
	\langle \psi_{m,0} | T_{tt} | \psi_{m, 0} \rangle_{FG} = \frac{1}{4 G} \Big( - \frac{1}{2} - \big( k_1 - 4(m+M) \big)  G_N \Big).
\end{eqnarray}
The function $b_3(r)$ can also be obtained by solving the 3rd equation in (\ref{toweree}) and fixing the constant using the $t\varphi$ component of the stress tensor. We would not require this to evaluate the corrections to the minimal area.

\section{Vector modes in BTZ Rindler and their quantization} \label{appenc}

In this appendix, we quantize the vector fields in the BTZ geometry.
The  massive Chern-Simons  equations in this background are given by 
\begin{eqnarray} \label{cseqrin}
\frac{g_{\tau\tau}}{\sqrt{-g} } ( \partial_\rho A_x - \partial_x A_\rho)  = - M A_\tau, \\ \nonumber
\frac{g_{\rho\rho}}{\sqrt{-g} } ( \partial_x A_\tau - \partial_\tau A_x) =  - M A_\rho, \\ \nonumber
\frac{g_{xx} }{\sqrt{-g} } ( \partial_\tau A_\rho - \partial_\rho A_\tau ) = - M A_x. 
\end{eqnarray}
Here, it is assumed that the vector fields are in the Rindler-BTZ coordinates with the metric (\ref{btzrinda}). Following the same steps as in the case of the global $AdS_3$ metric, as indicated in equations (\ref{mani1}) to (\ref{simeq2} ), the action of the  vector Laplacian in the BTZ coordinates on the component $A_\tau$ can be shown to reduce to 
\begin{eqnarray} \label{veceq1}
\nabla^2 A_\tau = \Box A_\tau  -2 A_\tau - 2 M A_x.
\end{eqnarray}
Here $\Box$ refers to the scalar Laplacian on BTZ. In arriving at this result, we would need to substitute the Christoffel symbols given in (\ref{chrisbtz}) for writing out the vector Laplacian. 
Similarly, the action of the vector Laplacian on the component $A_x$ reduces to 
\begin{eqnarray} \label{veceq2}
\nabla^2 A_x = \Box A_x  - 2 A_x - 2 M A_\tau.
\end{eqnarray}
The components of the vector field  in the BTZ background also obey the equation 
\begin{eqnarray}
\nabla^2 A_\mu = (M^2 - 2) A_\mu. 
\end{eqnarray}
Combining this with the action of the vector Laplacian in (\ref{veceq1}) and (\ref{veceq2}), we are led to the following coupled equations involving the scalar Laplacian
\begin{eqnarray}
\Box A_\tau - M^2 A_\tau - 2 M A_x =0, \qquad \Box A_x - M^2 A_x - 2M A_\tau =0.
\end{eqnarray}
We therefore consider the combinations
\begin{eqnarray} \label{aplusmbtz}
A_+ = A_\tau + A_x, \qquad A_-= A_\tau - A_x,
\end{eqnarray}
which obey the equations of two decoupled scalars 
\begin{eqnarray}
( \Box -  ( M+1)^2 + 1) A_+ = 0, \qquad   ( \Box -  ( M-1)^2 + 1) A_- = 0.
\end{eqnarray}
We can now proceed to solve the decoupled second-order equations. 
The directions $\tau, x$ are isometries of the metric, thus the solutions can be expanded as Fourier modes as follows
\begin{eqnarray}\label{aminuswave}
R_-^{(\omega, k )} &=&  A_-^{(\omega, k ) } e^{- i \omega \tau } e^{i k x}, \\ \nonumber
A_-^{(\omega, k)}&=& C_-^{(\omega, k )} 
 \rho^{-M} \Big(1-  \frac{1}{\rho^2}  \Big)^{- \frac{i \omega}{2} } 
 {}_2 F_1\Big( \frac{1}{2} \big( M - i ( k + \omega) \big), \frac{1}{2} \big( M + i ( k - \omega) \big) , M , \frac{1}{\rho^2} \Big).
\end{eqnarray}
These solutions satisfy the condition that they are normalizable at the boundary, and at the horizon, they have both ingoing and outgoing modes. Here $ k,\omega $ are continuous and take values 
from $-\infty$ to $\infty$ and $0$ to $\infty$ respectively. $C_-^{(\omega, k )} $ are constants for each $\omega, k $ which can be determined by demanding these solutions also satisfy the first order Chern-Simons equations, which will be done subsequently. Solving the equation for $A_+$, we obtain the following modes
\begin{eqnarray} \label{apluswave}
R_+^{(\omega, k ) } &=& A_+^{(\omega, k ) } e^{- i \omega \tau } e^{i k x}, \\ \nonumber
A_+^{(\omega, k)}&=& C_+^{(\omega, k )} 
\rho^{-(M+2) } \Big(1-  \frac{1}{\rho^2}  \Big)^{- \frac{i \omega}{2} } 
{}_2 F_1\Big( \frac{1}{2} \big( M +2  - i ( k + \omega) \big) , \frac{1}{2} \big( M +2 + i ( k - \omega) \big) , M +2 , \frac{1}{\rho^2} \Big).
\end{eqnarray}
Given the modes for $A_-^{(\omega, k ) } $ and $A_+^{(\omega, k ) }$, we determine the radial component $A_\rho$ by using the second equation in (\ref{cseqrin}). Then, demanding that the remaining first-order Chern-Simons equations are satisfied, we can fix the constants $C_+, C_-$. A simple way to do this is to examine the behaviour of the equations near the horizon $\rho \rightarrow 1$. 
\begin{eqnarray} \label{aminushor}
A_-^{(\omega, k ) }|_{\rho \rightarrow 1} = C_-^{(\omega, k ) } \Big[ (\rho - 1)^{- i \frac{\omega}{2} }  
n_-^{(\omega, k) } +  
(\rho -1)^{ i \frac{\omega}{2} } n_-^{(\omega, k) *} \Big]  + \cdots, \\ \nonumber
n_-^{(\omega,  k )} = - \frac{i 2^{- i \frac{\omega}{2} } \pi \Gamma (M)   \csch\pi \omega  
}{\Gamma ( 1- i \omega) \Gamma \big[\frac{1}{2} ( M -i ( k - \omega) ) \big] 
\Gamma\big[ \frac{1}{2} ( M + i ( k + \omega) ) \big]}.
\end{eqnarray}
Similarly 
\begin{eqnarray} \label{aplushor}
A_+^{(\omega, k ) }|_{\rho \rightarrow 1} = C_+^{(\omega, k ) } \Big[ (\rho - 1)^{- i \frac{\omega}{2} }  
n^{(\omega, k )}_+ +  
(\rho -1)^{ i \frac{\omega}{2} } n_+^{(\omega, k) *} \Big] + \cdots, \\ \nonumber
n_+^{(\omega, k) } = - \frac{i 2^{- i \frac{\omega}{2} } \pi \Gamma (M +2)   \csch\pi \omega  
}{\Gamma ( 1- i \omega) \Gamma \big[\frac{1}{2} ( M+2  -i ( k - \omega) )\big]
\Gamma\big[ \frac{1}{2} ( M +2 + i ( k + \omega) ) \big]}.
\end{eqnarray} 
From these limits, we can obtain the leading behaviours of $A_\tau, A_x$. Then, using the Chern-Simons equations, we obtain 
\begin{eqnarray}
A_\rho^{(\omega, k) } |_{\rho\rightarrow 1} = -i \frac{1}{2 M ( \rho - 1)} 
\big(  \omega A_x^{(\omega, k )}   +  k A_t^{(\omega, k ) }  \big) |_{\rho \rightarrow 1}.
\end{eqnarray}
This equation implies that the leading behaviour of $A_\rho$ is given by 
\begin{eqnarray}
A_\rho^{(\omega, k) } |_{\rho\rightarrow 1} \sim   B (\rho - 1)^{ - 1-i\frac{\omega}{2} }  + D  (\rho - 1)^{ - 1+i\frac{\omega}{2} }.
\end{eqnarray}
Now from the 3rd Chern-Simons equation in (\ref{cseqrin}), we note 
\begin{eqnarray}
-M A_x^{(\omega, k )}|_{\rho\rightarrow 1}  
\sim (  \partial_\tau A_\rho^{(\omega, k ) }  -\partial_\rho A_{\tau}^{(\omega, k ) } ) |_{\rho\rightarrow 1}.
\end{eqnarray}
Comparing the behaviours of the LHS and the RHS of this equation at the horizon, we see that the LHS behaves as $(\rho -1)^{\pm i \frac{\omega}{2}}$. Therefore, the coefficient of $(\rho -1)^{-1 -  i\frac{\omega}{2} } $ as well as $(\rho -1 )^{-1 +  i\frac{\omega}{2} } $  in the RHS of the equation must vanish independently. 
Requiring the coefficient of  $(\rho -1)^{-1 -  i\frac{\omega}{2} } $ to vanish leads to the following equation
\begin{eqnarray} \label{ratioeq}
i  ( C_+^{( \omega, k ) } n_+^{(\omega, k ) }  + C_-^{(\omega, k ) }  n_-^{(\omega, k ) }) 
= \frac{1}{M}  \Big( 
\omega  ( C_+^{( \omega, k ) } n_+^{(\omega, k ) }  - C_-^{(\omega, k ) }  n_-^{(\omega, k ) })  
+ k  ( C_+^{( \omega, k ) } n_+^{(\omega, k ) }  + C_-^{(\omega, k ) }  n_-^{(\omega, k ) })  \Big).
\nonumber
\\
\end{eqnarray}
This equation allows us to obtain the ratio of $C_+$ to $C_-$
\begin{eqnarray} \label{ratiobtzrin}
\frac{C_+^{(\omega, k ) }}{ C_-^{(\omega, k ) } } = -\frac{M^2  + ( k - \omega)^2 }{ 4 M ( 1+M) },
\end{eqnarray}
where we have used the expressions for $ n_-^{(\omega, k ) }$ and $ n_+^{(\omega, k ) }$ from (\ref{aminushor}) and (\ref{aplushor}). 
The fact that this ratio is real ensures that the analogous equation to (\ref{ratioeq}) arrived at by requiring the coefficient of  $(\rho -1)^{-1 +  i\frac{\omega}{2} } $ to vanish will also be satisfied. We have explicitly verified that it is indeed the case. 

Now that the ratio of the constants $ C_+$ to $C_-$ are determined, we can find all components of the vector $A_\mu$ in terms of $C_-^{(\omega, k) }$. It can then be verified that all three Chern-Simons equations are satisfied. We will determine the constant $C_-^{(\omega, k) }$ by ensuring the expansions of the vectors in terms of the modes obey the Dirac bracket commutation relations. To do that, it is useful to write down the expressions for $A_\rho$ and $A_x$ as $\rho \rightarrow 1$. 
\begin{eqnarray} \label{arhonear}
A_\rho^{(\omega, k ) }|_{\rho \rightarrow 1} &=&  C_-^{(\omega, k ) } \Big[ ( \rho - 1)^{- 1 - i \frac{\omega}{2} } n_\rho^{(\omega, k)} 
- ( \rho - 1)^{( -1 + i \frac{\omega}{2}) } n_\rho^{(\omega, k ) *}  \Big] + \cdots, \\ \nonumber
n_\rho^{( \omega, k ) } &=& \frac{ 2^{-2 - i \frac{\omega}{2} } \pi \omega\Gamma(M)   \csch\pi \omega }{
\Gamma( 1- i \omega) \Gamma\big[ \frac{1}{2} ( M - i ( k -\omega)  ) \big] \Gamma\big[ \frac{1}{2} ( M+2  +i ( k +\omega) )  \big] }.
\end{eqnarray}
To obtain this result, we have used the 2nd equation of motion  (\ref{cseqrin}), where $A_x, A_\tau$ are obtained using (\ref{aplusmbtz}) and the limiting behaviour (\ref{aminushor}), (\ref{aplushor}), together with the ratio given in (\ref{ratiobtzrin}). Similarly, the result expression for $A_x$  near the horizon is given by 
\begin{eqnarray}\label{axnear}
A_x^{(\omega, k) } |_{\rho\rightarrow 1} = 
 C_-^{(\omega, k ) } \Big[  ( \rho -1)^{-i \frac{\omega}{2} } n_x^{(\omega, k ) } 
 + ( \rho -1)^{i \frac{\omega}{2} } n_x^{( \omega, k) *}  \Big] + \cdots, \\ \nonumber
 n_x^{(\omega, k ) } = -\frac{ ( k - i M ) 2^{-1- i \frac{\omega}{2} }  \pi \Gamma( M)  \csch\pi \omega }{
 \Gamma ( 1- i \omega)  \Gamma\big[ \frac{1}{2} ( M - i ( k -\omega) ) \big] \Gamma\big[ \frac{1}{2} ( M+2  +i ( k +\omega) ) \big] }.
\end{eqnarray} 
Let us now discuss how boundary conditions are imposed on these modes. We are in the $\omega \neq 0$  sector, at the  `brick wall'  or the boundary $\gamma$, which is 
$\rho = 1+\epsilon$, we see that  from (\ref{pmc}) we need to impose $n^\mu A_\mu = 0$. Using the equation (\ref{arhonear}), we see that this can be done if we choose
\begin{eqnarray} \label{quantcond}
\omega \log( 2 \epsilon) + {\rm Arg} \left( 
\frac{ \Gamma( 1- i \omega) \Gamma\big[ \frac{1}{2} ( M  - i( k - \omega )  ) \big]
\Gamma\big[ \frac{1}{2} ( M  +2  +  i( k +  \omega )  ) \big] }
{ \Gamma( 1+ i \omega) \Gamma\big[ \frac{1}{2} ( M  + i( k - \omega )  ) \big]
\Gamma\big[ \frac{1}{2} ( M  +2  -   i( k +  \omega )  ) \big] } \right)  = 2\pi n  \nonumber \\
\end{eqnarray}
Though this is a transcendental equation, it is clear that for 
\begin{eqnarray} \label{discfreq}
\omega = \frac{ 2\pi n }{ |\log ( 2 \epsilon )|  } , \qquad n = 1, 2 ,3 \cdots, 
\end{eqnarray}
with  $n$ finite and for $\epsilon \rightarrow 0$, we can ensure that the equation (\ref{quantcond}) is satisfied. We have chosen positive $n$ and the absolute value $|\log ( 2\epsilon) |$ so that the solution to the frequency is positive. Similarly, let us examine the second condition of (\ref{pmc}), $n^\rho F_{\rho \tau} = 0$  on $\gamma$. From the last Chern-Simons equation (\ref{cseqrin}), we see that 
\begin{eqnarray}
n^\rho F_{\rho \tau} =   \rho \sqrt{\rho^2 - 1} M A_x.
\end{eqnarray}
Now from (\ref{axnear}) we see that 
\begin{eqnarray}
\rho \sqrt{\rho^2 - 1} M A_x^{(\omega, k ) } |_{\rho = 1 + \epsilon}  & = &  C_-^{(\omega, k ) }
\sqrt{\epsilon} \Big[  e^{- i \frac{\omega}{2} \log (\epsilon)} n_x^{(\omega, k ) } +    e^{+ i \frac{\omega}{2} \log (\epsilon)} n_x^{(\omega, k ) * }  \Big], \\ \nonumber
& \sim &  \sqrt{\epsilon}. 
\end{eqnarray}
This implies that as $\epsilon\rightarrow 0$ we see that the normal electric field at the `brick wall' vanishes. This concludes our discussion that demonstrates the boundary conditions (\ref{pmc}) can be satisfied as $\epsilon\rightarrow 0$ as long as the frequency is discrete and given by the equation (\ref{discfreq}).

\subsection*{Expansion of the vectors $A_+, A_-$ in modes}

Since we have found the solutions of the massive Chern-Simons equations in BTZ, we can expand the vector fields in terms of modes. The  mode expansion of the fields $A_+$ is given by 
\begin{eqnarray} \label{rinamexpa}
A_-  = \sum_{I \in L, R} \int_{\omega >0} \frac{d\omega dk}{4\pi^2} \left[ 
 b_{\omega, k , I } A_-^{(\omega, k ) I } e^{-i \omega \tau + i k x}
+  b_{\omega, k,  I }^\dagger   A_-^{(\omega, k ) I * } e^{i \omega \tau - i k x} \right].
\end{eqnarray}
The sum $I$ runs over the left and right halves of the Rindler BTZ, the wave functions $A_-^{(\omega, k ) I }$ for both left and right BTZ Rindler are given by (\ref{aminuswave}). In the next section, we will fix the overall normalization $C_-^{(\omega, k ) }$ using the fact that the oscillators $b_{\omega, k }$ should obey the canonical commutation relation once the vector fields are promoted to operators. Similarly, the expansion of $A_+$ is given by 
\begin{eqnarray} \label{rinapexpa}
A_+  = \sum_{I \in L, R} \int_{\omega >0} \frac{d\omega dk}{4\pi^2} \left[
  b_{\omega, k , I } A_+^{(\omega, k ) I } e^{-i \omega \tau + i k x}
+  b_{\omega, k,  I }^\dagger   A_+^{(\omega, k ) I * } e^{i \omega \tau - i k x} \right].
\end{eqnarray}
Here, the wave functions $A_+^{(\omega, k ) I}$ can be read out from (\ref{apluswave}). The constants $C_+^{(\omega, k ) }$ are determined in terms of $C_-^{(\omega, k ) }$ using the 
ratio (\ref{ratiobtzrin}). From the expansions in (\ref{rinamexpa}) and (\ref{rinapexpa}), we can find the expansion for $A_\rho$ using the  2nd Chern-Simons equation (\ref{cseqrin}). 

\subsection{Quantization of the vector modes}

We can repeat the analysis of Dirac quantization for $AdS_3$ in the BTZ background. The analysis proceeds on similar lines as global coordinates, for each step in section \ref{diracquant}, we can replace $t, r, \varphi$ with $\tau, \rho, x$. In the end, we arrive at the following Dirac bracket 
\begin{eqnarray} 
\{ A_\rho( \tau, \rho, x) , A_x( \tau, \rho', x') \}_{\rm DB}  = -\frac{1}{2} \delta ( \rho - \rho') \delta ( x-x').
\end{eqnarray}
To quantize these fields, we lift the fields as operators and promote the Dirac bracket to the following commutation relation
\begin{eqnarray} \label{dracbrakbtz}
[ A_x  ( \tau, \rho, x) , A_\rho ( \tau, \rho', x') ] =  \frac{i}{2} \delta ( \rho - \rho') \delta ( x-x'). 
\end{eqnarray}
The operators $b_{\omega, k , I} $ obey the oscillator algebra 
\begin{eqnarray}
[b_{\omega, k, I} , b_{\omega', k', J}^\dagger] = (2\pi)^2 \delta ( \omega -\omega') \delta(k-k') \delta_{IJ}.
\end{eqnarray}
We can substitute the mode expansions in the LHS of the equation (\ref{dracbrakbtz}), use the canonical commutations and determine the normalisation $C_-^{(\omega, k ) }$. For this, it is convenient to look at the commutation relation near the horizon. Using (\ref{axnear}), the mode expansions of the field $A_x$, near the horizon is  given by 
\begin{eqnarray}
A_x|_{\rho \rightarrow 1} =  \sum_{I \in L, R} \int_{\omega >0} \frac{d\omega dk}{4\pi^2} \left[ b_{\omega, k, I}^\dagger
C_- ^{(\omega, k )} \Big\{ ( \rho -1)^{-i \frac{\omega}{2} } n_x^{(\omega, k ) } 
+ ( \rho -1)^{i \frac{\omega}{2} } n_x^{( \omega, k) *}  \Big\} e^{-i \omega t + i k x}  + {\rm hc} \right],
\nonumber
\\
\end{eqnarray}
where ${\rm hc}$ refers to Hermitian conjugate. Similarly we have  from (\ref{arhonear})
\begin{eqnarray}
A_\rho|_{\rho \rightarrow 1} =  \sum_{I \in L, R} \int_{\omega >0} \frac{d\omega dk}{4\pi^2} \left[ b_{\omega, k, I}^\dagger
C_- ^{(\omega, k )} \Big\{ ( \rho -1)^{-1-i \frac{\omega}{2} } n_\rho^{(\omega, k ) } 
- ( \rho -1)^{-1+ i \frac{\omega}{2} } n_\rho^{( \omega, k) *}  \Big\} e^{-i \omega t + i k x}  + {\rm hc} \right].
\nonumber
\\
\end{eqnarray}
We now substitute this expansion in the LHS of (\ref{dracbrakbtz}), and use the canonical oscillator algebra to evaluate the commutator. We can determine the normalization by demanding that we obtain the algebra (\ref{dracbrakbtz}). This results in the condition 
\begin{eqnarray}
|C_-^{(\omega, k )}|^2 \left( n_x^{(\omega, k ) } n_\rho^{(\omega, k )*} - 
n_x^{(\omega, k )  * }  n_\rho^{(\omega, k )} \right)   = \frac{1}{4}.
\end{eqnarray}
To obtain this relation, we need to use 
\begin{eqnarray}
\delta \Big( \frac{1}{2} \log ( \rho -1) - \frac{1}{2} \log( \rho'-1) \Big) = 2 (\rho -1)  \delta ( \rho- \rho'). 
\end{eqnarray}
Also it can be seen that only terms involving products $(\rho -1)^{-i \frac{\omega}{2} }$  with $(\rho' -1)^{+i \frac{\omega}{2} }$ contribute to the delta function in the commutator. The rest of the terms involve an integral, which vanishes. Finally substituting $n_x^{(\omega, k ) }$ and $n_\rho^{(\omega, k )}$ from
(\ref{axnear}) and (\ref{arhonear}) we obtain 
\begin{eqnarray}
C_-^{(\omega, k ) } &=&  \frac{ \Big|\Gamma( 1+ i \omega) \Gamma\big[ \frac{1}{2} ( M - i ( k -\omega) ) \big] \Gamma\big[ \frac{1}{2} ( M+2  +i ( k +\omega) ) \big] \Big| } { \pi \csch(\pi \omega) \Gamma(M) \sqrt{M \omega} } \\ \nonumber
&\equiv& N_{\omega,  k}.
\end{eqnarray}
Here we have chosen $  C_-^{(\omega, k ) } $ to be real. This completes the quantization of the vectors in the BTZ Rindler background. 
 
\section{Bogoliubov coefficients} \label{append}
 
We have expressed the bulk entanglement entropy in terms of Bogoliubov coefficients relating oscillators in the global $AdS_3$ to those of the BTZ. In this section, we provide the details of computing these coefficients. We begin with the following two-point function evaluated at the boundary
\begin{eqnarray} \label{2ptfna}
\lim_{r \rightarrow \infty} \langle 0| A_-(t,r,\varphi) a_{m,n}^\dagger |0\rangle.
\end{eqnarray}
Here, the vacuum is the global $AdS_3$ vacuum. We look at this two-point function in two ways, first we expand the vector field in terms of the global $AdS_3$ oscillators as given in (\ref{aminusr}) and evaluate the resulting correlator. The second approach is to do the following: express the creation operator $a_{m,n}^\dagger$ in terms of the oscillators in Rindler-BTZ using the Bogoliubov coefficients, write the vector field using the coordinate transformation in (\ref{coordchange}) in terms of oscillators in BTZ, and relate the global vacuum to that of the BTZ  using (\ref{eq: therofield double})  and evaluate (\ref{2ptfna}). Equating the results from these two approaches enables us to obtain an equation for the Bogoliubov coefficients.

We will first give the details of the evaluation of Bogoliubov coefficients for the lowest energy state $|\psi_{1, 0}\rangle$ and then relate it to the Bogoliubov coefficients of the holomorphic descendants $|\psi_{m, 0 } \rangle$.

\subsection*{The lowest energy state $|\psi_{1, 0}\rangle$}

For the oscillator $a_{1, 0}^\dagger$ in (\ref{2ptfna}), we see that it is only the term involving the wave function $R^{(1, 0)}_-$ that contributes in the mode expansion. This wave function can be read out from \ref{low lying states} and is given by 
%\begin{align*}
%& A^{1,0}_-(t,r,\varphi)= \left[a_{1,0;\omega,k} R^{1,0}_-(r) e^{i m \varphi}e^{-i \omega t}+a^\dagger_{1,0;\omega,k} (R^{1,0}_-)^\star(r) e^{-i m \varphi}e^{i \omega t}\right].\\
%\end{align*}
%In ground state  $A_t(t,r,\varphi)=-A_\varphi(t,r,\varphi)$, hence this is the only non-vanishing lightcone field. 
%The radial part $R^{1,0}_-(r)$ can be obtained from \eqref{eq: Am r}.
\begin{align}\label{eq: wavefn m=1,n=0}
{R}^{(1,0)}_-(r)= C_{-}^{1,0}r \left(r^2+1\right)^{-\frac{M}{2}-\frac{1}{2}} =\sqrt{\frac{M+1}{\pi}} r \left(r^2+1\right)^{-\frac{M}{2}-\frac{1}{2}}.
\end{align}
Evaluating the two point function (\ref{2ptfna}) we obtain 
%We compute the following two-pt function in the boundary using \eqref{eq: wavefn m=1,n=0} and the standard commutation relation.
\begin{eqnarray}\label{eq: 2-pt bogo m=1 n=0}
\lim_{r \rightarrow \infty} \langle 0| A_-(t,r,\varphi) a_{m,n}^\dagger |0\rangle
&=& 
%& \lim_{r \rightarrow \infty}\nonumber \langle 0| A^{1,0}_-(t,r,\varphi) a_{1,0;\omega,k}^\dagger |0\rangle\\
\lim_{r\rightarrow \infty} \langle 0| a_{1,0} {R}^{1,0}_-(r) e^{i \varphi} e^{-i(M+1)t}a^\dagger_{1,0} |0 \rangle\\ 
\nonumber 
& =&\sqrt{\frac{M+1}{\pi}}\frac{e^{i\varphi} e^{-i (M+1) t}}{r^M}.
\end{eqnarray}
Now we need to evaluate the same two-point function in BTZ Rindler space. For that, we first transform the vector using the coordinate transformation  (\ref{coordchange}). This is given by 
{\small 
\begin{eqnarray}\label{eq: global to rindler vector}
\nonumber \begin{pmatrix}
A_r(t,r,\varphi)\\
A_t(t,r,\varphi)\\
A_\varphi (t,r,\varphi)\\
\end{pmatrix}
& =&\mathcal{J}^{-1}_{t,r,\varphi}(\tau,\rho,x)
\begin{pmatrix}
A_\rho'(\tau,\rho,x)\\
A_\tau'(\tau,\rho,x)\\
A_x' (\tau,\rho,x)\\
\end{pmatrix}\\
&=& \begin{pmatrix}
\frac{\partial r(\tau,\rho,x)}{\partial \rho} & \frac{\partial t(\tau,\rho,x)}{\partial \rho} & \frac{\partial \varphi(\tau,\rho,x)}{\partial \rho}\\
\frac{\partial r(\tau,\rho,x)}{\partial \tau} & \frac{\partial t(\tau,\rho,x)}{\partial \tau} & \frac{\partial \varphi(\tau,\rho,x)}{\partial \tau}\\
\frac{\partial r(\tau,\rho,x)}{\partial x} & \frac{\partial t(\tau,\rho,x)}{\partial x} & \frac{\partial \varphi(\tau,\rho,x)}{\partial x}
\end{pmatrix}^{-1}
\begin{pmatrix}
A_\rho'(\tau,\rho,x)\\
A_\tau'(\tau,\rho,x)\\
A_x'(\tau,\rho,x)\\
\end{pmatrix}.
\end{eqnarray}}
Here we have used the superscript $'$ to emphasize the fact that we are in the Rindler-BTZ coordinates. As the LHS of \eqref{eq: 2-pt bogo m=1 n=0} is computed near the boundary, we expand the elements of the Jacobian $\mathcal{J}_{t,r,\varphi}(\tau,\rho,x)$ in  \eqref{eq: global to rindler vector} around $\rho \rightarrow \infty$ and consider the term up to $\mathcal{O}(\rho^1)$. The rest of sub-leading terms are of $\mathcal{O}(\frac{1}{\rho})$
or $\mathcal{O}(\frac{1}{\rho^2})$, hence vanishes near the boundary. We write down the explicit form of \eqref{eq: global to rindler vector} below.
\footnotesize
\begin{align}
\nonumber
\lim_{r\rightarrow \infty}  & \begin{pmatrix}
A_r(t,r,\varphi)\\
A_t(t,r,\varphi)\\
A_\varphi (t,r,\varphi)\\
\end{pmatrix}\\
&=
\lim_{\rho\rightarrow\infty} \begin {pmatrix}
J(\tau,x,\eta) & 0 & 0 \\ 
-\rho\cosh\eta\sinh\tau\cosh x &\cosh\eta\cosh\tau\cosh x + \sinh\eta & \cosh\eta \sinh\tau\sinh x \\
- \rho\cosh\eta \cosh\tau \sinh x & \cosh\eta \sinh\tau \sinh x & \
\cosh \eta \cosh\tau\cosh x + \sinh\eta \\
\end{pmatrix}
\begin{pmatrix}
A_\rho'(\tau,\rho,x)\\
A_\tau'(\tau,\rho,x)\\
A_x' (\tau,\rho,x)\\
\end{pmatrix}.
\end{align}
\normalsize
where  
{\footnotesize
\begin{eqnarray}
J(\tau,x,\eta)=\frac {1} {\sqrt { ( \cosh \eta \cosh \tau + \sinh\eta \cosh x)^2 
+  \sinh^2x}}.
\end{eqnarray}}
The relation between the light-cone field components  between two coordinates is given by 
\begin{eqnarray}\label{eq: global to rindler lightcone}
\lim_{r\rightarrow \infty}A_+(t,r,\varphi)  &=& \lim_{\rho\rightarrow\infty} \Big\{ 
A_+'(\tau,\rho,x) \big[\cosh \eta \cosh(\tau+x)+\sinh \eta\big]- A_\rho'(\tau,\rho,x) \rho \cosh \eta \sinh(\tau+x) \Big\}, 
\nonumber \\
 \nonumber 
 \lim_{r\rightarrow \infty}
 A_-(t,r,\varphi)  
&= & \lim_{\rho\rightarrow\infty} \Big\{ 
A_{-}' (\tau,\rho,x) \big[\cosh \eta \cosh(\tau-x)+\sinh \eta \big]-A_\rho'(\tau,\rho,x) \rho \cosh \eta \sinh(\tau-x)
\Big\} .
\\
\end{eqnarray}
From the solutions in (\ref{aminuswave}) and using the Chern-Simons equation for $A_\rho$ in (\ref{cseqrin}), we see that 
\begin{eqnarray}
\lim_{\rho\rightarrow \infty }
A_-'(\tau, \rho, x) \sim \rho^{-M},  \qquad\qquad  \lim_{\rho \rightarrow \infty}  \rho A_\rho'( \tau, \rho,  x)  \sim 
\rho^{-M-2}.
\end{eqnarray}
This implies that we have 
\begin{align}
\lim_{r\rightarrow\infty} 
A_-(t,r,\varphi)
= \lim_{\rho\rightarrow\infty} A_-'(\tau, \rho, x) 
\big[\cosh \eta \cosh(\tau-x)+\sinh \eta \big].
\end{align}
Substituting this relation in the left hand side of (\ref{eq: 2-pt bogo m=1 n=0}), we obtain
\begin{align}\label{eq: 2-pt bogo m=1 n=0 final}
&\lim_{r \rightarrow \infty} \langle 0| A^{1,0}_-(t,r,\varphi) a_{1,0}^\dagger |0 \rangle 
%& \nonumber = \lim_{\rho \rightarrow \infty}-\langle 0|(A^{1,0}_\tau(\tau,\rho,x)-A^{1,0}_x(\tau,\rho,x))\left[\cosh \eta \cosh(\tau-x)+\sinh \eta\right]a_{1,0}^\dagger |0 \rangle\\
\\ \nonumber
& =\lim_{\rho \rightarrow \infty}\langle 0|(A^\prime_-(\tau,\rho,x)\big[\cosh \eta \cosh(\tau-x)+\sinh \eta\big]
a_{1,0}^\dagger |0 \rangle 
\\  \nonumber
& =\lim_{r \rightarrow \infty} \sqrt{\frac{M+1}{\pi}}\frac{e^{i\varphi(\tau,\rho,x)} e^{-i (M+1) t(\tau,\rho,x)}}{r(\tau,\rho,x)^M}.
\end{align}
This completes rewriting the two-point function \eqref{eq: 2-pt bogo m=1 n=0} in terms of Rindler coordinates. We substitute the expression of $A_-^\prime(\tau,\rho,x)$ using \eqref{rinamexp} and the commutation relation (\ref{rincomrel})  in \eqref{eq: 2-pt bogo m=1 n=0 final} to obtain
{\small \begin{align}\label{eq: bbgv eq}
%& \nonumber \lim_{\rho \rightarrow \infty}-\langle 0|(A^\prime_-)^{(1,0)}(\tau,\rho,x)\left[\cosh \eta \cosh(\tau-x)+\sinh \eta\right]a_{1,0;\omega,k}^\dagger |0 \rangle= \lim_{\rho \rightarrow \infty} \sqrt{\frac{M+1}{\pi}}\frac{e^{i\varphi(\tau,\rho,x)} e^{-i (M+1) t(\tau,\rho,x)}}{r(\tau,\rho,x)^M}\\
&\nonumber \Bigg\{  \lim_{\rho \rightarrow \infty}\langle 0| \big[ \cosh \eta \cosh(\tau-x)+\sinh \eta\big] \sum_{\omega,k}\left[ b_{\omega,k} A_-^{(\omega, k )} e^{i k x-i\omega \tau} +  b^\dagger_{1,0;\omega,k}  A_-^{(\omega, k ) *} e^{-i k x+i\omega \tau}\right]\\
&\times \sum_{\omega^\prime,k^\prime}\left[(1-e^{-2\pi \omega^\prime})\alpha_{1,0;\omega^\prime,k^\prime}b^\dagger_{\omega^\prime,k^\prime}+(1-e^{2\pi \omega^\prime})\beta_{1, 0;\omega^\prime,k^\prime} b_{\omega^\prime,k^\prime} \right]|0 \rangle \Bigg\} \nonumber \\
& 
\qquad\qquad =\lim_{\rho \rightarrow \infty}\sqrt{\frac{M+1}{\pi} } 
\frac{1}{r(\tau,\rho,x)^M} 
e^{i\varphi(\tau,\rho,x)} e^{-i (M+1) t(\tau,\rho,x)}.
\end{align} }
%where 
%\begin{align}\label{eq: radial wavefn rindler m1n0}
%f_{\omega,k}(\rho)= \rho^{-M} \left(\frac{1}{\rho^2}-1\right)^{-i \frac{\omega}{2}} {}_2F_1\left[ \frac{1}{2}(M-i(k+\omega)),\frac{1}{2}(M+i(k-\omega)),M;
%\frac{1}{\rho^2}\right].
%\end{align}
The integral over $\omega$ and $k$ has been replaced with sums as defined in (\ref{notatsum}). Furthermore, all the oscillators involved are oscillators that act on the right Rindler vacuum. 
%a short hand notation $\sum_{\omega,k}$ and the subscript $(1,0)$ has been removed from the \bbgv coefficients and creation(annihilation) operators for convenience.
To simplify \eqref{eq: bbgv eq}, we use the following non-zero correlation functions
\begin{subequations}
\begin{align}
\langle 0| b_{\omega,k} b^\dagger_{\omega^\prime,k^\prime}|0 \rangle=\frac{(2 \pi)^2 e^{2 \pi \omega}}{e^{2 \pi \omega}-1}\delta(\omega-\omega^\prime)\delta(k-k^\prime),\\ \nonumber
\langle 0| b^\dagger_{\omega,k} b_{\omega^\prime,k^\prime}|0 \rangle=\frac{(2 \pi)^2}{e^{2 \pi \omega}-1}\delta(\omega-\omega^\prime)\delta(k-k^\prime).
\end{align}
\end{subequations}
Here, it is understood that the global $AdS_3$ vacuum has been written as the thermo-field double using (\ref{eq: therofield double}). Now the relation (\ref{eq: bbgv eq}) becomes 
{\small \begin{align}\label{eq: bbgv eq2}
 &\lim_{\rho \rightarrow \infty} \Bigg\{ r(\tau,\rho,x)^M\langle 0|\left[\cosh \eta \cosh(\tau-x)+\sinh \eta\right]
\\ \nonumber
& \qquad\qquad \times  \sum_{\omega,k}\left[e^{-i \omega \tau}e^{i kx} A_-^{(\omega,k)} \alpha^*_{1,0;\omega,k}-e^{i \omega \tau}e^{-i kx}
 A_-^{(\omega,k)*} \beta_{1,0; \omega,k}\right] \Bigg\} \\ \nonumber
=& \lim_{\rho \rightarrow \infty} \sqrt{\frac{M+1}{\pi}}e^{i\varphi(\tau,\rho,x)} e^{-i (M+1) t(\tau,\rho,x)}. 
\end{align}}
From the coordinate transformation (\ref{coordchange}), we can obtain the following relations
\begin{align}\label{eq: global coordinate boundary limit}
& \lim_{\rho \rightarrow \infty} r(\tau,\rho,x)= \lim _{\rho \rightarrow \infty} \rho^M \left[ \sinh^2 x+(\cosh \eta \cosh \tau+\sinh \eta \cosh x)^2 \right]^{M/2},\\ \nonumber
& \lim_{\rho \rightarrow \infty} t(\tau,\rho,x)= \arctan \left[ \frac{\sinh \tau}{\cosh x \cosh \eta+ \cosh \tau \sinh \eta} \right]\equiv t(\tau,x),\\ \nonumber
& \lim_{\rho \rightarrow \infty}\varphi(\tau,\rho,x)= \frac{\theta}{2}+\arctan\left[\frac{\sinh x}{\cosh \tau \cosh \eta+ \cosh x \sinh \eta}  \right]\equiv \varphi(\tau,x),
\end{align}
and the radial wavefunction \eqref{aminusexpmain} satisfies the following relation
\begin{align}
\lim_{\rho \rightarrow \infty} \rho^M A_-^{(\omega,k)}(\rho)= N_{\omega, k } .
\end{align}
Substituting these relations in \eqref{eq: bbgv eq2} result in 
{\small \begin{align}
& \int \frac{d \omega d k}{(2\pi)^2}\Big[e^{-i \omega \tau}e^{i kx} N_{\omega, k }
\alpha^*_{1,0;\omega,k}-e^{i \omega \tau}e^{-i kx}\beta_{1,0;\omega,k} N_{\omega, k }^* \Big] \\  \nonumber 
& =\sqrt{\frac{M+1}{\pi}}\frac{e^{i\varphi(\tau,x)} e^{-i (M+1) t(\tau,x)}}{\left[\cosh \eta \cosh(\tau-x)+\sinh \eta\right]}\frac{1}{\left[ \sinh^2 x+(\cosh \eta \cosh \tau+\sinh \eta \cosh x)^2 \right]^{M/2}}.
\end{align} }
The integral over $\omega$ runs from $0$ to $\infty$, while the integral over $k$ runs from $-\infty$ to $\infty$. Let us define 
{\small \begin{align}\label{eq: B10}
\mathcal{B}_{1,0}(\tau,x)=\sqrt{\frac{M+1}{\pi}}\frac{e^{i\varphi(\tau,x)} e^{-i (M+1) t(\tau,x)}}{\left[ \sinh^2 x+(\cosh \eta \cosh \tau+\sinh \eta \cosh x)^2 \right]^{M/2}} \frac{1}{\left[\cosh \eta \cosh(\tau-x)+\sinh \eta\right]}.
\end{align} }
Performing the Fourier transform of LHS in \eqref{eq: bbgv eq2}, we obtain the expression of \bbgv coefficients.
\begin{eqnarray}  \label{ftbg}
 \alpha_{\omega,k} &=& \frac{1}{N_{\omega,k}^*}\int_{\infty}^{\infty} d\tau dx e^{-i \omega \tau}e^{i kx} \mathcal{B}_{1,0}^*(\tau,x),\\ \nonumber
 \beta_{\omega,k} &=& -\frac{1}{N_{\omega,k}^*}\int_{\infty}^{\infty} d\tau dx e^{-i \omega \tau}e^{i kx} \mathcal{B}_{1,0}(\tau,x).
\end{eqnarray}
We now provide the steps to carry out the Fourier transform in (\ref{ftbg}). First, we change the coordinate to a light-cone variable in order to evaluate the integral.
\begin{align}
x^+=\frac{x+\tau}{2}, \qquad\qquad \ x^-=\frac{x-\tau}{2}.
\end{align}
Second, we proceed to simplify the expression of $\mathcal{B}_{1,0}(\tau,x)$ in \eqref{eq: B10} with the following identities, 
{\small 
\begin{eqnarray}\label{eq: identities bogo}
& &\nonumber \arctan x=\frac{1}{2 i} \log \left( \frac{1+i x}{1-i x}\right),\\
& &\nonumber\sinh^2 x+(\cosh \eta \, \cosh \tau + \sinh \eta \, \cosh x  )^2= \sinh^2 \tau + (\cosh x \cosh \eta + \cosh \tau \sinh \eta)^2, \\
& &\nonumber \notag \cosh x \cosh \eta+ \cosh \tau \sinh \eta -i \sinh \tau \\
& & \nonumber \qquad\qquad = \frac{e^{-\eta}}{4} e^{-x^+-x^-}(e^\eta+ e^{\eta+2x^+}-i e^{2x^+}+i)(e^\eta+ e^{\eta+2x^-}+i e^{2x^-}-i),\\
& &\nonumber \notag \cosh \tau \cosh \eta+ \cosh x \sinh \eta +i \sinh x \\
&  &\nonumber\qquad\qquad= \frac{e^{-\eta}}{4} e^{-x^++x^-}(e^\eta+ e^{\eta+2x^+}+i e^{2x^+}-i)(e^\eta+ e^{\eta-2x^-}-i e^{-2x^-}+i), \\ \nonumber
%\end{align}
%The factor $\frac{1}{\left[\cosh \eta \cosh(\tau-x)+\sinh \eta\right]}$ can be manipulated as follows.
%\begin{align}
& &\nonumber \frac{1}{\left[\cosh \eta \cosh(\tau-x)+\sinh \eta\right]}\\
\nonumber &  &  \qquad\qquad =\frac{4}{e^{\eta}(e^{x^-}+ e^{-x^-})^2+e^{-\eta}(e^{x^-}- e^{-x^-})^2 }\\
%\nonumber = & \frac{4 e^{\eta}}{\left \lbrace e^\eta(e^{x^-}+e^{-x^-})+i(e^{x^-}-e^{-x^-})\right \rbrace \left \lbrace e^\eta(e^{x^-}+e^{-x^-})-i(e^{x^-}-e^{-x^-})\right \rbrace}\\
 & &  \qquad\qquad =  \frac{4 e^{\eta} e^{2 x^-}}{\left \lbrace e^{\eta}(e^{2 x^-}+1)+i(e^{2 x^-}-1)\right \rbrace \left \lbrace e^{\eta}(e^{2 x^-}+1)-i(e^{2 x^-}-1)\right \rbrace }.
\end{eqnarray} }
Using these identities, we decompose $\mathcal{B}_{1,0}(\tau,x)$ in \eqref{eq: B10} in $x^+$ and $x^-$ dependent parts.
\begin{align}\label{eq: B10 decomposed}
\mathcal{B}_{1,0}(x^+,x^-)= \sqrt{\frac{M+1}{\pi}}2^{2(M+1)}e^{\eta(M+1)}e^{\frac{i\theta}{2}}B_{1,0}^+(x^+) B_{1,0}^-(x^-),
\end{align}
where
{\small 
\begin{align} \label{eq: bogo B expression m=1 n=0}
\mathcal{B}^+_{1,0}(x^+)= &\left[\frac{e^{2 x^{+}}}{(e^\eta-i+ e^{2 x^+}(e^{\eta}+i) )^{2}}\right]^{\frac{M}{2}},\\
\nonumber \mathcal{B}^-_{1,0}(x^-)= & \left[\frac{e^{2 x^{-}}}{(e^\eta+i+ e^{2 x^-}(e^{\eta}-i) )^{2}}\right]^{\frac{M}{2}} \times \frac{e^{2x^{-}}(e^{\eta}+i)+e^{\eta}-i}{e^{2x^{-}}(e^{\eta}-i)+e^{\eta}+i}\\
& \nonumber \times \frac{e^{2 x^-}}{\left \lbrace e^{\eta}(e^{2 x^-}+1)+i(e^{2 x^-}-1)\right \rbrace \left \lbrace e^{\eta}(e^{2 x^-}+1)-i(e^{2 x^-}-1)\right \rbrace }\\ \nonumber
= & \left[\frac{e^{2 x^{-}}}{(e^\eta+i+ e^{2 x^-}(e^{\eta}-i) )^{2}}\right]^{\frac{M}{2}+1}.
\end{align} }
From the above expression, we note that we have a left propagating light-cone mode with dimension $\frac{M}{2}$ and a right propagating mode with dimension $\frac{M}{2}+1$. Therefore $\beta_{1,0;\omega,k}$ reduces to
{\small \begin{eqnarray}\label{eq: beta m=1 n=0a}
\beta_{1,0;\omega,k} &=&  2 \sqrt{\frac{M+1}{\pi}}2^{2(M+1)}e^{\eta(M+1)}e^{\frac{i\theta}{2}}\frac{1}{N^*_{\omega,k}} \\ \nonumber
& & \qquad \qquad \times \int_{-\infty}^{\infty}B_{1,0}^+(x^+) e^{i(k-\omega) x^+} dx^+ \int_{\infty}^{\infty} B_{1,0}^-(x^-) e^{i(k+\omega)x^-} dx^-\\ \nonumber 
  &  & = 2 \sqrt{\frac{M+1}{\pi}} \frac{1}{N_{\omega k }^* }2^{2(M+1)}e^{\eta(M+1)}e^{\frac{i\theta}{2}} I_1 \times I_2.
\end{eqnarray}
The factor of 2 is due to the Jacobian. We substitute $e^{2 x^+}=p$ and use the integral representation of the beta function 
\begin{eqnarray}
\int_0^\infty  \frac{u^{m}}{(1+u)^{m+n+2}}= \frac{\Gamma[m+1]\Gamma[n+1]}{\Gamma[m+n+2]}, 
\end{eqnarray}
to compute the integrals $I_{1}, I_{2}$. 
{\small 
\begin{align}
I_1=& \frac{1}{2}\frac{1}{(e^\eta-i)^M}\left( \frac{e^\eta-i}{e^\eta+i}\right)^{\frac{M}{2}+\frac{i(k-\omega)}{2}}\times \frac{\Gamma\left(\frac{M}{2}+\frac{i(k-\omega)}{2}\right)\Gamma\left(\frac{M}{2}-\frac{i(k-\omega)}{2}\right)}{\Gamma\left(M\right)},\\ \nonumber
I_2= & \frac{1}{2}\frac{1}{(e^{\eta}+i)^{M+2}}\left(\frac{e^\eta+i}{e^\eta-i)}\right)^{\frac{M}{2}+1+\frac{i(k+\omega)}{2}}\frac{\Gamma\left(\frac{M}{2}+1+\frac{i(k+\omega)}{2}\right)\Gamma\left(\frac{M}{2}+1-\frac{i(k+\omega)}{2}\right)}{\Gamma(M+2)}.
\end{align} }
Substituting the above integrals in \eqref{eq: beta m=1 n=0a} we obtain
{\small \begin{align}\label{eq: beta m=1 n=0}
\beta_{1,0;\omega,k}= \nonumber  & -\ha \frac{1}{N_{\omega,k}^*}\sqrt{\frac{M+1}{\pi}}2^{2(M+1)}e^{\eta(M+1)}e^{\frac{i\theta}{2}}\frac{1}{(e^{2 \eta}+1)^{M+1}}\left( \frac{e^\eta+i}{e^\eta-i}\right)^{ i \omega}\frac{1}{\Gamma\left(M\right) \Gamma\left(M+2\right)}\\
& \times \Gamma\left(\frac{M}{2}+\frac{i(k-\omega)}{2}\right)\Gamma\left(\frac{M}{2}-\frac{i(k-\omega)}{2}\right)\Gamma\left(\frac{M}{2}+1+\frac{i(k+\omega)}{2}\right)\Gamma\left(\frac{M}{2}+1-\frac{i(k+\omega)}{2}\right).
\end{align} }
Similarly one can compute $\alpha_{1,0;\omega,k}$ as
{\small \begin{align}\label{eq: alpha m=1 n=0}
\alpha_{1,0;\omega,k}= \nonumber  & \ha \frac{1}{N_{\omega,k}^*}\sqrt{\frac{M+1}{\pi}}2^{2(M+1)}e^{\eta(M+1)}e^{-\frac{i\theta}{2}}\frac{1}{(e^{2 \eta}+1)^{M+1}}\left( \frac{e^\eta-i}{e^\eta+i}\right)^{i \omega}\frac{1}{\Gamma\left(M\right)\Gamma\left(M+2\right)}\\
& \times \Gamma\left(\frac{M}{2}+\frac{i(k-\omega)}{2}\right)\Gamma\left(\frac{M}{2}-\frac{i(k-\omega)}{2}\right)\Gamma\left(\frac{M}{2}+1+\frac{i(k+\omega)}{2}\right)\Gamma\left(\frac{M}{2}+1-\frac{i(k+\omega)}{2}\right),
\end{align} }
where the normalization constant $N_{\omega,k}$ is given in \eqref{aminusexpmain}. From (\ref{largeeta}), we see that the short interval limit is obtained by taking the limits 
\begin{align} \label{smallangle}
\theta \rightarrow 0, \qquad \qquad \eta\rightarrow \infty.
\end{align}
Substituting the value of normalization $N^*_{\om,k}$ from \eqref{aminusexpmain}, we obtain the leading behaviour of the \bbgv coefficients of the lowest energy state in the short interval limit.
{\small \begin{align}\label{eq: alpha m=1 n=0 large eta}
& \notag \lim_{\eta \rightarrow \infty}\alpha_{1,0;\om,k}=\\
 & \notag \ha  \sqrt{\frac{M}{\om}}\sqrt{\frac{M+1}{\pi}}2^{(M+1)}(\cosh \eta)^{-M+1} \\ \nonumber 
 &\qquad\qquad\times  \frac{1}{\Gamma\left(M+2\right)}\left| \Gamma(1+ i \om) \Gamma\left(\frac{M}{2}+\frac{i(k-\omega)}{2}\right) \Gamma\left(\frac{M}{2}+1+\frac{i(k+\omega)}{2}\right)\right|\\\
& =(\cosh \eta)^{-M+1}F(\om,k).
\end{align} }
$F(\om,k)$ is defined as
{\small \begin{align}\label{eq: def F}
F(\om,k)=\ha \sqrt{\frac{M}{\om}}\sqrt{\frac{M+1}{\pi}}2^{(M+1)}\frac{1}{\Gamma\left(M+2\right)}\left| \Gamma(1+ i \om) \Gamma\left(\frac{M}{2}+\frac{i(k-\omega)}{2}\right) \Gamma\left(\frac{M}{2}+1+\frac{i(k+\omega)}{2}\right)\right|.
\end{align} }
Similarly we obtain
\begin{align}\label{eq: beta m=1 n=0 large eta}
\lim_{\eta \rightarrow \infty} \beta_{1,0,\om,k}=  - (\cosh \eta)^{-M+1}F(\om,k).
\end{align}

\subsection*{Tower of states: $|\psi_{m, 0} \rangle, m \geq 1$.}

We now evaluate the Bogoliubov coefficients relating the descendants $|\psi_{m, 0}\rangle$ to excitations in BTZ. The radial wave functions can be obtained from  \eqref{aminusm}, with the normalization from (\ref{neqzerom})
\begin{align}
R_-^{(m,0)}(r)=\sqrt{\frac{1}{\pi} \frac{\Gamma(M+m+1)}{\Gamma(m)\Gamma(M+1)}} r^m (1+r^2)^{-\ha(m+M)}, 
\qquad\qquad m \geq 1.
\end{align}
%We have replaced $|m|$ with $m$ as we have noted for $n=0$, $m<1$ quantum numbers are not allowed. 
Evaluating the two-point function, we obtain
\begin{align}
\lim_{r \rightarrow \infty} r^M \langle 0 | A_-^{(m,n)}(t,r,\varphi) a_{m,n}^\dagger|0 \rangle= \sqrt{\frac{1}{\pi} \frac{\Gamma[M+m+1]}{\Gamma[m]\Gamma[M+1]}} e^{-i(M+m)t}e^{i m \varphi}.
\end{align}
Proceeding on similar lines as done for the lowest energy states in \eqref{eq: B10 decomposed}, we decompose $\mathcal{B}_{m,0}(\tau,x)$ using light-cone coordinates $x^+$ and $x^-$. This results in 
\begin{align}\label{eq: Bm0 decomposed}
\mathcal{B}_{m,0}(x^+,x^-)= \sqrt{\frac{1}{\pi} \frac{\Gamma(M+m+1)}{\Gamma(m)\Gamma(M+1)}}2^{2(M+1)}e^{\eta(M+1)}e^{\frac{i\theta}{2}}B_{m,0}^+(x^+) B_{m,0}^-(x^-),
\end{align}
where
\begin{align} \label{eq: bogo B n=0}
& \mathcal{B}^+_{m,0}(x^+)= \left[\frac{e^{2 x^{+}}}{(e^\eta-i+ e^{2 x^+}(e^{\eta}+i) )^{2}}\right]^{\frac{M}{2}},\\
\nonumber & \mathcal{B}^-_{m,0}(x^-)= \left[\frac{e^{2 x^{-}}}{(e^\eta+i+ e^{2 x^-}(e^{\eta}-i) )^{2}}\right]^{\frac{M}{2}+1} \left[\frac{(e^\eta-i)+ e^{2 x^-}(e^{\eta}+i)}{(e^\eta+i)+ e^{2 x^-}(e^{\eta}-i)}\right]^{m-1} .
\end{align}
We note in large $\eta$ limit or the short interval limit, 
\begin{align} \label{limietab}
\left[\frac{(e^\eta-i)+ e^{2 x^-}(e^\eta+i)}{(e^\eta+i)+ e^{2 x^-}(e^\eta-i)}\right] =\left[\frac{(e^\eta+i)+ e^{2 x^+}(e^\eta-i)}{(e^\eta-i)+ e^{2 x^+}(e^\eta+i)}\right] \sim 1+ \mathcal{O}(e^{-\eta}).
\end{align}
Therefore, we can use the results obtained for the Bogoliubov coefficients of the lowest energy states in this limit. This leads to the following expression of \bbgv coefficients \footnote{We have taken the large $\eta$ limit before the Fourier transform of $\mathcal{B}^-_{m,0}(x^-)$. We have seen, case by case, that performing the Fourier transform first and then taking the large $\eta$ limit yields the same result. }
\begin{align}
& \notag \beta_{m,0;\om,k}= -\alpha_{m,0;\om,k},\\ 
\lim_{\eta\rightarrow\infty} \alpha_{m,0;\om,k}= \notag & \ha \frac{1}{N^*_{\om,k}}\sqrt{\frac{1}{\pi} \frac{\Gamma[M+m+1]}{\Gamma[m]\Gamma[M+1]}} 2^{(M+1)}(\cosh \eta)^{-M+1}\frac{1}{\Gamma\left(M+2\right)}\\
& \times \left| \Gamma(1+ i \om) \Gamma\left(\frac{M}{2}+\frac{i(k-\omega)}{2}\right) \Gamma\left(\frac{M}{2}+1+\frac{i(k+\omega)}{2}\right)\right|.
\end{align}
This implies that in the large $\eta$ limit, the  Bogoliubov coefficients  $\alpha_{m,0;\omega,k}, \beta_{m,0;\omega,k}$ 
are proportional to the  Bogoliubov coefficients of the lowest energy state:  $\alpha_{1,0;\omega,k}, \beta_{1,0;\omega,k}$ respectively. This phenomenon was also observed for descendants of the scalar field in \cite{Chowdhury:2024fpd}. Thus we have 
\begin{eqnarray}
\label{ratiohigherb} 
\lim_{\eta\rightarrow \infty} \frac{ \alpha_{m,n;\omega,k} }{ \alpha_{1,0;\omega,k} }
= \frac{ \beta_{m,n;\omega,k} }{\beta_{1,0;\omega,k}} =\sqrt{\frac{\Gamma(M+m+1)}{\Gamma(m)\Gamma(M+2)} }. 
\end{eqnarray} 

\subsection*{Bogoliubov coefficients in the zero mode sector}

In this section, we evaluate the Bogoliubov coefficient in the zero modes sector. This coefficient contributes to the correction to entanglement entropy due to the edge modes. Again, we start with the 2-point function 
\begin{eqnarray}
\lim_{r\rightarrow\infty} \langle 0 | A_-( t , r, \varphi) a_{1, 0}^\dagger |0\rangle\Big|_{\omega =0}.
\end{eqnarray}
Going through similar steps as leading up to equation (\ref{eq: bbgv eq}), we arrive at 
{\small \begin{align}\label{eq: bbgv eqz}
&\nonumber \Bigg\{  \lim_{\rho \rightarrow \infty}\langle E \sum_{k, I }\left[ b_{0,k, I } A_{-, I }^{(0, k )} e^{i k x} +  b^\dagger_{0, k, I }  A_{-, I}^{(0, k ) *} e^{-i k x }\right]\\
&\times \sum_{k^\prime}\left[ \hat\alpha_{1,0;  - k^\prime}^* ( \hat a_{k', R} - \hat a_{-k', L }) \right]
|E \rangle \Bigg\} \nonumber \\
& 
\qquad\qquad = \frac{1}{L} \int_{-\frac{L}{2} }^{\frac{L}{2}}   d\tau   \times \Bigg\{ \\ \nonumber 
&   \qquad\qquad \lim_{\rho \rightarrow \infty}\sqrt{\frac{M+1}{\pi} } 
\frac{1}{r(\tau,\rho,x)^M} 
e^{i\varphi(\tau,\rho,x)} e^{-i (M+1) t(\tau,\rho,x)} \times \frac{1}{\big[ \cosh \eta \cosh(\tau-x)+\sinh \eta\big] } \Bigg\}.
\end{align} }
Recall that we had denoted the global coordinate operators and \bbgv coefficients in the edge sector using \textit{hat} symbol in the main text. The sum over $I$ runs over $L, R$ and we have used (\ref{globalzerocon}) and (\ref{statebc1}) to write the oscillator $a^\dagger_{1, 0}$. We projected to the zero frequency sector by integrating $\tau$. Since there is a cut-off in $\tau$, the integral on the LHS results in the factor 
\begin{eqnarray} \label{defL}
 L = | \log (2\epsilon) |, 
\end{eqnarray}
which we brought to the denominator of the RHS. We write the oscillators $b$ in terms of the momentum and position operators using (\ref{posmomop}) and the 
%\begin{eqnarray}
%b_k  = \frac{ q_{-k} - i a_k}{\sqrt{2} }, \qquad  b_{-k}^\dagger  = \frac{ q_{-k} + i a_{k} }{\sqrt{2}}
%\end{eqnarray}
%Substituting this and using
reality property of the wave functions (\ref{expansionsh})
\begin{eqnarray}
A_{-, I}^{(0, k ) *} = A_{-, I}^{(0, - k ) * }=A_{-, I}^{(0, k )} =A_{-, I}^{(0, -k ) },
\end{eqnarray}
to get  
{\small \begin{align}\label{eq: bbgv eqzz}
%& \nonumber \lim_{\rho \rightarrow \infty}-\langle 0|(A^\prime_-)^{(1,0)}(\tau,\rho,x)\left[\cosh \eta \cosh(\tau-x)+\sinh \eta\right]a_{1,0;\omega,k}^\dagger |0 \rangle= \lim_{\rho \rightarrow \infty} \sqrt{\frac{M+1}{\pi}}\frac{e^{i\varphi(\tau,\rho,x)} e^{-i (M+1) t(\tau,\rho,x)}}{r(\tau,\rho,x)^M}\\
&\nonumber \Bigg\{  \lim_{\rho \rightarrow \infty}\langle E 
\sum_{k, I }\left[ \sqrt{2} \hat q_{0,k, I } A_{-, I }^{(0, k )} e^{- i k x}  \right]\\
&\times \sum_{k^\prime}\left[ \hat \alpha_{1,0;  - k^\prime}^* ( \hat a_{k', R} - \hat a_{-k', L }) \right]
|E \rangle \Bigg\} \nonumber \\
& 
\qquad\qquad = \frac{1}{L} \int_{-\frac{L}{2} }^{\frac{L}{2}}   d\tau   \times \Bigg\{ \\ \nonumber 
&   \qquad\qquad \lim_{\rho \rightarrow \infty}\sqrt{\frac{M+1}{\pi} } 
\frac{1}{r(\tau,\rho,x)^M} 
e^{i\varphi(\tau,\rho,x)} e^{-i (M+1) t(\tau,\rho,x)} \times \frac{1}{\big[ \cosh \eta \cosh(\tau-x)+\sinh \eta\big] } \Bigg\}.
\end{align} }
Using the definition of the edge state in (\ref{defedgesup}),  after some straightforward manipulations, we obtain 
\begin{eqnarray}
& &\int \frac{dk}{2\pi} \frac{i}{\sqrt{2}} C_-^{(0, k ) } \hat  \alpha_{1,0;  - k^\prime}^*   e^{-i k x} 
=  \\ \nonumber
& &
\qquad\qquad = \frac{1}{L} \int_{-\frac{L}{2} }^{\frac{L}{2}}   d\tau   \times\Bigg\{ \\ \nonumber 
&&    \qquad\qquad \lim_{\rho \rightarrow \infty}\sqrt{\frac{M+1}{\pi} } 
\frac{1}{r(\tau,\rho,x)^M} 
e^{i\varphi(\tau,\rho,x)} e^{-i (M+1) t(\tau,\rho,x)} \times \frac{1}{\big[ \cosh \eta \cosh(\tau-x)+\sinh \eta\big] } \Bigg\}.
\end{eqnarray}
Note that the factor of $i$ is due to the commutation relation of the momentum and position oscillators, the factor of $1/2$ is because of the additional contribution since the edge states are weighted by the probability  $\sqrt{p( \{ \varepsilon \},  \{ \varepsilon^*  \} )}$. We can now perform the inverse Fourier transform to get 
\begin{eqnarray}
 &&\hat \alpha_{1,0;   k^\prime}^*  =  -i 
 \frac{ \sqrt{2} }{L C_-^{(0, k )  } }\int_{-\frac{L}{2} }^{\frac{L}{2} }   d\tau  dx e^{- i k x}    \times \Bigg\{ \\ \nonumber 
&&    \qquad\qquad \lim_{\rho \rightarrow \infty}\sqrt{\frac{M+1}{\pi} } 
\frac{1}{r(\tau,\rho,x)^M} 
e^{i\varphi(\tau,\rho,x)} e^{-i (M+1) t(\tau,\rho,x)} \times \frac{1}{\big[ \cosh \eta \cosh(\tau-x)+\sinh \eta\big] } \Bigg\}.
\end{eqnarray}
It is important to observe that the combination  $L C_-^{(0, k ) }$ is finite from (\ref{zeromodnorm}) and (\ref{defL}). Therefore, we push the limits of the integration to infinity. Comparing with the earlier calculation in the bulk sector, we see that the integral is just the result of the  $\omega \rightarrow 0$ limit in (\ref{eq: alpha m=1 n=0}), which leads to 
{\small 
\begin{eqnarray}
& &  \hat \alpha_{1,0;  k} = \frac{i }{ \sqrt{2}  L C_-^{(0, k ) } } \times \sqrt{\frac{M+1}{\pi}}2^{(M+1)} ( \cosh\eta)^{-(M+1)} 
 % e^{-\frac{i\theta}{2}}\frac{1}{(e^{2 \eta}+1)^{M+1}}
  \frac{1}{\Gamma\left(M\right)\Gamma\left(M+2\right)} \nonumber \\
& & \times \Gamma\left(\frac{M}{2}+\frac{i k}{2}\right)\Gamma\left(\frac{M}{2}-\frac{i k }{2}\right)\Gamma\left(\frac{M}{2}+1+\frac{i k }{2}\right)\Gamma\left(\frac{M}{2}+1-\frac{i k }{2}\right).
\end{eqnarray} }
Where we have taken the small interval limit as specified in (\ref{smallangle}). 
Substituting in  the value of $C_-^{(0, k ) }$ from (\ref{zeromodnorm}), we obtain
 \begin{eqnarray} \label{finalbgza}
& &  \hat\alpha_{1,0;  k}  =i \sqrt{  \frac{|k| }{2} }   \times 
    \sqrt{\frac{M+1}{\pi}}2^{(M+1)}e^{\eta(M+1)}
   (\cosh \eta)^{-( M+1)}
     \frac{1}{ \Gamma\left(M+2\right)} \nonumber \\
     & &  \qquad\qquad\qquad  \times
     \left|  \Gamma\left(\frac{M}{2}+\frac{i k}{2}\right)  \Gamma\left(\frac{M}{2}+1+\frac{i k }{2}\right) 
     \right|.
 \end{eqnarray}
Observe that this Bogoliubov coefficient is of $O(1)$ when the brick wall cut-off is taken to zero, i.e.,  $\epsilon \rightarrow 0$.

\bibliographystyle{JHEP}
\bibliography{references} 

\end{document}